\documentclass{sig-alternate-05-2015}
\usepackage{amsmath}
\usepackage{amsfonts}
\usepackage{amssymb}
\usepackage{bbm}

\makeatletter
\newcommand\footnoteref[1]{\protected@xdef\@thefnmark{\ref{#1}}\@footnotemark}
\makeatother

\usepackage{algorithm}
\usepackage[noend]{algorithmic}

\usepackage{booktabs}
\usepackage{rotating}
\usepackage{multirow}
\usepackage{array}

\usepackage{graphicx}
\usepackage{subfig}
\usepackage{epstopdf}
\DeclareGraphicsExtensions{.eps,.png,.jpg,.pdf,.tif,.tiff,.ps}
\graphicspath{{./img/}{./}}

\usepackage{url}
\usepackage{hyperref}
\usepackage[svgnames]{xcolor}
\definecolor{navy}{rgb}{.25,.25,.75}
\definecolor{ruby}{rgb}{.70,.20,.20}
\definecolor{magenta}{rgb}{1.0,.0,1.0}
\newcommand{\surl}[1]
{
	\urlstyle{same}\url{#1}
}

\usepackage{soul}
\usepackage[final]{pdfpages}

\newcommand{\eat}[1]{}

\newcommand{\revOLD}[1]{{#1}}
\newcommand{\rev}[1]{{#1}}
\newcommand{\rva}[1]{{#1}}
\newcommand{\secmoveup}{\vspace{-0mm}}

\newcommand{\textmoveup}{\vspace{-0.mm}}         	%{\vspace{-0.08in}}
\newcommand{\eqmoveup}{\vspace{-0.mm}}                %{\vspace{-0.16in}}
\newcommand{\captionmoveup}{\eqmoveup\vspace{-0.mm}}   %{\vspace{-2.4mm}}
\hypersetup
{
pdftitle={Evolution of Privacy Loss in Wikipedia},%
pdfauthor={Marian-Andrei Rizoiu, Lexing Xie, Tiberio Caetano and Manuel Cebrian},%
pdfkeywords={online privacy, de-anonymization, temporal loss of privacy}%
}

\begin{document}
%
% --- compulsory camera-ready info
\setcopyright{rightsretained}
\conferenceinfo{WSDM'16,}{February 22--25, 2016, San Francisco, CA, USA.}
\isbn{978-1-4503-3716-8/16/02.}
\doi{http://dx.doi.org/10.1145/2835776.2835798}
%Authors, replace the red X's with your assigned DOI string.

\clubpenalty=10000
\widowpenalty = 10000
% --- End of Author Metadata ---

\title{Evolution of Privacy Loss in Wikipedia}
%\subtitle{[Extended Abstract]
%\titlenote{A full version of this paper is available as
%\textit{Author's Guide to Preparing ACM SIG Proceedings Using
%\LaTeX$2_\epsilon$\ and BibTeX} at

\numberofauthors{4} %  in this sample file, there are a *total*
\author{
% 1st. author
%% crammed version
%\alignauthor
%Marian-Andrei Rizoiu\titlenote{\label{note1}Corresp. author: Marian-Andrei.Rizoiu@nicta.com.au.}, Lexing Xie, Tiberio Caetano, Manuel Cebrian\\
%       \affaddr{NICTA}\\
%      \affaddr{Australian National University}\\
%       %\affaddr{7 London Circuit}\\
%       %\affaddr{Australia}\\
%% 2nd. author
\alignauthor
Marian-Andrei Rizoiu\titlenote{Corresponding author: marian-andrei@rizoiu.eu}\\
       \affaddr{NICTA, ANU}\\
       %\affaddr{7 London Circuit}\\
%       \affaddr{Canberra, Australia}\\
% 2nd. author
\alignauthor Lexing Xie\\
       \affaddr{ANU, NICTA}\\
       \affaddr{Canberra, Australia}
%       \email{lexing.xie@anu.edu.au}
% use '\and' if you need 'another row' of author names
% 3rd. author
\alignauthor Tiberio Caetano\\
       \affaddr{Ambiata, ANU, UNSW}\\
%       \affaddr{University of New South Wales}\\
       \affaddr{Sydney, Australia}\\
%       \email{Tiberio.Caetano@nicta.com.au}
%\and
% 4th. author
\alignauthor Manuel Cebrian\\
       \affaddr{NICTA}\\
%       \affaddr{University of Melbourne}
       \affaddr{Melbourne, Australia}\\
%       \email{manuel.cebrian@nicta.com.au}
}

% There's nothing stopping you putting the seventh, eighth, etc.
% author on the opening page (as the 'third row') but we ask,
% for aesthetic reasons that you place these 'additional authors'
% in the \additional authors block, viz.
%\additionalauthors{Additional authors: John Smith (The Th{\o}rv{\"a}ld Group,
%email: {\texttt{jsmith@affiliation.org}}) and Julius P.~Kumquat
%(The Kumquat Consortium, email: {\texttt{jpkumquat@consortium.net}}).}
\date{10 Dec 2015}
% Just remember to make sure that the TOTAL number of authors
% is the number that will appear on the first page PLUS the
% number that will appear in the \additionalauthors section.

\maketitle
\begin{abstract}
The cumulative effect of collective \rev{online participation} has an important and adverse impact on individual privacy.
As an online system evolves over time, new digital traces of individual behavior may uncover previously hidden statistical links between an individual's past actions and her private traits.
\eat{Furthermore, this de-anonymization trend may not be observable when analyzing short or medium time-span snapshots of data.}%%LX: this claim is not supported, suggest remove
To quantify this effect, we analyze the evolution of individual privacy loss by studying the edit history of Wikipedia \rva{over 13 years}, including more than 117,523 different users performing 188,805,088 edits.  
We trace each Wikipedia's contributor using apparently harmless features, such as the number of edits performed on predefined broad categories in a given time period (\textit{e.g.} Mathematics, Culture or Nature).  
We show that even at this unspecific level of \eat{identification}\rva{behavior description}, it is possible to use off-the-shelf machine learning algorithms to uncover usually undisclosed \rva{personal} traits, such as gender, religion or education.  
%%LX: private traits here is not well-defined (unlike facebook) suggest using personal traits.
We provide empirical evidence that the prediction accuracy for almost all private traits consistently improves over time.  
\eat{ %%LX: this description does not seem to match the paper content, "Safe" period is unclear+undefined
Moreover, we observe that the system also shows improved prediction for users who participated in the system during ``safe'' periods --- periods where a given individual's private traits could not be \rev{inferred} --- showing that de-anonymization threats are hard to foresee as online systems evolve.
}
\rva{Surprisingly, the prediction performance for users who stopped editing 
after a given time still improves. The activities performed by new users seem to have 
contributed more to this effect than additional activities from existing (but still active) users.}
Insights from this work should help users, system designers, and policy makers understand 
\eat{and debate the design}%%LX: debate is too vague
and \rva{make long-term design choices} in online content creation systems. 
\end{abstract}

%% MAR: no longer needed
%\vspace{1.5mm}
%\noindent {\bf Category and Subject Descriptors}
%CCS - {\sc Security and privacy}-{Human and societal aspects of security and privacy}
%%I.2.6 {\sc Artificial Intelligence} Learning --- Knowledge acquisition

\vspace{0.5mm}

%%LX: tighter spacing for indexing terms above
\eat{
\category{CCS}{Security and privacy}{Human and societal aspects of security and privacy}
%%A category including the fourth, optional field follows...
%\category{D.2.8}{Software Engineering}{Metrics}[complexity measures, performance measures]

}

%!TEX root = WSDM_2015_wikipedia.tex

\secmoveup
\section{Introduction}
\label{sec:intro}

%\textcolor{red}{One of our main contributions not clear enough in the introduction: privacy loss due to TEMPORAL effect in the online environment. 
%Privacy loss (\textit{i.e.}, predicting increasingly well private traits) is due to two factors: 
%i) effect of more data, \textit{e.g.}, more users with disclosed private traits are available for learning and a better model can be inferred from more ``data points'', and 
%ii) the temporal effect, \textit{i.e.}, the privacy loss due to a user's own activity.
%As far as we known, this is the first study to look at the temporal effect at this scale and the relations between the two effects.}

Privacy is a relatively new concern of modern society~\cite{Meyer2015}.
Historically, compromising one's privacy was a difficult task, being mainly achieved by using constant physical surveillance, costly by nature, and easy to thwart. 
\eat{This has changed in recent years, most notably with the advent of the online environment: }
\rva{The advent of the online environment has changed the privacy landscape:}
\rva{users of} social network, blogging, microblogging platforms \rva{willingly or unwillingly} share information with the public and with organizations. %corporations and institutions. 
%This has raised important questions among researchers and policy makers alike. 
The general public are already aware~\cite{Barth-Jones2015,Montjoye2015} that information inadvertently left online can hurt privacy, and researchers showed that~\cite{KOS13} personal attributes can be predicted from these online behavioral traces.
%The general public is already aware~\cite{Barth-Jones2015,Montjoye2015} that leaving information behind in the online environment hurts privacy.
However, the longitudinal change of privacy loss is not well understood -- namely, how information collected over several years can compromise privacy, and how the predictability of private attributes evolve.
In this paper, we set out to answer such challenging questions by curating a novel large-scale behavioral trace dataset, and by measuring the predictability of personal traits in a number of ways.

%But having a principled estimate of the rate at which this happens in different platforms and the factors that influence it are the next step in this scientific endeavor.
\eat{
Therefore, in this paper we study the relatively unexplored problem of \emph{the evolution of privacy loss in online knowledge repositories}. 
We are interested in how the cumulative effect of time affects the ability to infer private data. 
We hypothesize and prove that temporal data creation patterns intrinsically embed precious information about an individual's gender, education level, social status or religious view. 
}

We construct a new dataset from all editing activities in and around Wikipedia -- the largest  encyclopedia to date collaboratively constructed by hundreds of thousands of users. 
We use as input each user aggregated editing activities in a number of broadly defined content and community categories, and the target output are personal traits from \rev{Wikipedia} {\em badges}, i.e., what users choose to disclose on their personal pages. 
This problem and system setting allows us to make several key observations: 
(1) We show that Wikipedia editors' private traits can be inferred using off-the-shelf machine learning algorithms, and that the prediction performance consistently improves over our prediction period from 2007 to 2013. 
In particular, our results include predicting an individual's gender, educational status and religious views.
%%LX: revise result summary re: accuracy, we do not need F1, simplify mention of ePR
\eat{We show that
\rev{certain attributes are particularly sensitive to being uncovered from raw editing data:
the editors gender (F-Measure $F1 = 0.84$, equal Precision Recall $ePR = 0.79$) or their religious views (\textit{muslim} ($F1 = 0.85$, $ePR = 0.9$) and \textit{jewish} ($F1 = 0.94$, $ePR = 0.91$)).% are predicted quite accurately.
}}
\rva{Among the different personal attributes, a subset of showed high prediction accuracy as measured by the equal-precision-recall metric -- namely, the editors' gender at 0.79, practicing muslim religion at 0.9 or being jewish at 0.91.}
(2) We quantify the effect of different features using a \rva{temporal measure} called information transfer. We observe that while the marginal utility of newer features decreases over time, the new users consistently add additional information for the prediction tasks. 
(3) We show that the prediction of private attributes \rev{continues} to improve for users who \rev{exited} the system  -- or stopped editing after 2007. 
The continued loss of privacy for these users \rev{seems to be associated with} two quantifiable factors: the information learned from a user's own activity (or {\em online breadcrumbs}) and the activity of other editors. %, and we study each factor independently. 

To the best of our knowledge, this is the first work to quantify \rva{longitudinal change} in privacy loss, \rva{carried out} on a large dataset and over more than \rev{13} years. 
Not only do we show that private traits can be predicted increasingly well with time, we also provide several methods to quantify the value of new information over time, and the different source of information loss - from more activity or more users. 
%but new insights can be gained even for users who have retired from the editing activity in the early stages of the platform. 
%
%Furthermore, we show using measures inspired from economics (\textit{i.e.}, the \textit{marginal utility} measure) that most of the information gain is obtained in the beginning of the studied period.
%Whatsoever, the accuracy of the prediction increases as data of longer temporal extent is available for learning, proving that the online privacy continues to degrade with time, while increasingly slower. 
%We show that 
Our findings suggest privacy continues to erode for all users, even after one stops publishing data online. 
Our findings \eat{\rev{have} implications in} \rva{also can help design} data storage and retention policies, \rva{can make} users aware of the implications of seemingly harmless online activities, and adds to the very lively research \rva{topic} about online privacy.  %%LX: style edits, avoid repeating word (implication) and use smaller words
%This happens because a more precise model can be constructed using the data of newly entered individuals and it can be applied on the retired user's old data, in order to obtain better predictions.

%!TEX root = WSDM_2015_wikipedia.tex

\secmoveup
\section{Related \revOLD{Work}}
\textmoveup

\textbf{The value of privacy.}
Social scientists \rev{have} been interested in how individuals perceive and values their privacy. 
Acquisiti et al.~\cite{ACQ13} revealed that the perceived value of privacy, while not entirely arbitrary, is highly malleable.
For example, there is a large gap between the amount of money that individuals would accept to disclose private information and the amount of money they would pay to protect it.
%The same study concludes that, while not ready to sell/give away privacy completely, individuals are reluctant to invest value into protecting it, or to stop from using the platforms that release such information. 
Furthermore, certain categories seem particularly vulnerable to the online privacy issue. 
Young adults have been shown\rev{~\cite{boyd2011}} to be as worried about their privacy as older adults, especially in what concerns giving personal information to businesses, having photos of them uploaded to the internet or the legislation protecting privacy.
%\rev{While debated, certain studies~\cite{HOO10} indicate that} 
\rva{On the contrary, Hoofnagle et al.~\cite{HOO10} found that}
young adults tend to expose themselves more, especially on popular social networks, because they are less aware of the risks, less informed about the protection given by law and more prone to social peer pressure.
%Our work joins the effort to sensibilise the public opinion about the issue of privacy, 
%while adding an additional warning about the dangers of the online environment for privacy.
Our work can contribute to the public understanding of privacy risks, 
by studying on public open data and with quantifiable outcomes. 

%% LX - this statement should be in the Conclusion, not here. 
%We advocate that new means should be found to tackle the issue of privacy online, especially considering our finding that data published online continues to provide new insights long after the users stopped posting.

\textbf{Privacy Loss and inferring private traits.} 
\rev{One definition of} Personal Identifiable Information \rev{(PII) is} private information relating to a person, which can be deductively identified, based on the person's public profile~\cite{NAR10}.
This is a source of concern particularly in the context of the Open Data effort of governments, in which anonymous datasets are released publicly, after removing private attributes such as name and contact information.
% are removed (\textit{e.g.}, names, contact information).
The literature shows many applications in which private information, which was never intended to be publicly released, can be inferred from apparently harmless data.
In one notorious example, the medical condition of an American politician was inferred starting from anonymized medical records released to the public~\cite{SWE02}.
% LX: "individual tracking" is vague
%Furthermore, individual tracking has been simplified: 
Researchers found that de-anonymization can be carried out in large-scale. 
The 2011 IJCNN Social Network Challenge
%, dealing with research on real-world link prediction, 
was won by de-anonymizing the identity of the Flikr users, including those in the test set~\cite{NAR11}. 
More recently, De Montjoye et al.~\cite{MON13} showed that only four spatio-temporal points are enough to identify 95\% of individual trajectories using mobile carrier's antenna information, while the same group~\cite{DeMontjoye2015} show that 90\% of individuals can be re-identified using their credit card transactions trajectory.

%% flipping the narrative of the two cites work (respect their temporal order)
Another salient source of private informations are patterns of online behavior.  
Kosinski, Stillwell and Graepel showed~\cite{KOS13} how private traits like gender, sexual orientation, ethnic origin and even the fact that a user's parents have divorced before her twenty-first birthday can be accurately inferred from apparently harmless, naturally revealed public data, such as Facebook likes.
In another data domain, the pattern of an individual's online or phone activity have been shown~\cite{SAR14} to reveal precious information about her habits and preferences.
Furthermore, computer-based evaluations of human personality have been shown more accurate than those of close friends~\cite{Youyou2014}.
%% LX -- we don't need these narrative
%Profiling users' patterns has been increasingly used by 
%Various online media services uses such patterns to tailor what is shown to each user,  
%the information in order to provide a better quality content. Whatsoever, 
%while the same techniques can be used to infer private traits that a user does not want to share. 
%Recently, it has been shown by

We show that private information can be \rev{extracted} not only from structured anonymized datasets (such as~\cite{SWE02}) or datasets rich in social information (such as~\cite{KOS13,SAR14}), but even from %apparently completely benign 
data traces left for the public good, such as Wikipedia.
Ramachandran and Chaintreau~\cite{Ramachandran2015} recently studies how the structure of locally connected individuals affects privacy loss.
Our focus is in \rev{the time dimension} -- in quantifying the temporal evolution of privacy loss.

%Furthermore, de-anonymization can have unsuspected applications. 
%The 2011 IJCNN Social Network Challenge, dealing with research on real-world link prediction, was won by de-anonymizing the identity of the Flikr users, including those in the test set~\cite{NAR11}.

\textbf{Editing behavior in Wikipedia.}
Wikipedia is the largest online collaborative \rev{encyclopedia}. 
In its early years, Wikipedia \revOLD{showed rapid} %has shown %an exponential 
%% LX - i'm always cautious in describing someting as expoential when you've not quantified that
growth. 
%in the number of \revOLD{new} %added 
%articles and revisions. 
%Some 
Initial studies~\cite{ALM07} explained the growth as driven by the rapidly increasing user base. 
%Whatsoever, the growth of the English Wikipedia peaked in 2007, 
%% LX -- slowed vs peaked, suggest not to have the readers reason over flipped semantic
The growth of the English Wikipedia slowed after 2007, with fewer new editors joining, and fewer new articles created. 
A few studies~\cite{Gibbons2012,Suh2009} explain this dynamic as Wikipedia editors face increasingly limited opportunities to make novel contributions, with the {\em easy} articles already been created, leaving only more difficult topics to write about. 
%In order t %%novel, 
To make useful contributions to the site, editors must also meet an increasingly high bar of expertise in their field.
\rev{In addition,} 
Halfaker and McNeil~\cite{Halfaker2012} \rev{consider} that Wikipedia's mechanisms for managing quality and consistency 
%in the face of its massive growth %%LX simplify this sentence, do not need this info
deterred newcomers.

%Our  the same \revOLD{trends} in Wikipedia's evolution.
%Furthermore, 
In our profile of Wikipedia's growth and decline 
(Sect.~\ref{sec:data}), we were surprised to see that
\eat{the trends}
\rva{the changes of activity} over time
are not uniform across content and personal demographic attributes. 
There is a rise in site maintenance, and some user groups showed slower decline (\textit{PhD}), or even a rise (\rva{self-identified} \textit{muslism} users) in editing activities.
%\revOLD{editors' personal infor}%the personal editor information 
%to analyze in more detail the decline of editorship per editor category and we detect the rise of maintenance.
%Furthermore, while Wikipedia has been used before to detect social roles~\cite{WEL11} or language interactions~\cite{DAN12}
%social roles: Simply counting the number of revisions over a predefined set of categories has been shown to provide descriptive power in the  task of detection social roles in Wikipedia~\cite{WEL11}
Finally, while social interactions in Wikipedia have been studied before~\cite{DAN12,WEL11}, to the best of our knowledge, this is the first study of private traits that can be inferred from editing activities.

%!TEX root = WSDM_2015_wikipedia.tex

\begin{table*}[htbp]
\caption{Features to describe user editing patterns.
(A) The {\em basic} feature set quantifying the number of edits to all Wikipedia articles and various community and user pages.
%CONTENT feature quantifies \revOLD{the number of} contributions to the Wikipedia articles. Other categories deal with interaction between editors, personal pages and the organization of Wikipedia; 
(B) additional features in the {\em extended} feature set, encoding edits to the thematic categories within Wikipedia {\tt CONTENT}.}
\small
\begin{tabular}{lcp{3cm}p{10.6cm}}
\toprule
\multicolumn{1}{c}{\textbf{A.}} & \textbf{Feature name} & \textbf{Wikipedia namespace codes} & \textbf{Feature description} \\ \midrule
\multirow{6}{*}[-1.5em]{\begin{sideways}\textbf{Basic feature set}\end{sideways}} & \texttt{CONTENT} & \multicolumn{1}{c}{0, 6} & Revisions made to the body of the Wikipedia articles. Corresponds to the actual creation of information. \\ 
	& \texttt{TALK-C} & \multicolumn{1}{c}{1, 7} & Discussions on the talk pages of the Wikipedia articles. Corresponds to the overhead around the creation of \revOLD{encyclopeadic} information. \\ 
	& \texttt{USER} & \multicolumn{1}{c}{2} & Revisions made to \revOLD{user pages}, \revOLD{including a} user's own page or another user's page. \revOLD{Similar to profile edits and posts in online social networks}. \\ 
	& \texttt{TALK-U} & \multicolumn{1}{c}{3} & Revisions on the talk page corresponding to \revOLD{user pages}. \revOLD{This is similar to social discussions}, e.g., writing to a user's wall in a social network. \\
	& \texttt{WIKI} & \multicolumn{1}{c}{4, 5} & Revisions on community pages, help desk, village pump, and related talk pages. \\ 
	& \texttt{INFRA} & 8, 9, 10, 11, 12, 13, 14, 15, 100, 101 & Revisions on pages that provide infrastructure for other tasks in Wikipedia; template, categories and portals. \\ \bottomrule
 &  &  &  \\ \toprule
\multicolumn{1}{c}{\textbf{B.}} & \multicolumn{ 3}{c}{\textbf{{\em Extended} feature set: the 23 thematic features added to the {\em basic} set}} \\ \midrule
\multicolumn{4}{p{17.5cm}}{\texttt{AGRICULTURE}, \texttt{APPLIED-SCIENCES}, \texttt{ARTS}, \texttt{BELIEF}, \texttt{BUSINESS}, \texttt{CHRONOLOGY}, \texttt{CULTURE}, \texttt{EDUCATION}, \texttt{ENVIRONMENT}, \texttt{GEOGRAPHY}, \texttt{HEALTH}, \texttt{HISTORY}, \texttt{HUMANITIES}, \texttt{LANGUAGE}, \texttt{LAW}, \texttt{LIFE}, \texttt{MATHEMATICS}, \texttt{NATURE}, \texttt{PEOPLE}, \texttt{POLITICS}, \texttt{SCIENCE}, \texttt{SOCIETY}, \texttt{TECHNOLOGY}} \\ \bottomrule
\end{tabular}\captionmoveup
\label{tab:basic-extended-features}
\end{table*}

\secmoveup
\section{Wikipedia activities and user traits}
\label{subsec:construct-datasets}

\textbf{Why Wikipedia?}
Wikipedia is an ideal data source for studying longitudinal predictability of private traits, due to the following three reasons.
Firstly, it is an apparently harmless dataset, whose purpose is to be a reservoir of knowledge, with little or no focus on personal or social information. 
Unlike online social networks centered on users' profiles, Wikipedia is not intended to record any individual contributor's personal and social interactions. 
Secondly, Wikipedia's entire edit history is publicly accessible.
Wikipedia provides the longitudinal editing history for individuals, spanning over a decade -- 
2001--2013 at the time of our snapshot.
Such a unique long temporal extent allows the study of the effect of time in online privacy.
Finally, Wikipedia 
contributors are from many geographic locations and numerous social, 
religious, educational and political backgrounds -- providing a rich and diverse 
sample for activities and candidate personal traits. 
%This allows us access to a rich and diverse environment 
%% LX -- there is no basis to claim this
%and limit the bias in our conclusions.
%% LX - this was and will be said elsewhere, redundant
%We hypothesize that, given its sheer size, even benign activities such as making a contribution to a Wikipedia article can unknowingly reveal private information about its editors.
We show that as Wikipedia accumulates user data and editing activity, increasing amounts of private information can be inferred (see Sect.~\ref{sec:results}). 
%% LX - this should belong to later in this section
%We choose gender, education and religion as exponents of the private traits which can be deduced from public activity.

\textbf{\revOLD{The Wikipedia dataset}.}
Our dataset contains 13 years of edit history, from the beginning of Wikipedia in January 2001
\revOLD{to} July 2013.
188,805,088 revisions are performed by 117,523 editors to 22,172,813 pages.
A revision is a Wikipedia term referring to an atomic edit of a page by a user
\revOLD{with an associated timestamp}.
All data used in this study \revOLD{are} %has been 
obtained from July 2013 public Wikipedia dump, 
\revOLD{more details about data processing are}
in the supplemental material (SI)~\cite{supplemental}.
%\footnote{Supplementary Information: \url{https://db.tt/VaLqIvtN}}.
%For reasons of having enough active users, in our analysis in 
\revOLD{For predicting private traits (Sect.~\ref{subsec:measure-privacy-loss} and~\ref{subsec:edit-wikipedia}) we limit the studied period between 01/2007 and 07/2013 in order to have sufficient numbers of users.}

%\textbf{Profiling users.}
\textbf{\revOLD{User activity profiles}.}
We encode a user's activity using two sets of features. 
In the \emph{basic} set, we count the number of revisions performed in a given period of time, over six predefined categories of the edited pages. 
The intuition behind these features is to capture the intent of a user's editing effort.
For example, the \texttt{CONTENT} feature captures edits made to main Wikipedia articles and can be associated with the knowledge creation effort.
Similarly, the \texttt{WIKI} and \texttt{INFRA} features quantify the effort put into organizing the editing effort (\textit{i.e.}, community pages, help desks), while \texttt{USER} and \texttt{TALK-U} captures the social components, such as constructing a personal page and talking to other users.
Features in the {\em basic} set are based on the Wikipedia namespaces and \revOLD{a summary of their respective meanings} is in Table~\ref{tab:basic-extended-features}.
Wikipedia namespaces are organizational categories, to encode the intended purpose of a page (details in SI~\cite{supplemental}).
%A more detailed description of the namespaces, alongside with technical details concerning the construction of the dataset can be found in the Supplementary Information
The second set of features, \textit{i.e.} the \emph{extended} set, is constructed by adding to the {\em basic} set 23 new categories based on Wikipedia's top-level category hierarchy.
A Wikipedia page can be assigned by its editors to one or more of the 23 thematic categories such as History, Geography, \textit{etc.}
The {\em extended} set provides a more detailed profiling of a users' activity, by capturing their editing interests. 
Details of feature construction are described in Sect.~\ref{subsec:measure-privacy-loss}.

\textbf{The editors' personal information.}
Many Wikipedia users keep a user page (resembling a social network profile), on which they distribute information about themselves, their interests or the causes they support.
Some %equally 
distribute information \revOLD{typically considered} 
%which can be considered 
as private, \revOLD{such as} 
their gender, ethnic origin, religion, education, or even sexual preferences.
We retrieve these records using public APIs\footnote{\rva{\surl{https://www.mediawiki.org/wiki/API:Users}}}, and use them as target personal traits.
For the purpose of this study, we selected three private traits \revOLD{with sufficient user bases}: gender (declared by 6936 users), education (\textit{undergrads}, \textit{grads} and \textit{Phd}, declared by 9224 users) and religion (\textit{christian}, \textit{muslim}, \textit{atheist} or \textit{jewish}, declared by 7685 users).

%!TEX root = WSDM_2015_wikipedia.tex

\secmoveup
\section{Measuring privacy loss}
\label{subsec:measure-privacy-loss}
\textmoveup

We study the loss of privacy by modeling it as a \rva{prediction} problem: how well can we predict 
%can we explain %LX -- predict not explain
a given a class variable $Y$ (\textit{i.e.}, gender, education or religion) using descriptive features \revOLD{$X$} in the {\em basic} or {\em extended} set.
By following a set of users through time, we \revOLD{observe} the dynamics of \revOLD{the predictive \eat{power}\rva{performance}}. %power of explanation.
We define the temporal loss of privacy as \eat{the process of} better explaining a variable linked to a private trait, as we observe \rva{editing behaviors for} longer periods of time.
In this work we use two sets of tools. %, detailed in the f sections.
% LX -- flipped order 
The first is a predictive approach: we predict the private traits of a hold-out set of users on increasingly longer activity history on Wikipedia, and we observe the change in prediction accuracy.
The second approach uses {\em information transfer}, a measure from physics and economics, to quantify the uncertainty in $Y$ explained by feature $X$ over time.

%\subsection{Embedding temporal information} 
\secmoveup
\subsection{Encoding activities over time}
\label{subsec:encoding-features}

We denote a feature $X_i^u$ as computed in the timeframe $i$ for user $u$, with $X \in \{\texttt{CONTENT} $, $\texttt{TALK-C}$, $\texttt{USER}$, $\texttt{TALK-U}$, $\texttt{WIKI}$, $\texttt{INFRA}\}$ for the {\em basic} set, and encoded similarly for the {\em extended} set.
%% LX -- this is repetitive over multiple other places, delete
%In order to capture the effect of time on the predictability of the private traits, 
% LX - go directly to the description, "temporally incremental" is vague
%we construct temporally incremental datasets, 
%each capturing the users activity during \revOLD{a 3-month period}. %subsequent periods of 3 months.
We construct a series of temporal datasets, each \rev{having a 3-month period in addition to} the previous.
A user appears in a temporal dataset if she/he has performed at least one revision during or before the last 3-months.
We construct two kinds of features over time, the instantaneous features $f_i^u$ for user $u$ in the $i^{th}$ 3-month period alone, and the longitudinal 
%% LX -- it's better to be clear upfront that F is a series
feature $F_i^u = [f_{1:i}^u]$ for user $u$ 
-- containing the series up to (and including) timeframe $i$.
% LX -- feature description needs to be accurate enough to allow someone to re-do this
%\rva{\bf Discussion: replace $f\rightarrow x$ and $F\rightarrow X$ for better readability?} 
%\rev{\textbf{MAR: later on (when defining our information  theory method), we use X for an feature. Keeping as is keeps consistency.}}
Features are constructed by counting the number of revisions performed by $u$ during the given timeframe, over the predefined categories, \textit{e.g.}, $f_i^u = ( \texttt{CONTENT}_i^u $, $\texttt{TALK-C}_i^u$, $\texttt{USER}_i^u$, $\texttt{TALK-U}_i^u$, $\texttt{WIKI}_i^u$, $\texttt{INFRA}_i^u )$ for the \emph{basic} set. 
%% LX - this is redundant once we defined the series above
%To obtain $F_i^u$, we concatenate $f_i^u$ with the \revOLD{cumulative} feature set from the previous timeframe: $F_i^u = (F_{i-1}^u, f_i^u)$, where $F_1^u = f_1^u, \forall u$.
Naturally, features $F_i^u$ %are of increasing size and they %contain features to 
% LX - again two redundant mentions of 'increasing size/info', keep one
describe both the past and the current activities,  
%Therefore, the obtained feature sets embed 
and contain temporally increasing quantities of information.
%We take into account 
In addition, for newly joined editors, we explicitly encode the missing values in previous timeframes.  
This is done by including a binary missing feature flags for each activity category, a value of 0 means 
%in order to make the difference between missing activity and a real zero revision count (\textit{e.g.}, 
an editor has joined Wikipedia (even if she is on a pause during timeframe $i$), %one of the timeframes
and 1 means the editor has not joined Wikipedia, i.e. missing.
For the {\em basic} set, 
%we add an 
this results in additional six binary features, i.e., 
%, each signaling that the concerned features has a missing value, for the given user. 
%Therefore, the basic feature set contains 12 features: 
$f_i^u = (\texttt{CONTENT}_i^u$, 
$\texttt{p\_C}_i^u$, 
$\texttt{TALK-C}_i^u$, 
$\texttt{p\_TC}_i^u$, 
$\texttt{USER}_i^u$, 
$\texttt{p\_U}_i^u$, 
$\texttt{TALK-U}_i^u$,
$\texttt{p\_TU}_i^u$, 
$\texttt{WIKI}_i^u$, 
$\texttt{p\_W}_i^u$, 
$\texttt{INFRA}_i^u$, 
$\texttt{p\_W}_i^u )$, where features prefixed with $\texttt{p\_}$ are the missing flags of the preceding feature. 
For the {\em extended} set, additional features for \rva{the} 26 categories are constructed in the same manner and appended to each $f_i^u$.
\rva{We tried other schemes to encode user activity, such as cumulative features to encode activities from the beginning of until timeframe $i$, and found them to have lower performance. 
Therefore the rest of the paper presents only the incremental scheme.}
%% LX: active voice seem to work better here, more succinct
\eat{Other user activity encoding schemes have been tried (cumulative features in $f_i^u$ -- features counting the activity of $u$ from the beginning of her activity until timeframe $i$ -- and cumulative features over time). These have provided lower performance and, therefore, they are not presented in the rest of this manuscript.}

\begin{figure}[tb]
	\centering
	\includegraphics[width=0.25\textwidth]{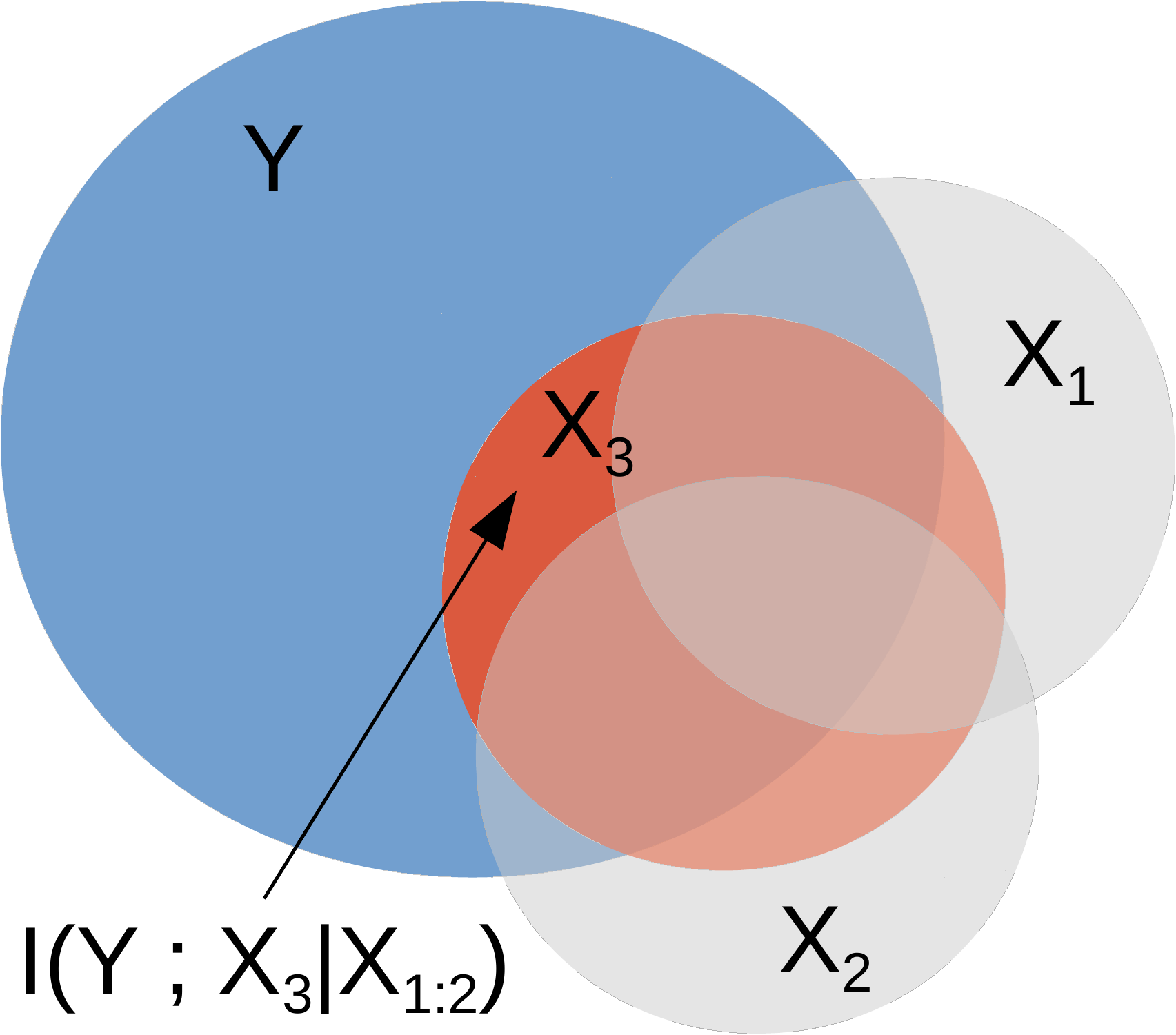}
	\caption{
	Venn diagram illustration of the Information Transfer measure between \revOLD{target} variable $Y$ and %the descriptive % LX - features are all 'descriptive', redundant
	feature $X$, %instantiated at %LX - cut the big words
	over three successive time points $X_1$, $X_2$ and $X_3$.
	Even if $X_3$ explains a large portion of the information in $Y$, most of $Y$ was already explained by $X_1$ and $X_2$, and the new information$X_3$ brings is given by the conditional mutual information $I(Y; X_3|X_{1:2})$.
	}
	\label{fig:information-transfer-schema}\captionmoveup
\end{figure}

\subsection{Predicting personal attributes}
\label{subsec:experiment-setup}
%In this approach, 
We evaluate how well a private trait can be predicted by setting up a set of binary classification tasks. 
%the descriptive 
%features through means of learning a model and predicting the private trait for new users.
%We train a logistic regression model for each temporal dataset.
Multi-class target variables (\textit{i.e.}, religion and education) are transformed to binary \rva{prediction tasks} in a one-vs.-all fashion (\textit{e.g.}, \textit{christian} vs. non-\textit{christian}). 
%The target population of users is divided into a training set and a testing set.
We use \eat{66\%}\rva{2:1} stratified splits to construct the training and test user subsets -- 66\% of the editors are randomly selected to be part of the training set, and the \rva{remaining} 33\% are used for testing, with the random sampling preserving class priors.
%Consequently, in both the training and the testing sets, each class (\textit{i.e.}, \textit{male}/\textit{female}) are represented in the same proportion as in the target population.
One model is learned for each class and each time period, using logistic regression classifier with \rev{L1 regularization} that favors sparse feature weights, with the hyperparameter obtained by cross-validation~\cite{James2013}.
\rva{We also tried the L2 regularizer and observed lower performances.}
\eat{The L2 regularizer has also been tested and it has shown lower performances.}
The performance is evaluated using the AUC metric (the area under the ROC curve)~\cite{James2013}, intuitively random guess classifiers have an AUC of $0.5$, and perfect classification has $1.0$.
\rev{Accuracy and F-score for predicting each private trait are given in the SI~\cite{supplemental}.}
We independently sample the train/test split 10 times, and record the mean and standard deviation of the AUC.
%The graphic in Fig. 6 is obtained slightly differently: all the editors, except the 650 users which stopped activity from 01.01.2008, form the training set. At each timeframe, the test set contains the target 650 users.
\rev{While other classifiers can be used, we have not tried them in our experiments 
%in Sect.~\ref{sec:results} 
since our interest lies in the evolution of prediction performance and not its absolute value.
}

\subsection{Information transfer over time}
Information theory measures are useful for capturing feature relevance in prediction tasks~\cite{mackay2003information}. 
One particular measure, Information Transfer, was recently used~\cite{VerSteeg2012} to uncover hidden links in social media.
Intuitively, we capture the uncertainty of private information with the entropy of the target variable $Y$.
The quantity of private information explained by \eat{a single descriptive } feature $X$ is then given by the \eat{Mutual Information}\rva{mutual information} $I(Y ; X)$. %LX: lower case
Since feature $X$ takes different values for each 3-months timeframe, it is useful to quantify the additional information contained in time period $X_t$ that were not already contained by earlier features $X_{1:t-1}$.  
The {\em Information Transfer} measure, $I(Y; X_t|X_{1:t-1})$, is designed for this purpose.
%% LX - define the measure first, then explain. 
%some of the information revealed at later moments of time might have been already uncovered at earlier moments.
Fig.~\ref{fig:information-transfer-schema} illustrated the intuition behind  $I(Y; X_t|X_{1:t-1})$ using a Venn diagram. 
% (\textit{i.e.}, $X$ at time 3) % LX -- no need to repeat, 
% LX -- explain the example properly quantifying each technical term/part
In this example, $X_3$ contains quite a lot of information about $Y$ -- expressed as \rev{the mutual information} $I(Y; X_3)$, or the large intersection between red and blue circles. 
However, the {\em new} information that $X_3$ provides, in addition to $X_{1:2}$, is much smaller -- expressed as $I(Y; X_3|X_{1:2})$, as $I(Y; X_3)$ minus the part already covered by $I(Y; X_1)$ and $I(Y; X_2)$. 
%which lies inside the intersection of $Y$ with $X_1$ and $Y$ with $X_2$ is already known prior to time 3.
%Only a small portion of the information contained in $X_3$ is actually novel.
%We define, therefore, the \emph{Information Transfer} between a feature $X$ at time $t$ and the class variable $Y$ as the marginal utility of $X_t$ compared to $X_{1:t-1}$.
%% LX - already defined, illustrated, now explain more
Intuitively, information transfer is the
%Numerically, we compute it as the 
Conditional Mutual Information of $Y$ and $X_t$ given $X_{1:t-1}$, or
%I(Y ; X_t | X_{1:t-1})$.
the amount of uncertainty that will be reduced after observing $X_t$:
\begin{equation*}
	I(Y ; X_t | X_{1:t-1}) = H(Y | X_{1:t-1}) - H(Y | X_{1:t}) \enspace. 
	\eqmoveup
\end{equation*}
The relationship above follows from the definition of mutual information and conditional entropy~\cite{mackay2003information}.
%remaining uncertainty after observing $X_{1:t}$ is the conditional entropy $H(Y | X_{1:t})$.
We implemented information transfer using the \texttt{infotheo} toolbox in R~\cite{Meyer2014}, which computes conditional entropies and in high-dimensional input spaces by quantizing the input. 
We use Information Transfer $I(Y ; X_t | X_{1:t-1})$ and conditional entropies $H(Y | X_t)$ and $H(Y | X_{1:t})$ to answer two key questions: which features are most important in the disclosure of private information, and which time periods are critical to privacy loss.

%!TEX root = WSDM_2015_wikipedia.tex

% figure 2
\begin{figure*}[htbp]
	\centering
	\subfloat[] {
		\includegraphics[width=0.32\textwidth]{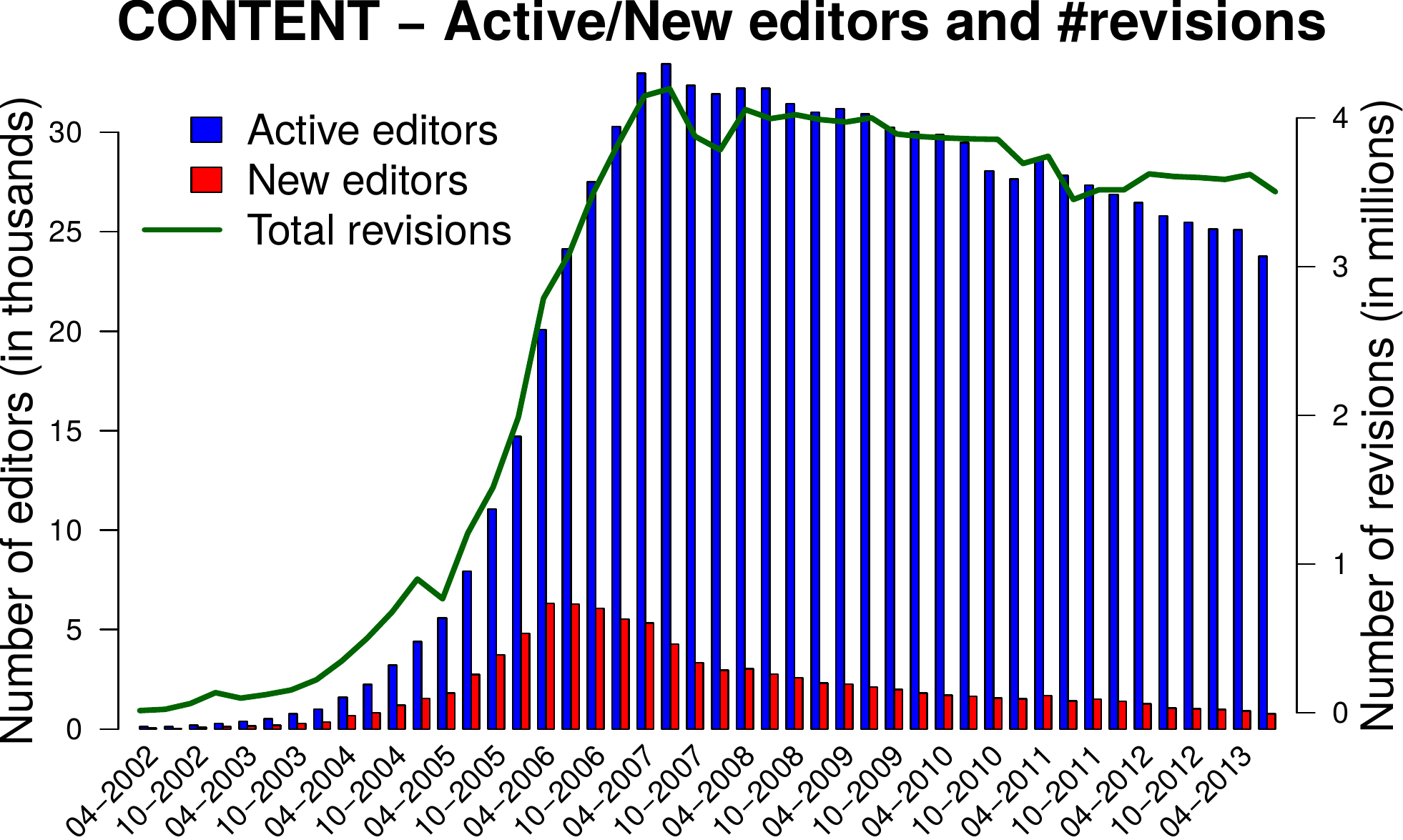}%
		\label{subfig:profile-content-decrease}
	}
	\hfill
	\subfloat[]{
		\includegraphics[width=0.32\textwidth]{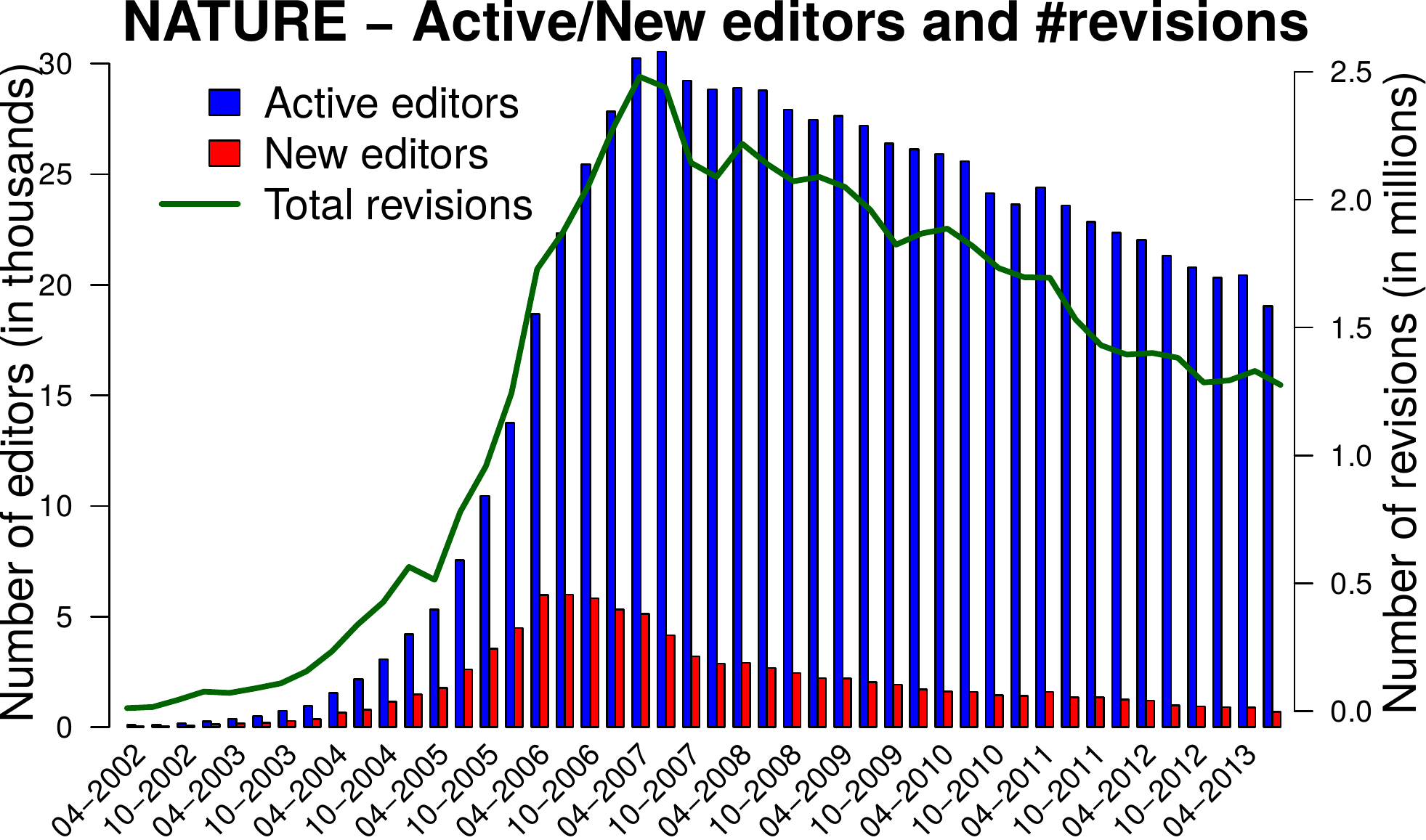}%
		\label{subfig:profile-NATURE-decrease}
	}	
	\hfill
	\subfloat[]{
		\includegraphics[width=0.32\textwidth]{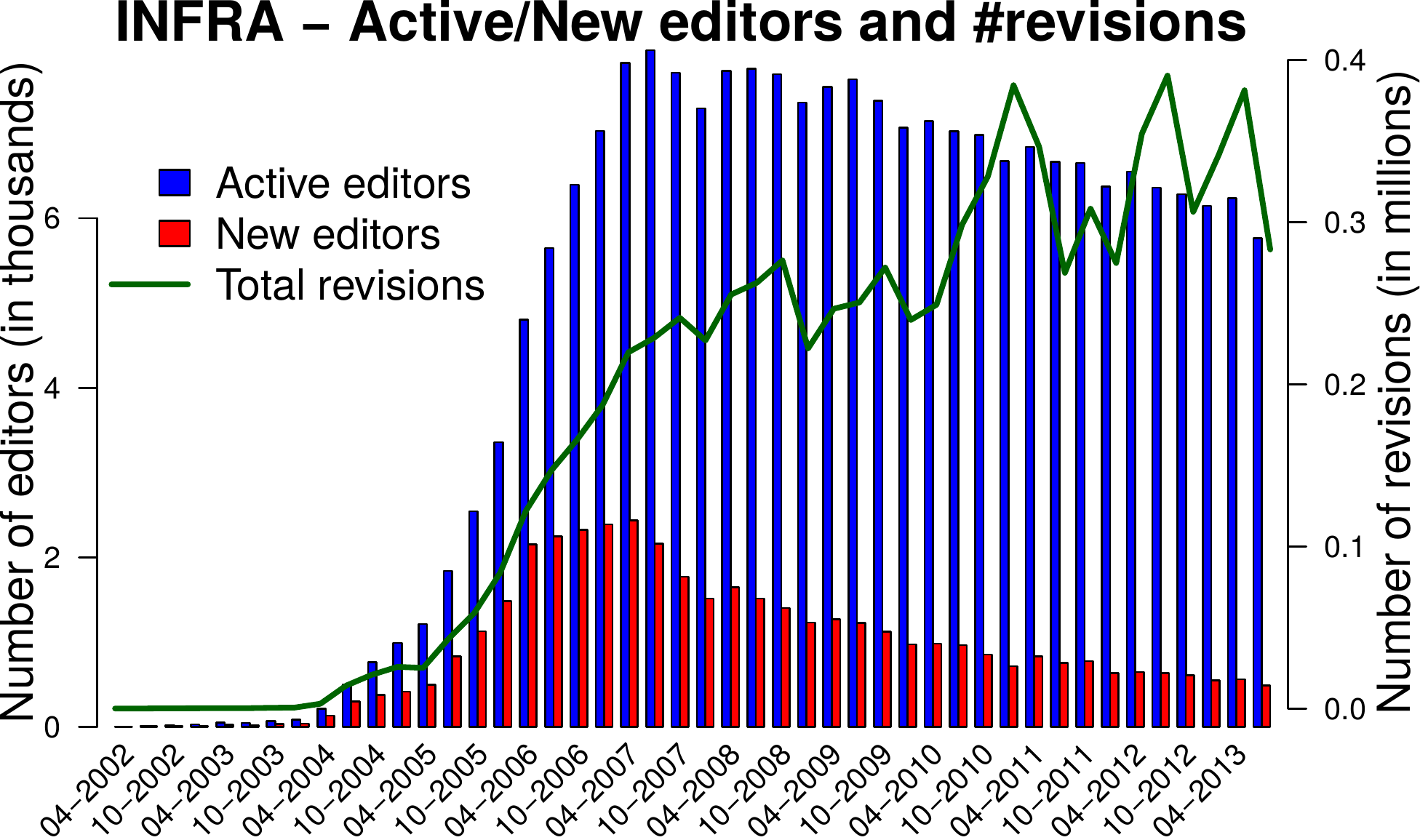}%
		\label{subfig:profile-infra-increase}
	}

	\caption{Wikipedia growth slowing down. 
	(a) The decrease of the number of active editors, new editors and the total number of revisions for \texttt{CONTENT} (a) and thematic features (shown here \texttt{NATURE}, others in the SI~\cite{supplemental}) (b). 
%	Similarly, the number of active editors, as well as newly entered editors, are also decreasing.
	(c) The maintenance effort (INFRA revisions) 
	needed to internally handle the bulk of Wikipedia 
	is increasing.
%	(b) The revisions on all the thematic features (in the \emph{extended} set) have identical trends (shown here \texttt{NATURE}, others in SI).
%	(c) The population of editors dealing with the organization of Wikipedia (INFRA revisions) is also decreasing, but the number of revisions needed to internally organize and handle the bulk of Wikipedia is constantly increasing. 
	}
	\label{fig:wiki-profiling}
%	\vspace{-0.2in}
\end{figure*}

% figure 3
\begin{figure*}[htbp]
	\centering
	\subfloat[]{
		\includegraphics[width=0.32\textwidth]{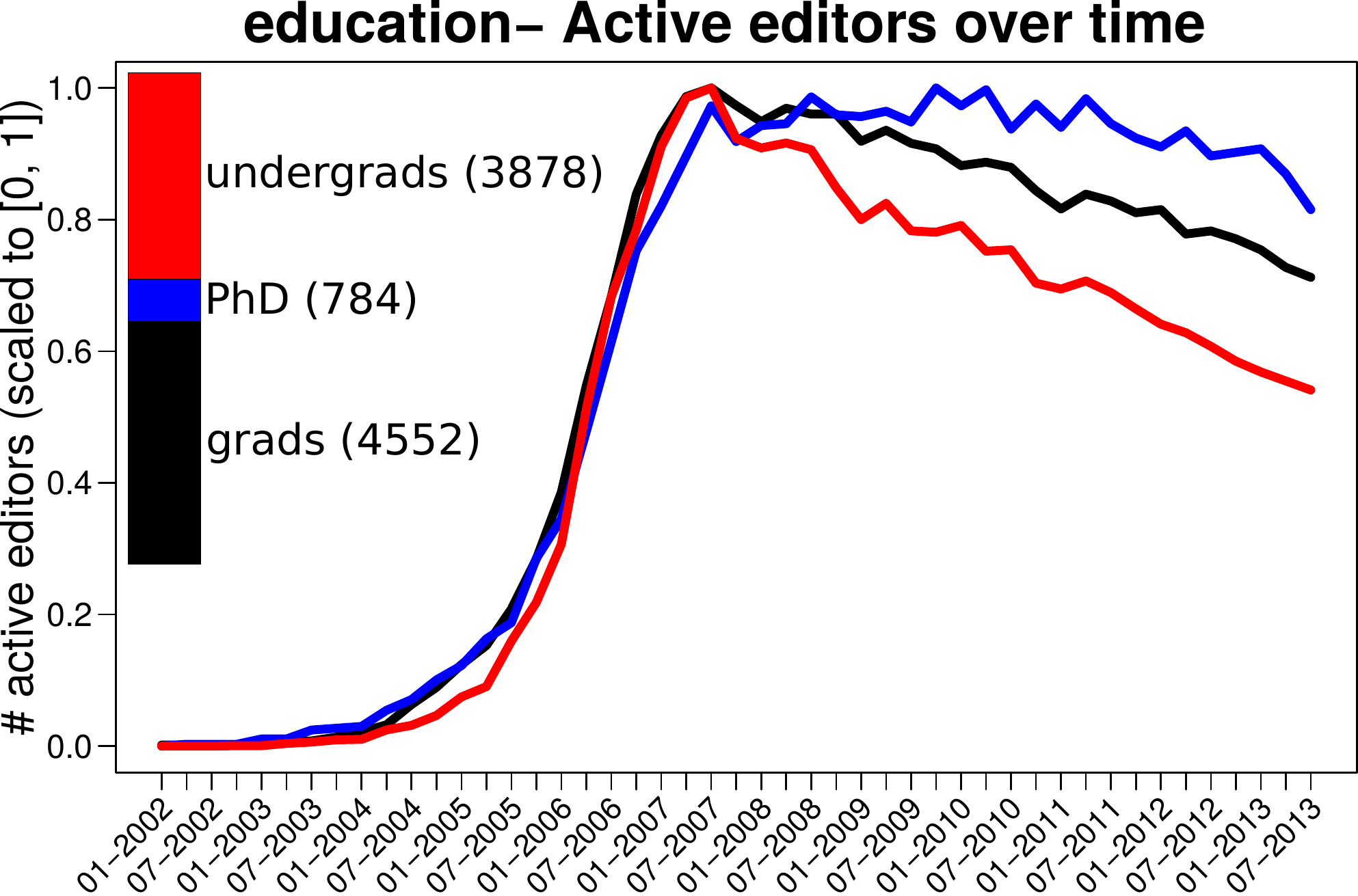}%
		\label{subfig:profile-phd-stay-active}
	}
	\hfill
	\subfloat[]{
		\includegraphics[width=0.32\textwidth]{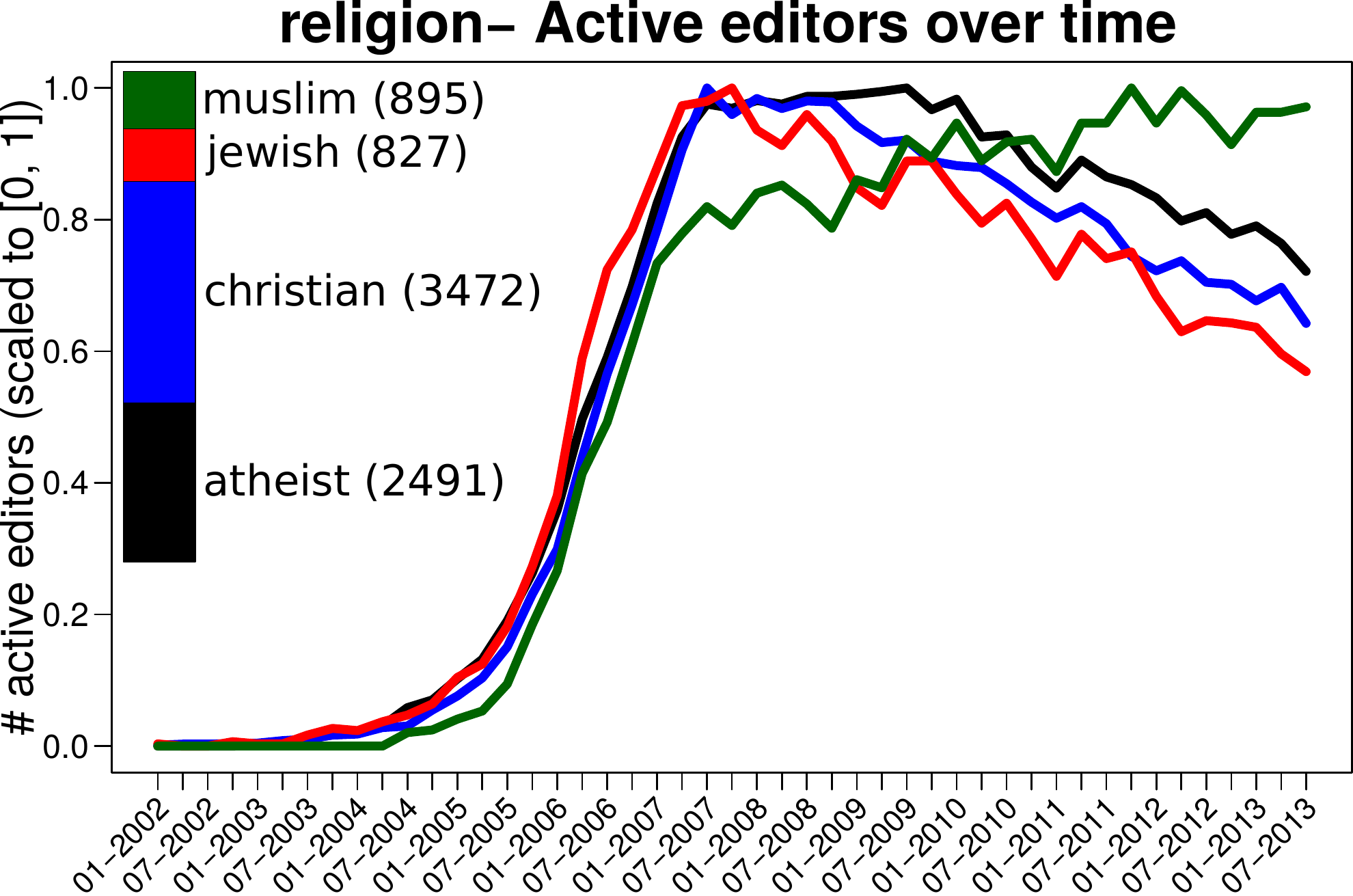}%
		\label{fig:profile-muslim-population-increases}
	}
	\hfill
	\subfloat[]{
		\includegraphics[width=0.32\textwidth]{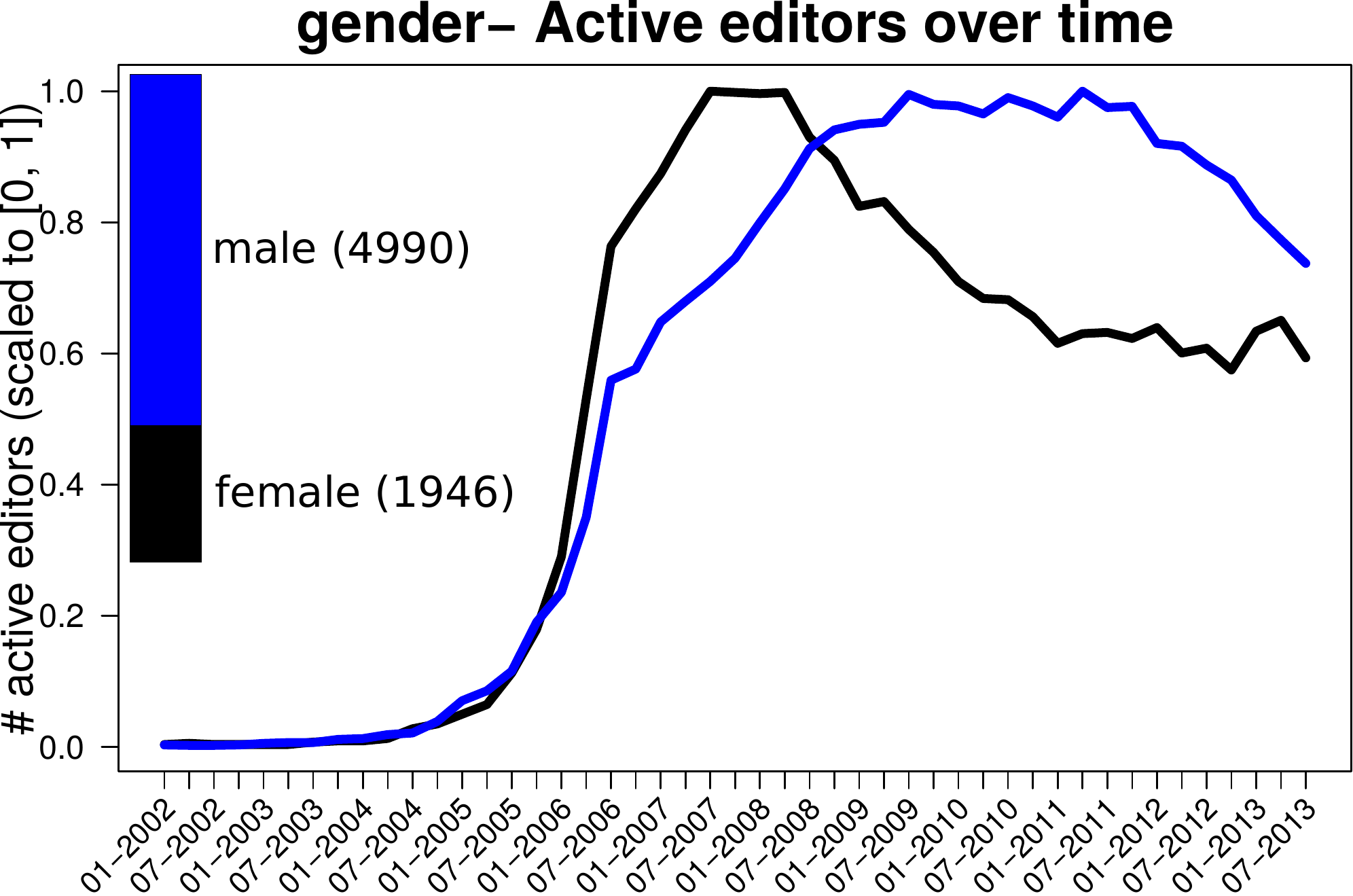}%
		\label{subfig:profile-gender}
	}
	\caption{\rva{The population of active editors over time, }\eat{Dynamics of evolution of the population of active editors, }%%LX: "Dynamics of evolution" -- redundant
	broken down by (a) gender, (b) education and (c) religion. Magnitudes for all classes are scaled from 0 to 1. 
	Barplots show the relative effectives of classes, absolute effectives are given in parenthesis.	
%	\rva{\bf TODO: include class-prior: i.e. breakdown of the several classes being shown here, at one point in time e.g. 07-2013}
%	Muslims are the only ones to continuously increase in number in the active population.
	}

\end{figure*}

%figure 4
\begin{figure*}[htbp]
\centering
	\subfloat[] {
		\includegraphics[height=0.15\textheight]{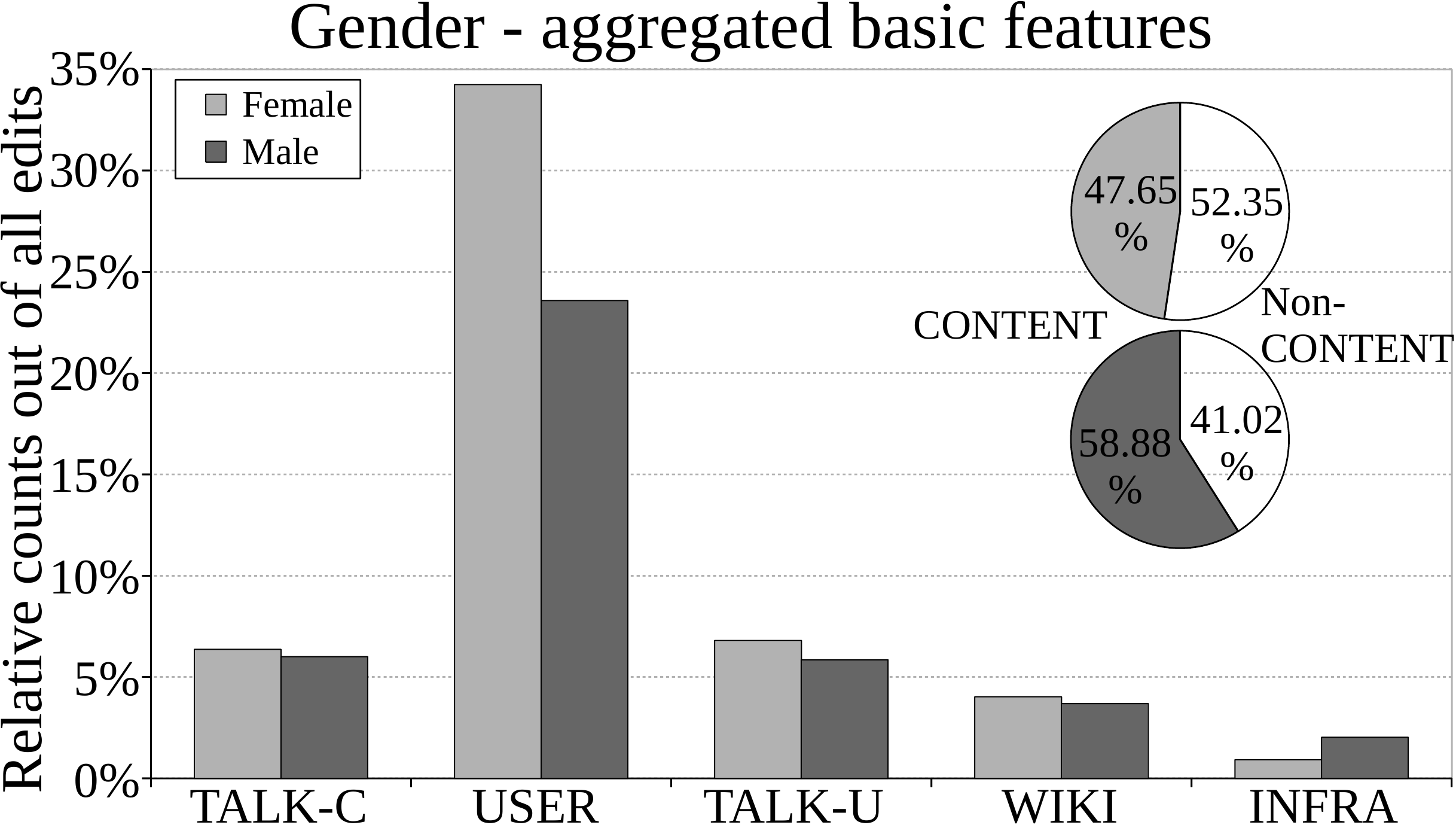}%
		\label{subfig:gender-over-features-basic}
	}
	\quad
	\subfloat[]{
		\includegraphics[height=0.15\textheight]{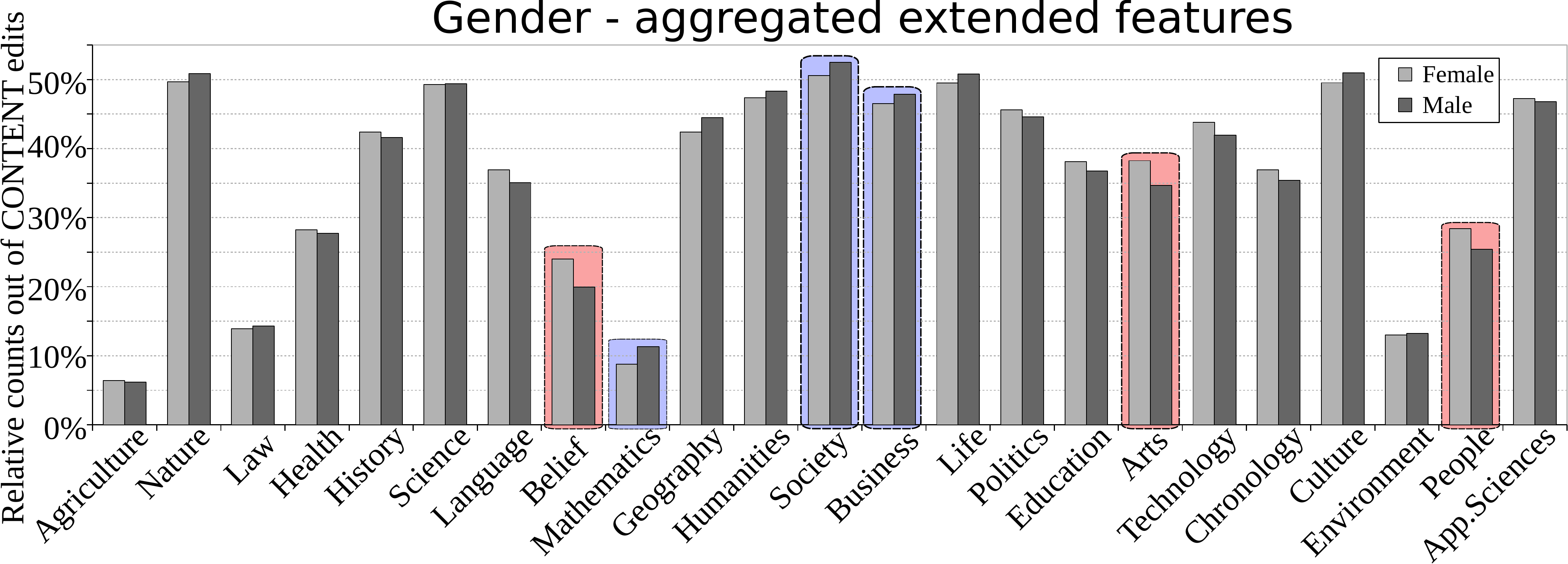}%
		\label{subfig:gender-over-features-extended}
	}
	\hfill
	\subfloat[]{
		\includegraphics[width=0.4\textwidth]{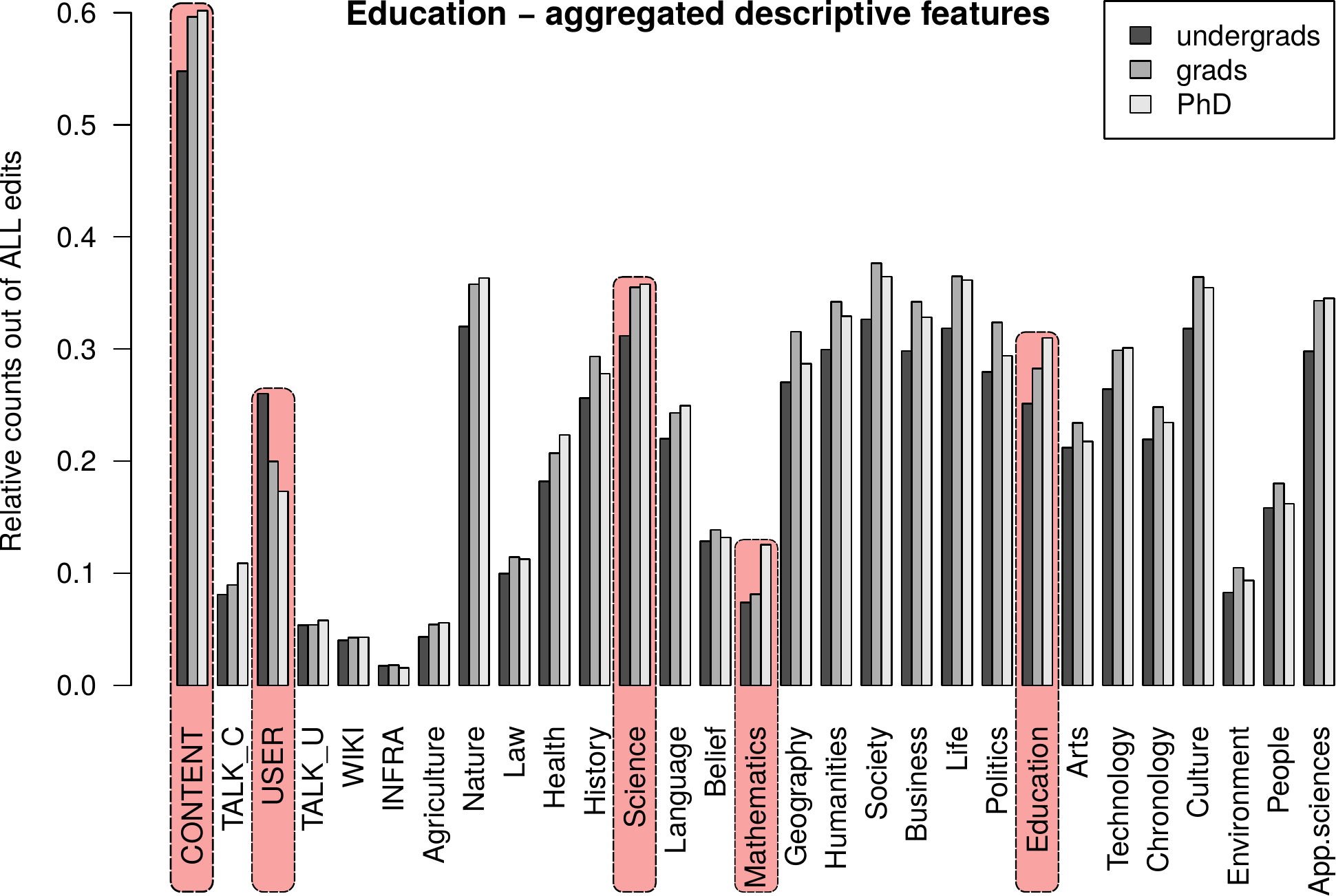}%
		\label{subfig:education-aggr-features}
	}
	\quad
	\subfloat[]{
		\includegraphics[width=0.4\textwidth]{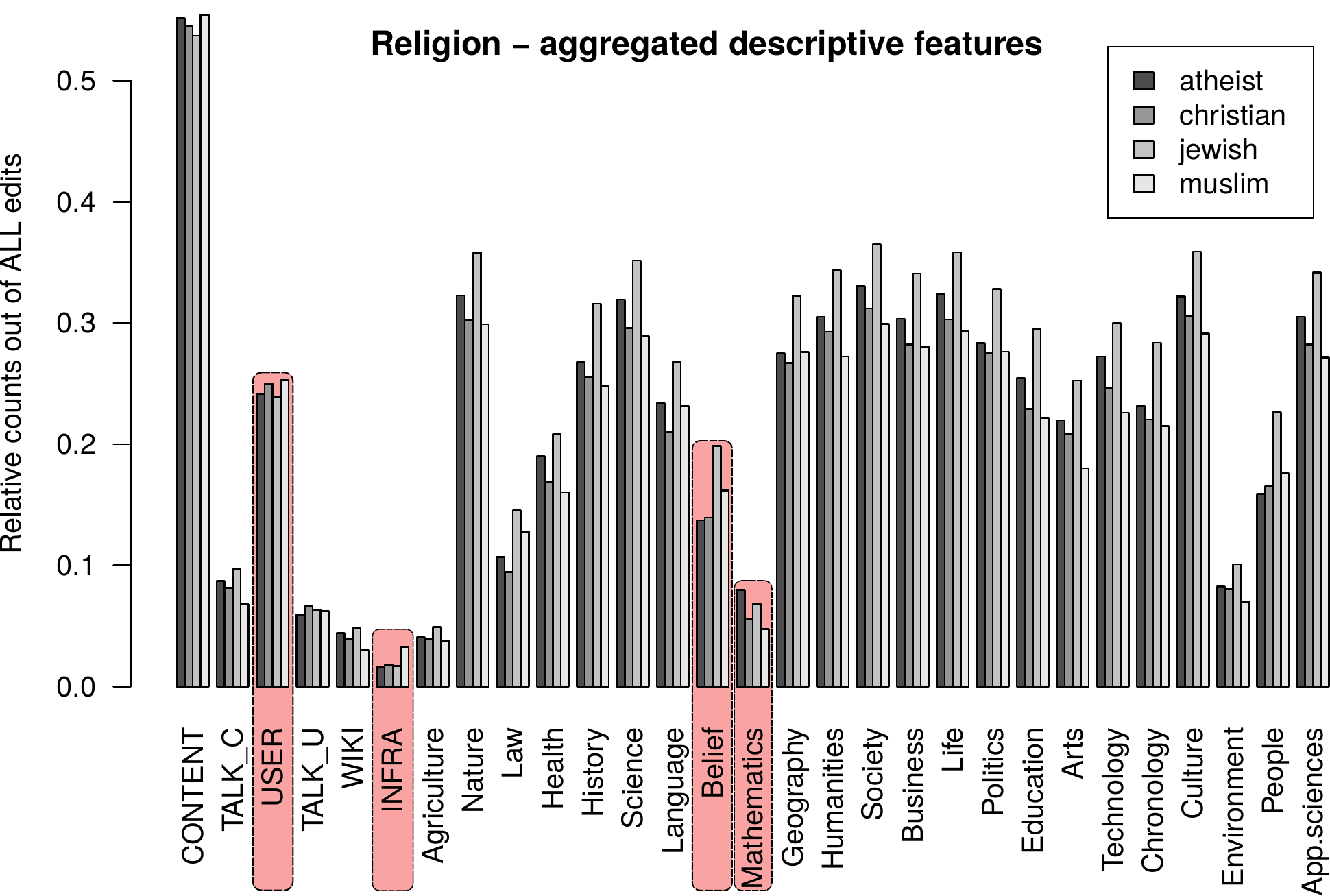}%
		\label{subfig:religion-aggr-features}
	}

	\caption{Aggregated descriptive features show differences in editing patterns, when tabulated per gender (a) and (b), education (c) and religion (d). 
	Features were computed as percentages out of the total revision count
%	(a) gender. For each editor, the \emph{basic} f and the mean is presented. 
%	(b) gender. Similarly to the \emph{basic} set, the \emph{extended} features were computed as percentages out of the \texttt{CONTENT} revisions.
%	The same features are aggregated by education level (c) and religion (d). 
	Particularly interesting features (\textit{i.e.}, features on which the separation is clearer or some patterns are inverted) are highlighted.	
	}
	\label{fig:private-traits-over-features}
%	\vspace{-0.2in}
\end{figure*}

%figure 5
\begin{figure*}[htbp]
	\centering
	\subfloat[]{
		\includegraphics[width=0.32\textwidth]{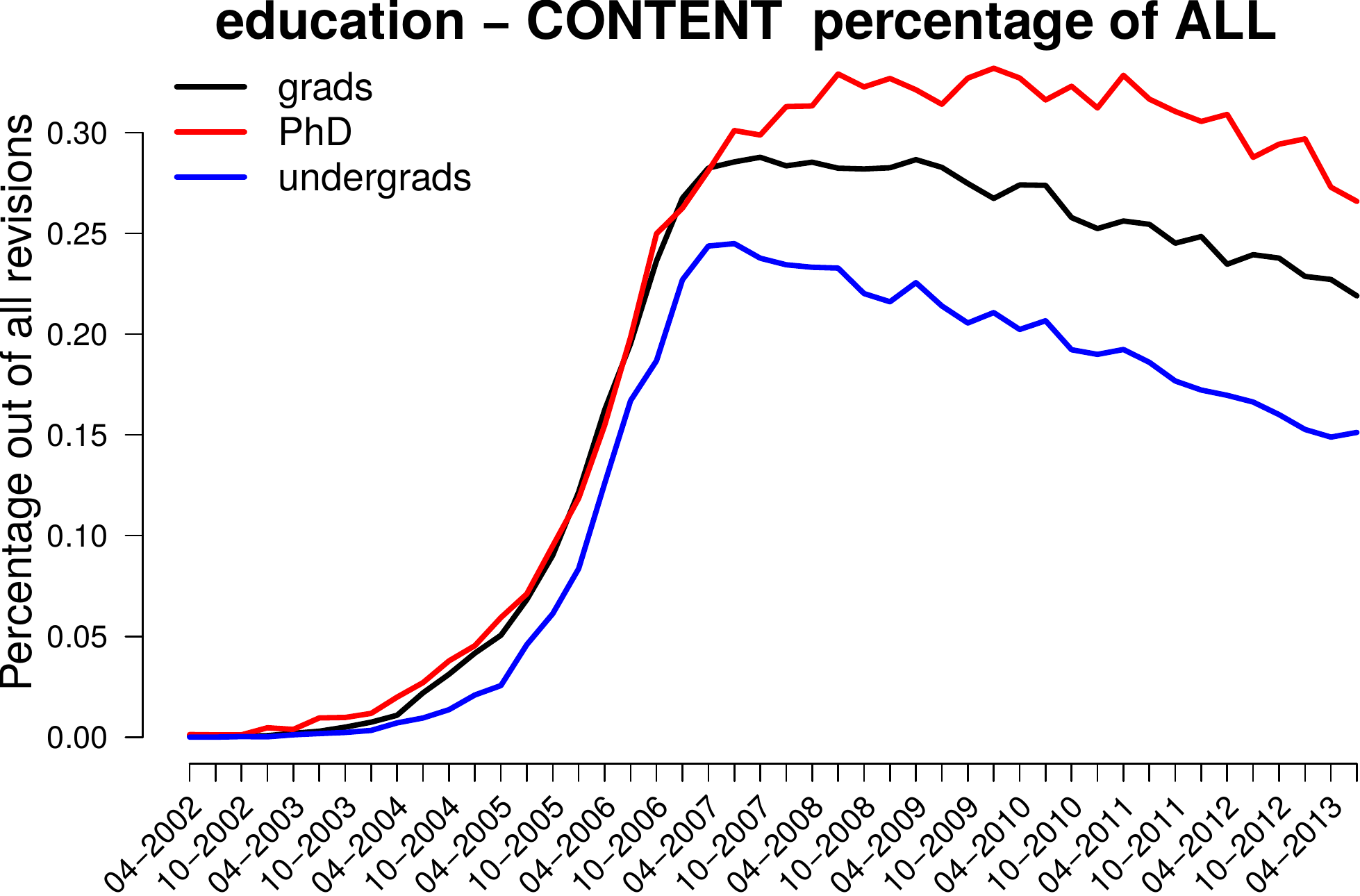}%
		\label{subfig:education-temp-feat}
	}
	\hfill
	\subfloat[] {
		\includegraphics[width=0.32\textwidth]{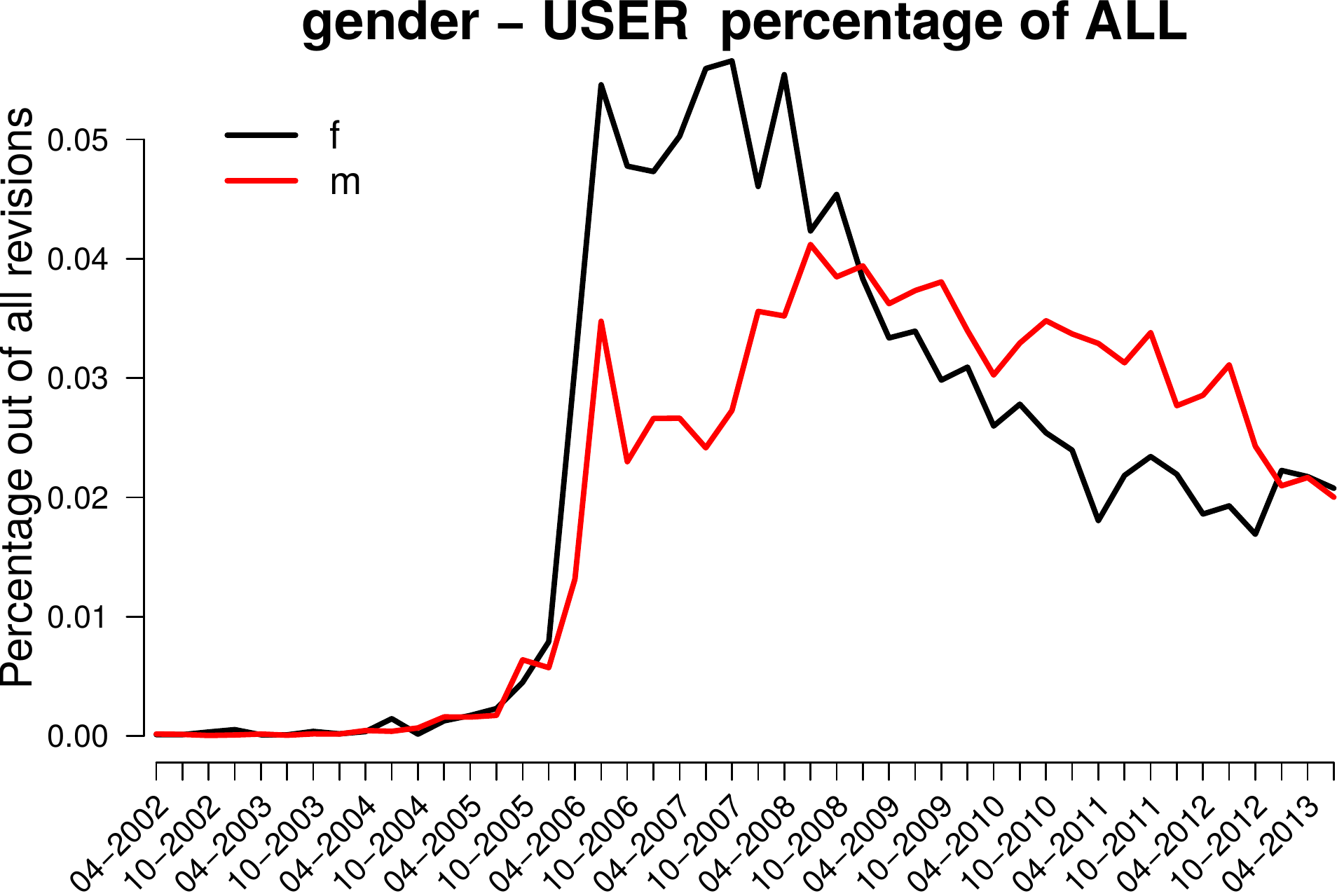}%
		\label{subfig:gender-temp-feat}
	}
	\hfill
	\subfloat[]{
		\includegraphics[width=0.32\textwidth]{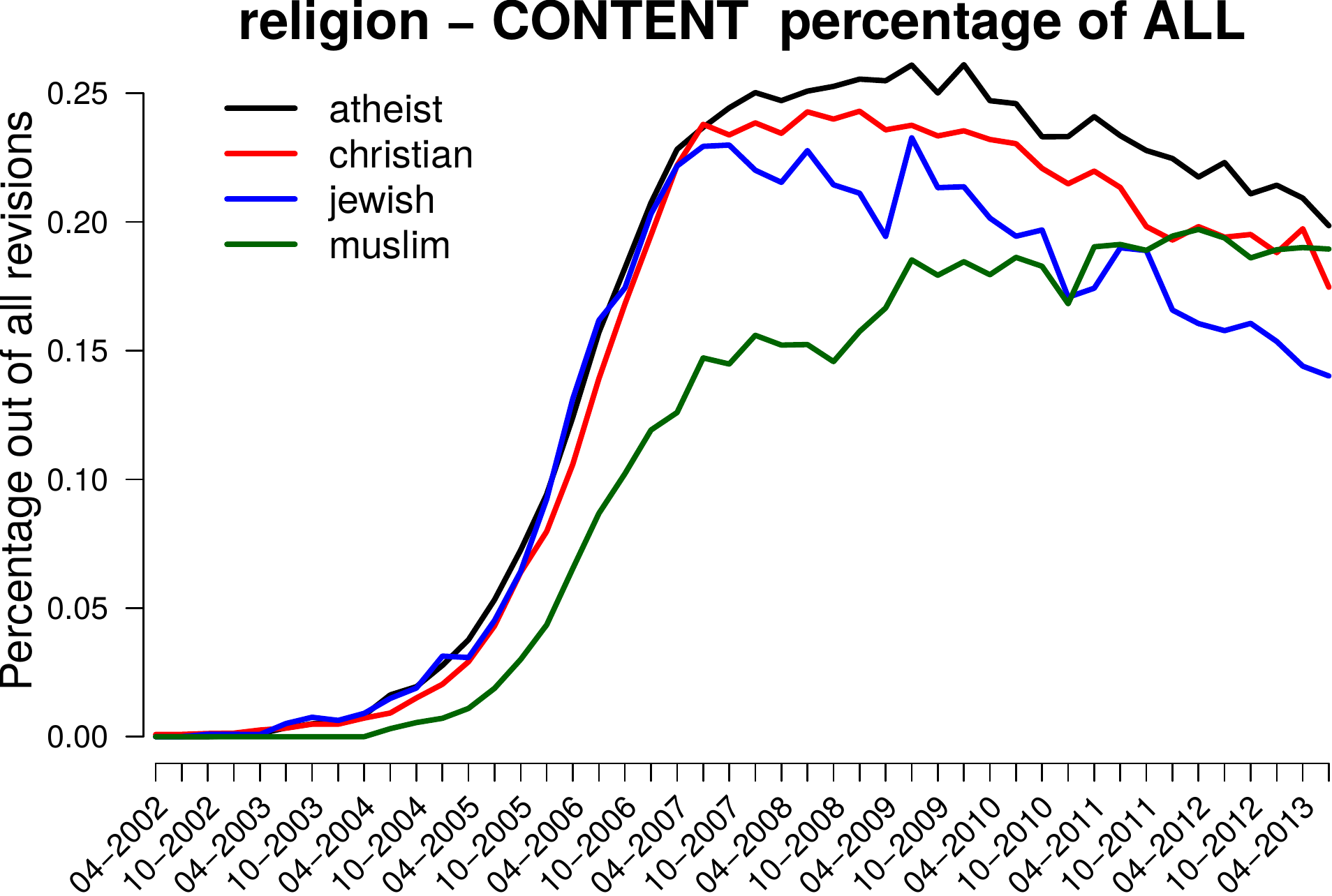}%
		\label{subfig:religion-temp-feat}
	}

	\caption{Temporal evolution of mean values of features, broken down by classes in each private trait.
	We present three selected examples of pairs (feature, trait), other graphics are in the SI~\cite{supplemental}: (a) (\texttt{CONTENT}, education), (b) (\texttt{USER}, gender ) and (c) (\texttt{CONTENT}, religion).
	All values are computed as the number of revisions on the given category during a timeframe and expressed as percentages of the total number of revisions.
	The mean value over all users is presented.}
	\label{fig:traits-temp-feat}\captionmoveup
%	\vspace{-0.2in}
\end{figure*}

\secmoveup
\section{Editing behaviour over time}
\label{sec:data}
\textmoveup

\rva{In this section, we present a profile of wikipedia editing behaviour over time. 
While our profile concur with the \rev{slowdown of} Wikipedia~\cite{Gibbons2012,Halfaker2012,Suh2009}, our analysis \rev{detects the rise of maintenance effort and} shows that different user groups contribute differently to various functional and topical sections of the encyclopedia.}

%\subsection{Wikipedia data profiles}
%\label{subsec:data-profiling}

\textbf{The decline of editorship and rise of maintenance.}
%After an initial period of rapid growth, starting from 2007 
\rva{It has been observed~\cite{Gibbons2012, Halfaker2012, Suh2009} that} Wikipedia's growth slowed since 2007, with fewer new editors joining, and fewer new articles created.
%This trend has been detected and studied previously. 
Our profiling \rva{shows the same phenomenon}.
Fig.~\ref{fig:wiki-profiling} \rva{shows the number of active editors, new editors, and number of edits over time.}\eat{shows the initial rapid growth tendency in the populations of new and active users.}
%%LX: first describe the plot, then describe the observation
A user is considered active in a time interval if she has submitted at least one revision in the given period. 
A user is considered a new users (or a newcomer) if she made her first revision in the given time interval. 
\eat{After the start of }\rva{We can see that}
the slowdown \rva{started in 2007}, with both the active population, and the total number of revisions\eat{related to the creation of knowledge }
decreasing steadily -- \rva{as seen in the volume of} \texttt{CONTENT} revisions in Fig.~\ref{subfig:profile-content-decrease}, and revisions to the {\tt Nature} section in Fig.~\ref{subfig:profile-NATURE-decrease}.
\eat{In addition to the conclusions of the previous studies, we detect that }
\rva{We can also see that} \rev{while} the \rva{overall} growth rate is slowing, the increasing amounts of accumulated information require an increasing effort to organize.
Fig.~\ref{subfig:profile-infra-increase} shows that the number of infrastructure-related revisions (\texttt{INFRA}) continues to increase, \rva{made by a decreasing number of users.}\eat{, despite the decreasing user population performing them.} 
\rev{To the best of our knowledge, this is the first work to detect and quantify this \emph{rise of maintenance}.}
%\rva{\bf Q: Is the increase of maintenance a *NEW* observation, or have been said before??}

%\textbf{The editor specialization hypothesis.}
%\rva{\textbf{Increased specialization of editors.}}
%%LX: this sec. started w. specialization, but is more general than that (e.g. male vs female)
\rva{\textbf{Different growth trends across editor demographics.}}
One explanation~\cite{Gibbons2012} for \rva{the} slowdown \rev{of Wikipedia} is that\eat{ all of} the easy articles have already been created. %%LX: quantify the statement, ``all of'' seem too strong here.
\rva{This means that} in order to make a novel, useful contribution to the site, editors must meet an increasingly high bar of expertise in \rva{the}\eat{their} field.
%%LX: swap sentences. again, first describe the plot, then state observations
Fig.~\ref{subfig:profile-phd-stay-active} plots \eat{the size of }the active population \rva{size}\eat{ of the} \rva{for users with a} declared education level.
%Our analysis of the number of active user broken down per levels of specialization seems to confirm that this hypothesis is one of the factors of the slowdown.
The three curves corresponding to \textit{undergrads}, \textit{graduates} and \textit{PhD} have been scaled in $[0, 1]$ to render them comparable.
All three population show the expected initial rapid increase.
The \textit{undergrads} and \textit{graduates} reach maximum at the same time as the general population \eat{and decrease, with the \textit{undergrads} observing a faster decay.}%%LX: not clear to me that you can say this. 
The more specialized \textit{PhD} population seems to peak much later, in early 2010,
\eat{, and to decreased only in recent times.}%%LX: this is vague and not entirely clear from the plot, best left out
\rev{which seems to confirm the hypothesis that the required increase in the specialization of editors is one of the factors responsible for the slowdown of Wikipedia.}
\eat{The phenomena seems to be modulated also by the editors' religion.}%%LX: ``modulated'' is vague
\rva{We can also see differing demographic trends across different groups in editors' religion and gender.}
In Fig.~\ref{fig:profile-muslim-population-increases}, the \rva{number of }\textit{christian}, \textit{jewish} and \textit{atheist} \eat{populations}\rva{editors} start to decrease \rva{around 2007}\eat{ about the same time and follow approximately the same decay rate}. On the other hand, the \rva{self-declared }\textit{muslim} population seems to continuously increase, at a slower pace than during Wikipedia's initial growth 2001-2007.
\rva{Fig.~\ref{subfig:profile-gender} plots} the number of active editors by gender, \rva{we can see} that the number of active female \rva{editors} started decreasing \eat{much} earlier \rva{ than that of male editors}.
%This agrees with the conclusions of~\cite{Halfaker2012} concerning 
\rev{While} gender imbalance in Wikipedia \rev{has been previously studied~\cite{Halfaker2012}, no other discussion of the evolution across time of gender, religion and education is present in prior literature.}
%\rva{\bf Need discussion: how do these observations differ or concur with [Gibbons2012, Halfaker2012]? is there anything new that people should take away?}

%\textbf{Aggregate analysis of features show potential to disclose private traits.}
{\bf Aggregated edit counts correlate with private traits.}
%We introduce a formal means to %% LX - redundant
We describe a user's editing activity by aggregating her revision counts over a number of predefined categories. 
%Is it possible to infer a public trait from this type of description? 
%Are these features informative about an editor's gender, religion or education level? 
%As a proof of concept, we first 
We conduct a\eat{straight-forward} exploratory analysis by presenting the averages of the features, with consideration for each of the private traits. 
%In 
Fig.~\ref{subfig:gender-over-features-basic}\eat{, the aggregate plot} shows differences between the average male and female behavior: 
%males edit more the \texttt{CONTENT} (article pages of Wikipedia) compared to females: 
59\% \rev{of all the revisions performed by males} are \texttt{CONTENT}, compared to 48\% for females. 
%However
Females tend to socially relate more, by writing\eat{ on their own user page and on others'} \rev{more on \texttt{USER}} pages (35\% of all revisions for females, \eat{compared to }less than 25\% for males).
\rev{Similarly, Fig.~\ref{subfig:gender-over-features-extended} presents average male (highlighted in blue) and female (highlighted in red) behavior, over features in the \emph{extended} set.}
%A similar analysis, over the \emph{extended} feature set, is presented in Fig.~\ref{subfig:gender-over-features-extended}.
Females edit more subjects like \texttt{Agriculture}, \texttt{Health}, \texttt{History}, \texttt{Language}, \texttt{Belief}, \texttt{Arts} and \texttt{People}, while males edit more \texttt{Mathematics}, \texttt{Society}, \texttt{Business}, \texttt{Geography} and \texttt{Culture}.
\rev{While the absolute differences between average behavior on gender tend to be rather small, they indicate a separability of the two classes.}
We perform\eat{ the same aggregation} \rev{a similar analysis} for education (Fig.~\ref{subfig:education-aggr-features}) and religion (Fig.~\ref{subfig:religion-aggr-features}):
%Editing patterns seem to emerge: 
\textit{undergrads} \rev{create less} \texttt{CONTENT} \rev{revisions}  and\eat{make up with the revisions on the social side (} \rev{more} \texttt{USER} \rev{revisions}.
\textit{graduates} and \textit{PhD} populations both dedicate a higher percentage of revisions to \texttt{CONTENT}.
The \textit{PhD} are more active on technical categories, such as \texttt{Science}, \texttt{Mathematics}, \texttt{Education}, \texttt{Technology}, \texttt{Health} and \texttt{Applied Sciences}, and the \textit{graduates} edit more subjects like \texttt{People}, \texttt{Environment}, \texttt{Culture}, \texttt{Society} and \texttt{Life}.
When aggregating per religion, Fig.~\ref{subfig:religion-aggr-features} shows that the \textit{jewish} editors are the most prolific in all thematic sections (the \emph{extended} feature set), except \texttt{Mathematics}.
\textit{muslim} editors dedicate higher attention to \texttt{Belief}, \texttt{Language}, \texttt{Law} and \texttt{INFRA}, and lower attention to \texttt{Arts}, \texttt{Education} and \texttt{Society}.
\textit{Atheist} users dedicate more time editing \texttt{Mathematics}, \texttt{Science}, \texttt{Nature} and \texttt{Culture}, and less time to \texttt{People} and \texttt{Law}.
\rev{This static analysis of mean behavior suggests that there are regularities in the editing patterns for each population.
These could be exploited for training a classifier and predicting weather a new user belongs to any of these classes.}

\begin{figure*}[htb]
	\centering
	\subfloat[] {
		\includegraphics[width=0.315\textwidth]{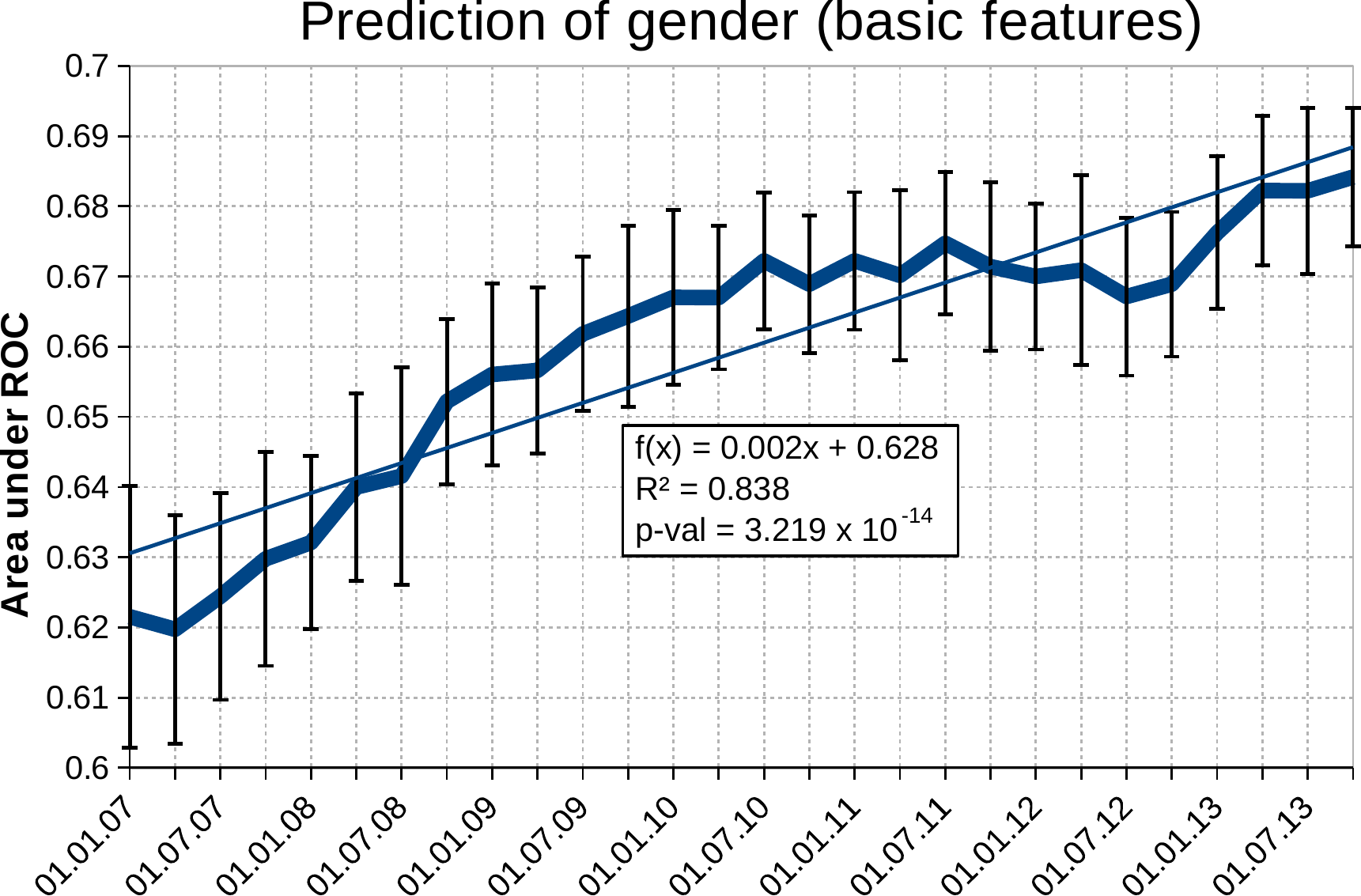}%
		\label{subfig:temporal-privacy-loss-gender}
	}
	\hfill
	\subfloat[]{
		\includegraphics[width=0.315\textwidth]{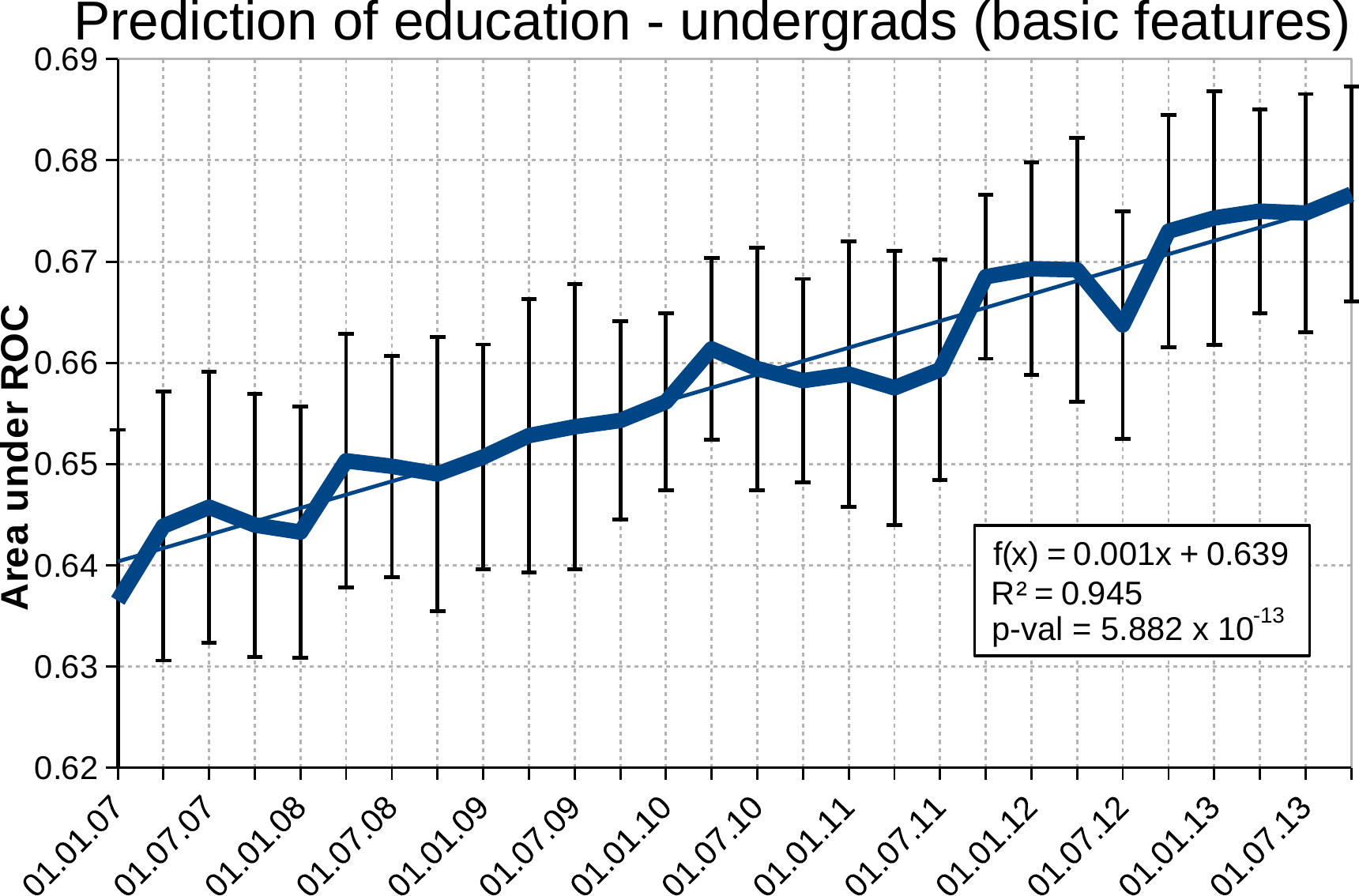}%
		\label{subfig:temporal-privacy-loss-education}
	}
	\hfill
	\subfloat[]{
		\includegraphics[width=0.315\textwidth]{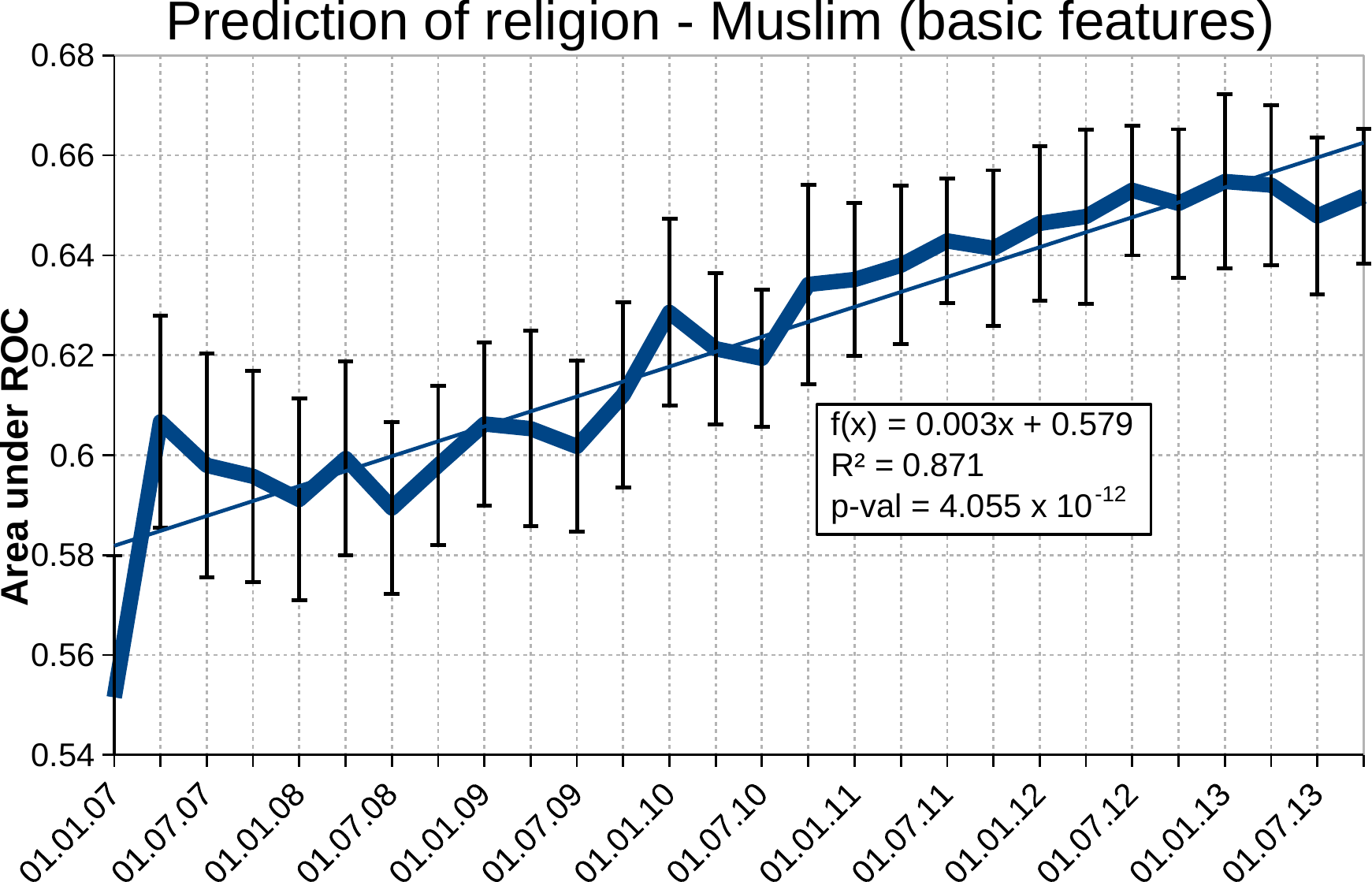}%
		\label{subfig:temporal-privacy-loss-religion}
	}

	\caption{\rev{Temporal evolution of privacy loss, measure using mean AUC value over 20 executions (error bars denote standard deviation)}.
	Result of inferring, using binary predictors on the \emph{basic} feature set, of gender (a), education/\textit{undergrads} (b) and religion/\textit{muslim} (c). 
	\rev{The results for all the other binary predictors are given} in the SI~\cite{supplemental}.}
	\label{fig:temporal-privacy-loss} \captionmoveup
%	\vspace{-0.2in}
\end{figure*}

\textbf{Evolution of editing patterns.}
\rev{We further study how editing patterns evolve over time.}
%Do these patterns evolve with time?
%Does the evolution increase or decrease the separability between classes?
%To answer these questions, We perform a similar \rev{mean behavior analysis for each timeframe} constructed as shown in Sec.~\ref{subsec:construct-datasets}.
%Similar conclusions can be drawn by performing the analysis from a temporal point of view. 
\rev{For each feature, we compute the mean number of revisions over each timeframe, broken down by class.}
%The number of revisions\eat{ performed by editors} during each\eat{ quarterly} timeframe on each feature are computed as percentages of the total number of revisions, and the mean value over all users is presented.
This value is still an aggregate measure over an entire subpopulation, \rev{but} it evolves temporally, therefore hinting changes in editing patterns.
Fig.~\ref{fig:traits-temp-feat} shows examples of temporal evolution \rev{for three} selected pairs (feature, private trait).
More examples are presented in the SI~\cite{supplemental}.
Fig.~\ref{subfig:education-temp-feat} shows the feature \texttt{\rev{CONTENT}} \rev{differentiated} over levels of education.
\rev{The static pattern shown in Fig.~\ref{subfig:education-aggr-features} \rva{is a result of change in }editing patterns over time}: 
%\textit{PhD} edit more than \textit{graduates}, who in turn edit more than \textit{undergrads}.
\rva{editors with a PhD edit \texttt{CONTENT} more as Wikipedia matures, as also shown in the editor population breakdown in Figure~\ref{subfig:profile-phd-stay-active}.}
%For \rev{other} features the editing patterns \rev{evolve with time (\textit{e.g.} the \texttt{USER} section, \rev{differentiated by} gender in Fig.~\ref{subfig:gender-temp-feat})}.
%The static analysis \rev{in Fig.~\ref{subfig:gender-over-features-basic}} showed that females \rev{in average} edit \texttt{\rev{USER}} pages \rev{more} than males, but the temporal analysis shows this to be true only until early 2009, after which the two series take comparable values.
\rva{For other features the editing pattern evolves over time.  
Contrasting Fig~\ref{subfig:gender-temp-feat} and Fig~\ref{subfig:gender-over-features-basic}, we can see the differentiation of edits to the USER section by gender -- female users edit more overall than male users, but this is only true until 2009.}
\rev{\texttt{TALK-C} and \texttt{BELIEF} also present the same temporal pattern shift, as shown in the SI~\cite{supplemental}.}
Finally, some \rev{features present} 
%not only a relative evolution between the different classes, but even 
unexpected trends.
%For example, 
Fig.~\ref{subfig:religion-temp-feat} unveils that, unlike the general trend of decreasing number of contributions, \rev{\texttt{CONTENT} related edits increase for \textit{muslim} editors}.
\rev{This temporal analysis reinforces the hypothesis that user editing patterns, as well as their evolution, \rva{are differentiated along different user traits}.}
%
%This aggregated analysis confirms that the proposed descriptive features have the potential to predict an editor's gender, religion or level of education. 
%The next step is to study \rev{this} link\eat{ between a user description and the private traits} at the individual level.
%Furthermore, we study how to infer the gender, religion or education of new users.
%
%The temporal analysis reinforces the conclusions of the aggregated static analysis about the connection of private traits and editing patterns.

%!TEX root = WSDM_2015_wikipedia.tex

\secmoveup
\section{Prediction results}
\label{sec:results}
\textmoveup

%\rva{TODO: add+revise a short pre-amble.} 
%We study the predictability of personal traits over time: 
%We further \rev{study whether and how} do data of increasingly longer temporal extent provide more private information about the users.
%Our hypothesis is that two factors contribute to the privacy loss: 
%i) \rev{the behavioral patterns learned from the} new users \rev{who} enter the population and 
%ii) the information revealed by a user's own behavior over time (``online breadcrumbs'').
\rva{We predict personal traits of editors using behavioral features described in Sec.~\ref{subsec:encoding-features}. 
We report the prediction performance over time using different features. We also perform feature relevance analysis using the information transfer metric to pin point the source of performance gain over time. }

%Detection of the temporal loss of privacy}
%Predicting private traits over time
%Degradation of privacy over time
\subsection{\rva{Predicting personal attributes over time}}

\label{subsec:edit-wikipedia}

\begin{figure}[tbp]
	\centering
	\includegraphics[width=0.485\textwidth]{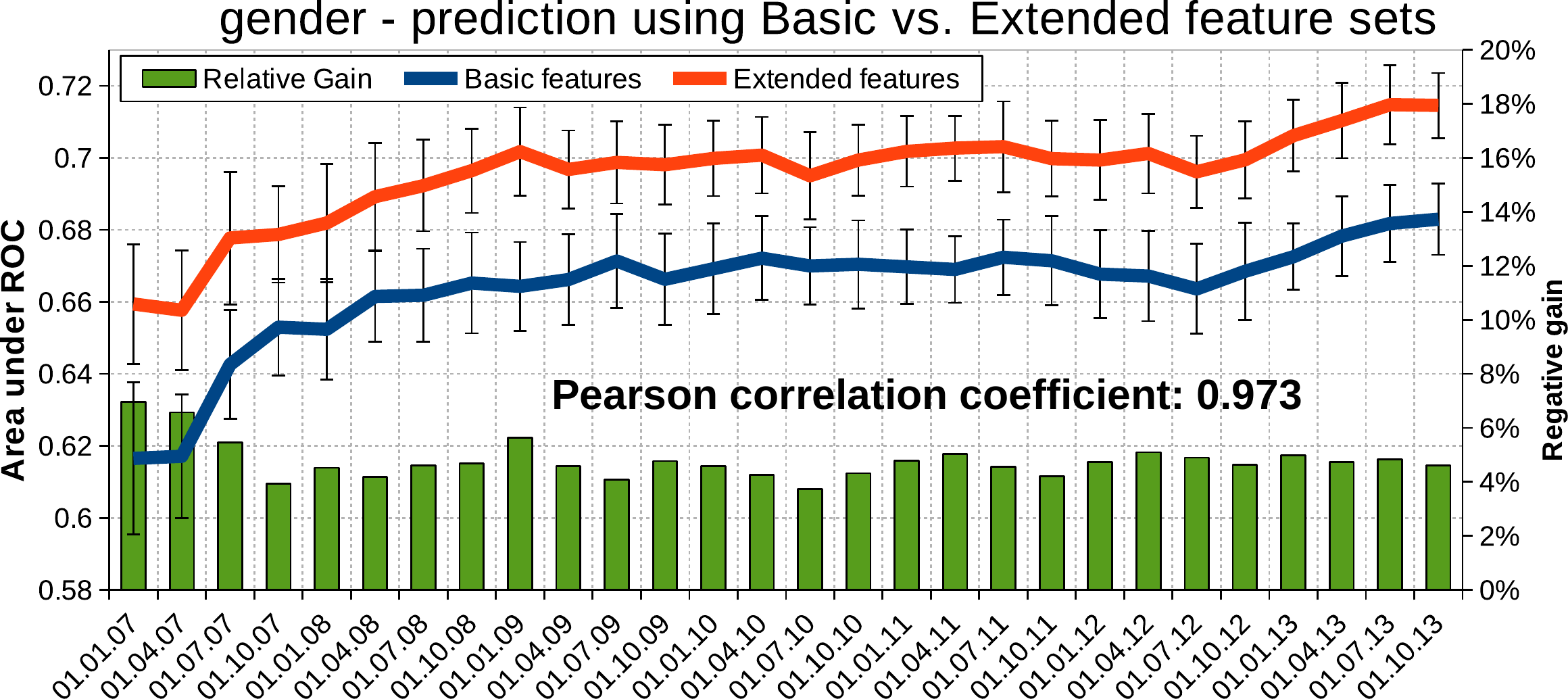}%
	
	\caption{Comparison of the temporal evolution for the privacy loss on gender for the \emph{basic} and \emph{extended} feature sets. 
	The \emph{extended} feature set consistently provides better performances, \rev{while} the AUC series of the predictors trained on the two feature sets are highly correlated and present the same trends.}
	\label{fig:temporal-privacy-basic-vs-extended} \captionmoveup
\end{figure}

%\textbf{Editing Wikipedia discloses private information.}
\textbf{\rev{Predictability of private traits improves over time.}}
We train binary predictors for every class of every private trait in our study, on datasets \rva{with} increasing amounts of history \rev{(as shown in Sec.~\ref{subsec:experiment-setup})}. 
The editors' activity\eat{ in each temporal dataset are } \rev{is} described using the \emph{basic} feature set. 
Fig.~\ref{fig:temporal-privacy-loss} shows the AUC \rev{over time for} three selected \rev{examples} of private traits (\rev{all remaining classes} are in the SI~\cite{supplemental}).
\rev{The graphics show the performance of predicting} the gender of editors (Fig.~\ref{subfig:temporal-privacy-loss-gender}), whether they are \textit{undergrads} (Fig.~\ref{subfig:temporal-privacy-loss-education}) or \rev{of} \textit{muslim} \rev{religion} (Fig.~\ref{subfig:temporal-privacy-loss-religion}).
\rev{The AUC measure increases over time, \rva{roughly} following a linear trend (coefficient of determination $R^2 > 0.83$ for all three examples).
The AUC differences\eat{ of performance} between the first and the last timeframes are statistically highly significant (t test $p < 0.001$, details and results in the SI~\cite{supplemental}).
We interpret this steady increase of performance over time as \emph{loss of privacy}:} as more historical information is available, the learning algorithm \rev{infers more accurately user traits which are potentially private}.

\rev{\rva{We also report the model performance on an intuitive measure called} {\em equal Precision and Recall} (ePR), \rva{defined as where a 45 degree line from the origin intersect with the precision-recall curve.}
%lx: explain why we like this measure, then define it
%alternatively evaluate the accuracy of our prediction using the ``equal Precision and Recall'' (ePR) measure.
%%LX: we do not need to do a line search
%We perform a line search for the cutoff point where the Precision and Recall are equal.
%LX: the sentence below is redundant
%Each model ranks the individual in the test set according to the probability of them belonging to the target class.  
For the three classes in Fig.~\ref{fig:temporal-privacy-loss}, in the last timeframe, we obtain a mean ePR of $0.791$ (for gender), $0.535$ (for education/\textit{undergrads}) and $0.9$ (for religion/\textit{muslim}).
All the other classes and the evolution of ePR over time are found in the the SI~\cite{supplemental}).
The ePR measure allows to quantify how accurate are the predictions of certain private traits.
We find that certain religious \eat{features}\rva{attributes %are predicted particularly accurate 
have notably high prediction performance} 
(\textit{muslim} $ePR = 0.9$ and \textit{jewish} $ePR = 0.913$) \rva{from behavioral features}}. 
%All graphics show a clear increasing trend, 
%is capable of inferring increasingly more accurately 
%The evolution of privacy loss 
%the relative gain between the first and the last timeframe being up to $18.1\%$ for predicting \textit{muslim}.

\begin{figure}[tbp]
	\centering
	\includegraphics[width=0.485\textwidth]{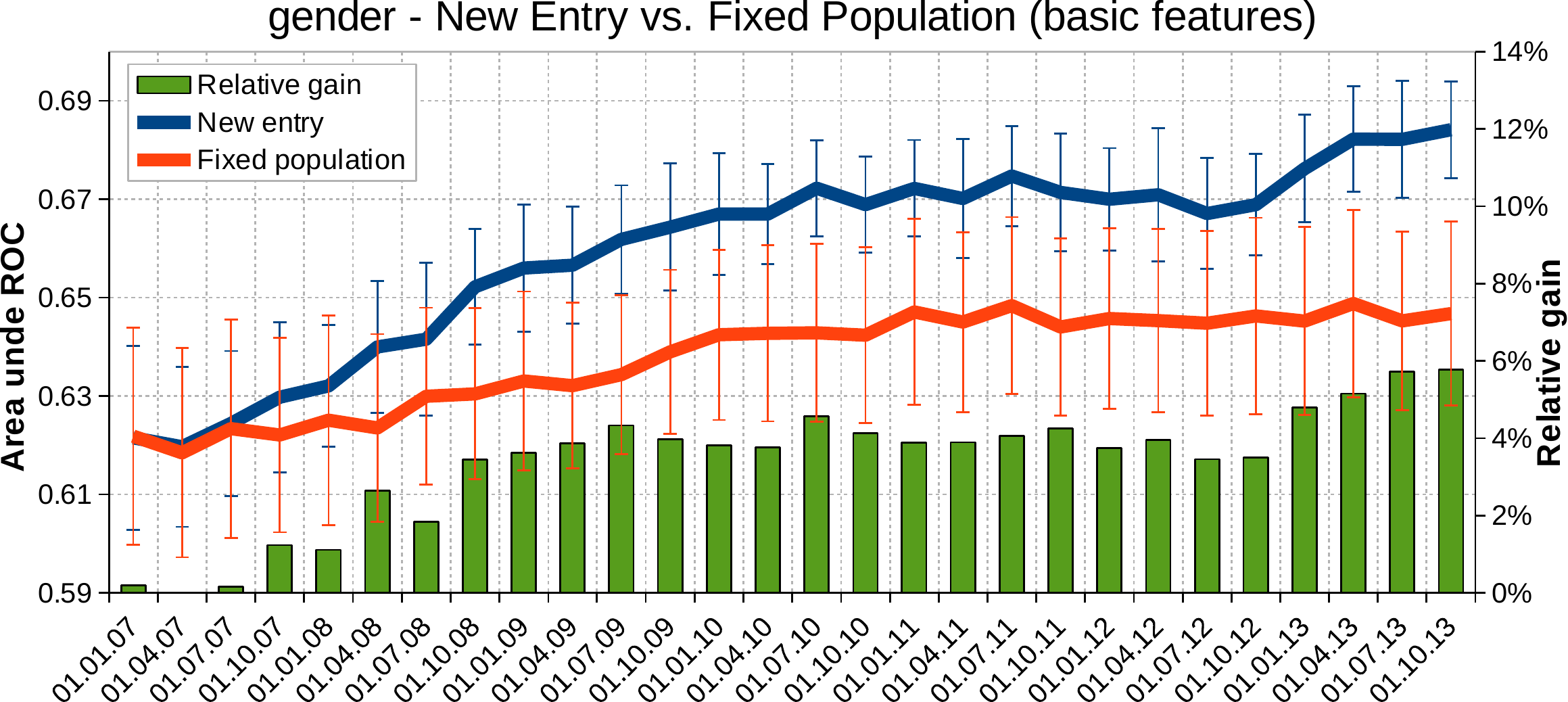}%
	
	\caption{Evolution of privacy loss for the population fixed to its component in the first quarter of 2007 (\textit{i.e.}, no newcomers) and a population in which new users can enter. 
	We quantify the privacy loss due to newcomers as the relative gain of learning performance for the two populations.}
	\label{fig:new-entry-vs-fixed} \captionmoveup
\end{figure}

\rev{In a similar setup, we predict user gender using the \emph{extended} feature set and we plot the AUC over time in Fig.~\ref{fig:temporal-privacy-basic-vs-extended}.
Alongside, we produce the results for} the \emph{basic} feature set and the relative gain between the two, for each timeframe.
%We perform the same evaluation, using the \emph{extended} feature set. 
%The evolution of the prediction of gender is plotted in Fig.~\ref{fig:temporal-privacy-basic-vs-extended}, alongside the evolution for the \emph{basic} feature set and the relative gain between the two, for each timeframe. 
%As expected, 
\rev{Predictions using} the \emph{extended} features consistently \rev{outperform those using} the \emph{basic} \rev{features}, \rev{showing} that knowledge about thematic editing patterns\eat{ of users} is
%more 
informative about \rev{user} private traits. 
\rev{The AUC over time series} for the two types of features sets
%, the AUC increasing trend is very similar and their AUC score series 
are highly correlated (Pearson correlation of $0.973$).
This shows that, while adding the thematic information improves the absolute value of the prediction accuracy, it \rev{seems to have} little influence on the evolution of the privacy loss. 
We speculate that \rev{the evolution of} privacy loss is not linked to the way the data is described, but it is rather intrinsic to the online social environment.
\rev{To the best of our knowledge, this is the first study to highlight and quantify this \emph{intrinsic cumulative effect} of time over privacy.}
%
%\textcolor{red}{confirming our initial intuition about the adverse cumulative effect of time over privacy.}

\begin{figure*}[htbp]
	\centering
	\subfloat[] {
		\includegraphics[height=0.165\textheight]{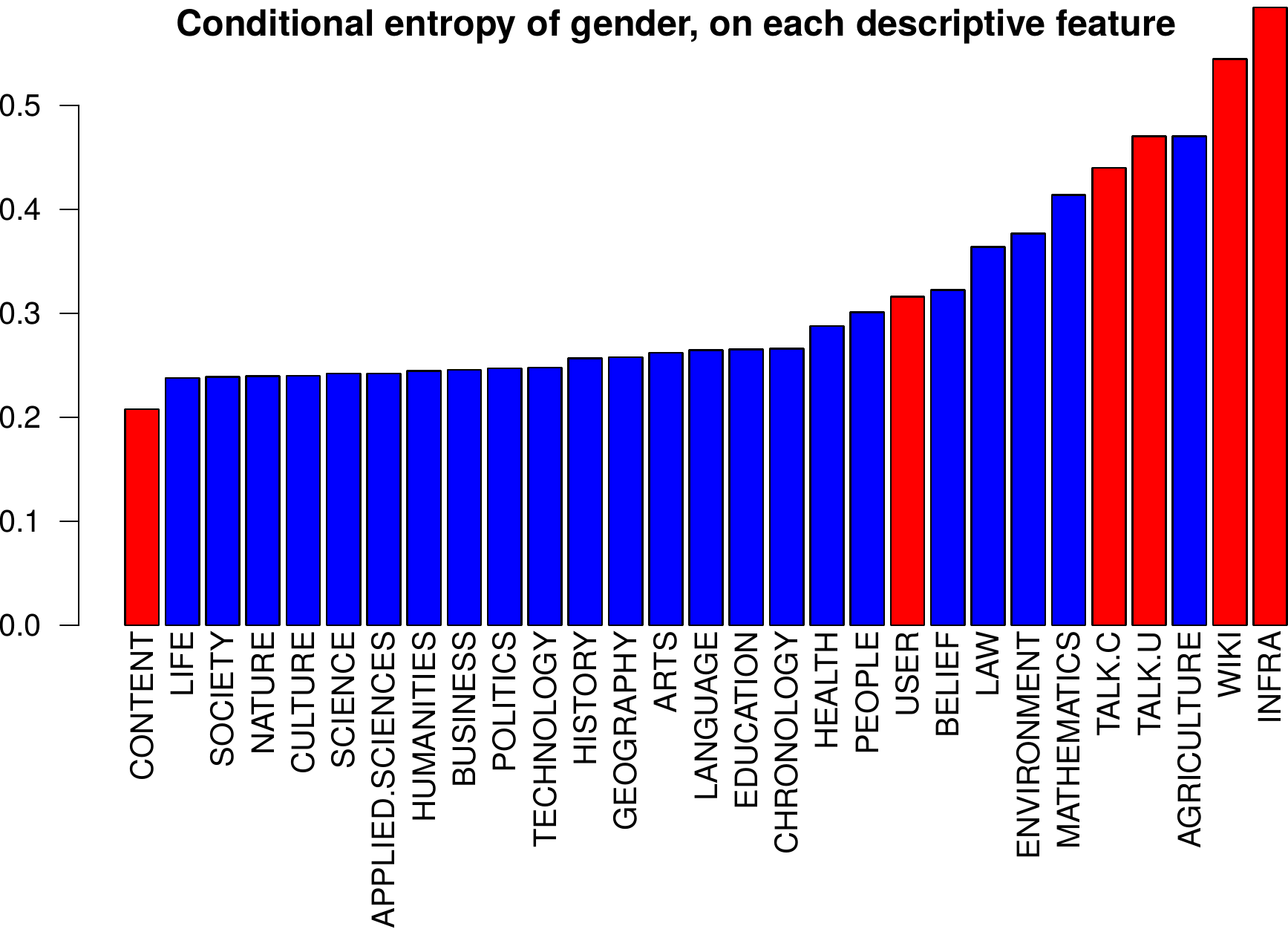}%
		\label{subfig:info-theo-total-remaining-unexplained-entropy-new-entry}
	}
	\hfill
	\subfloat[]{
		\includegraphics[height=0.165\textheight]{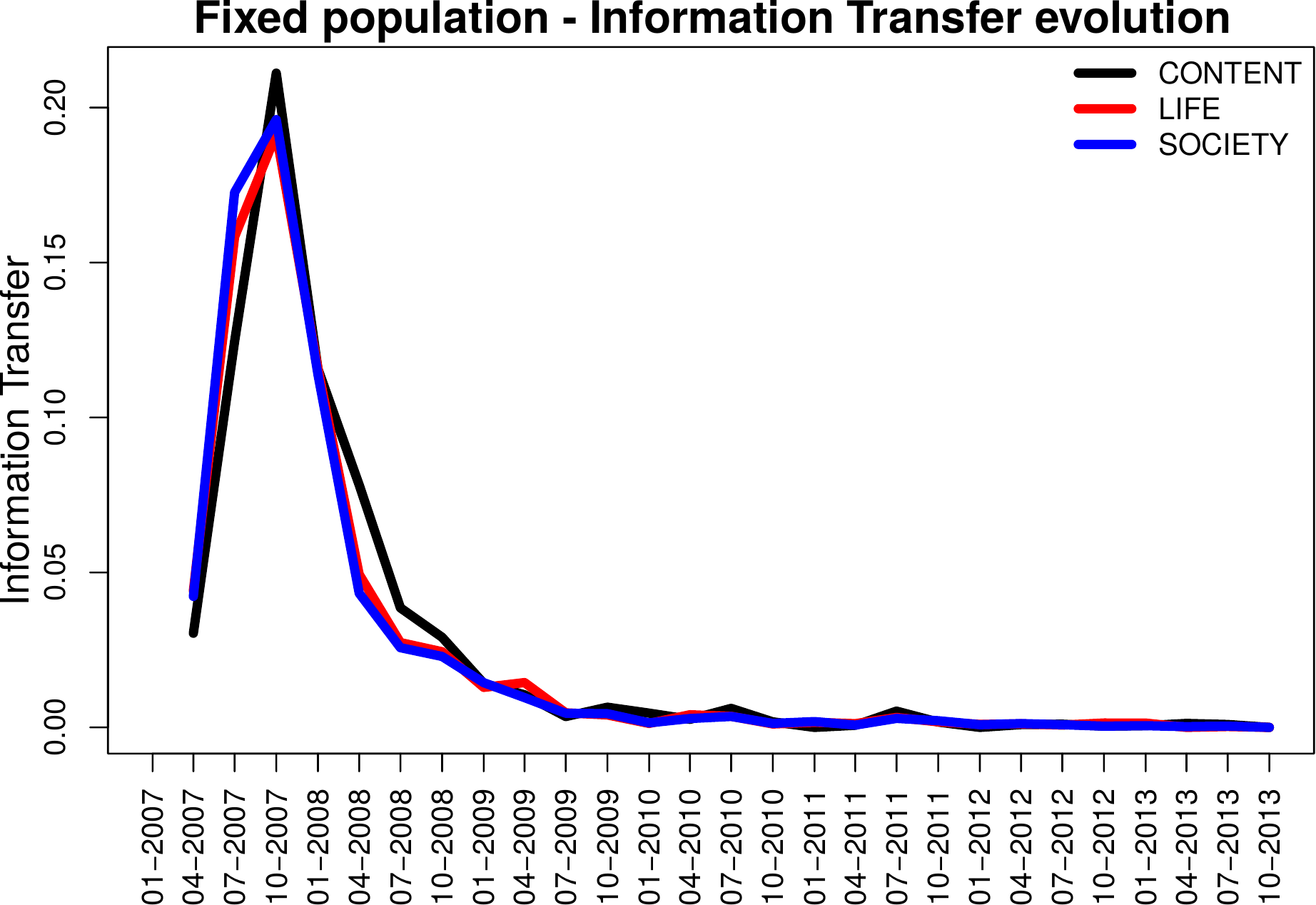}%
		\label{subfig:info-theo-information-transfer-evolution-fixed-population}
	}
	\hfill
	\subfloat[]{
		\includegraphics[height=0.165\textheight]{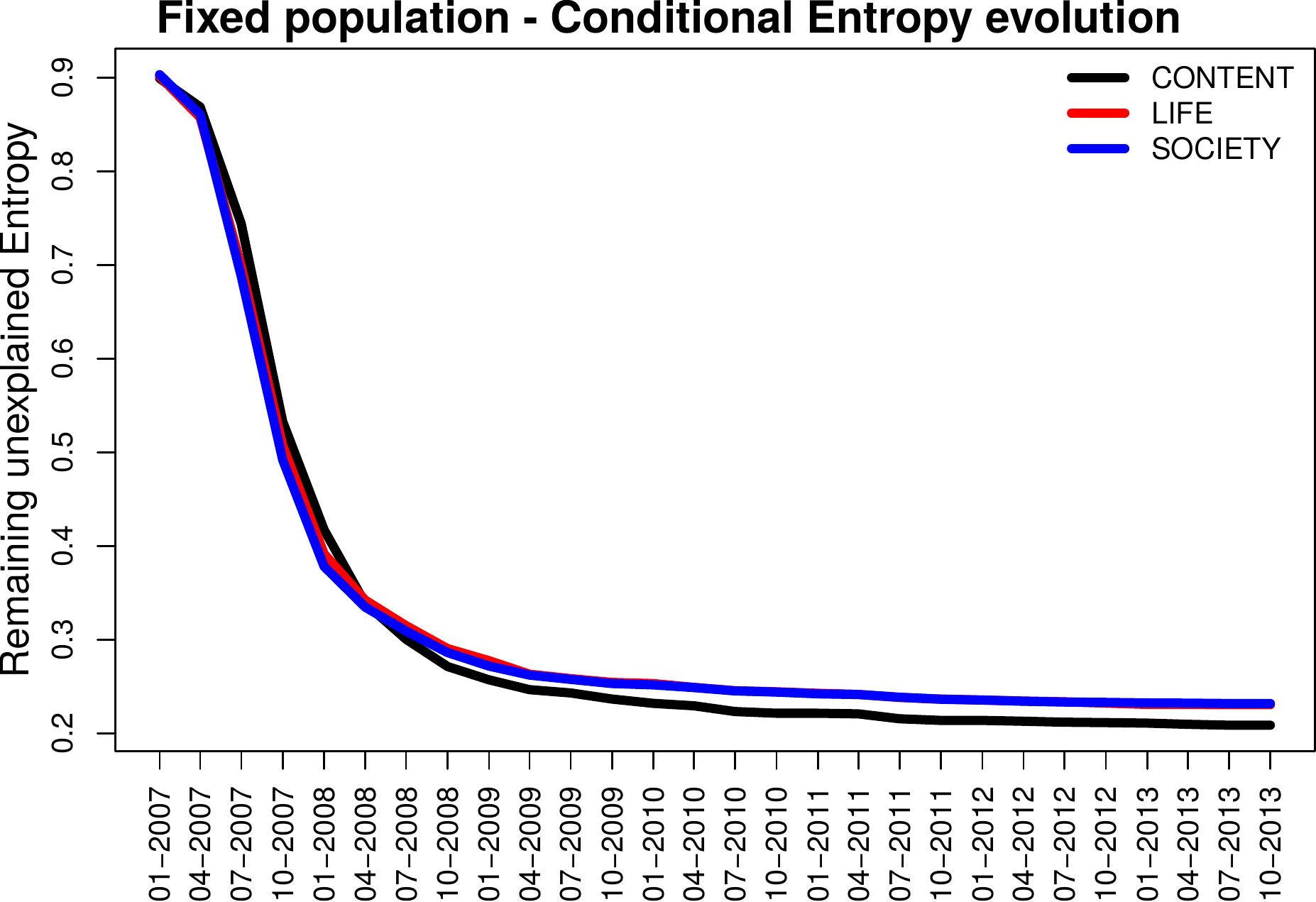}%
		\label{subfig:info-theo-remaining-unexplained-entropy-evolution-fixed-population}
	}
	\hfill
	\subfloat[] {
		\includegraphics[height=0.163\textheight]{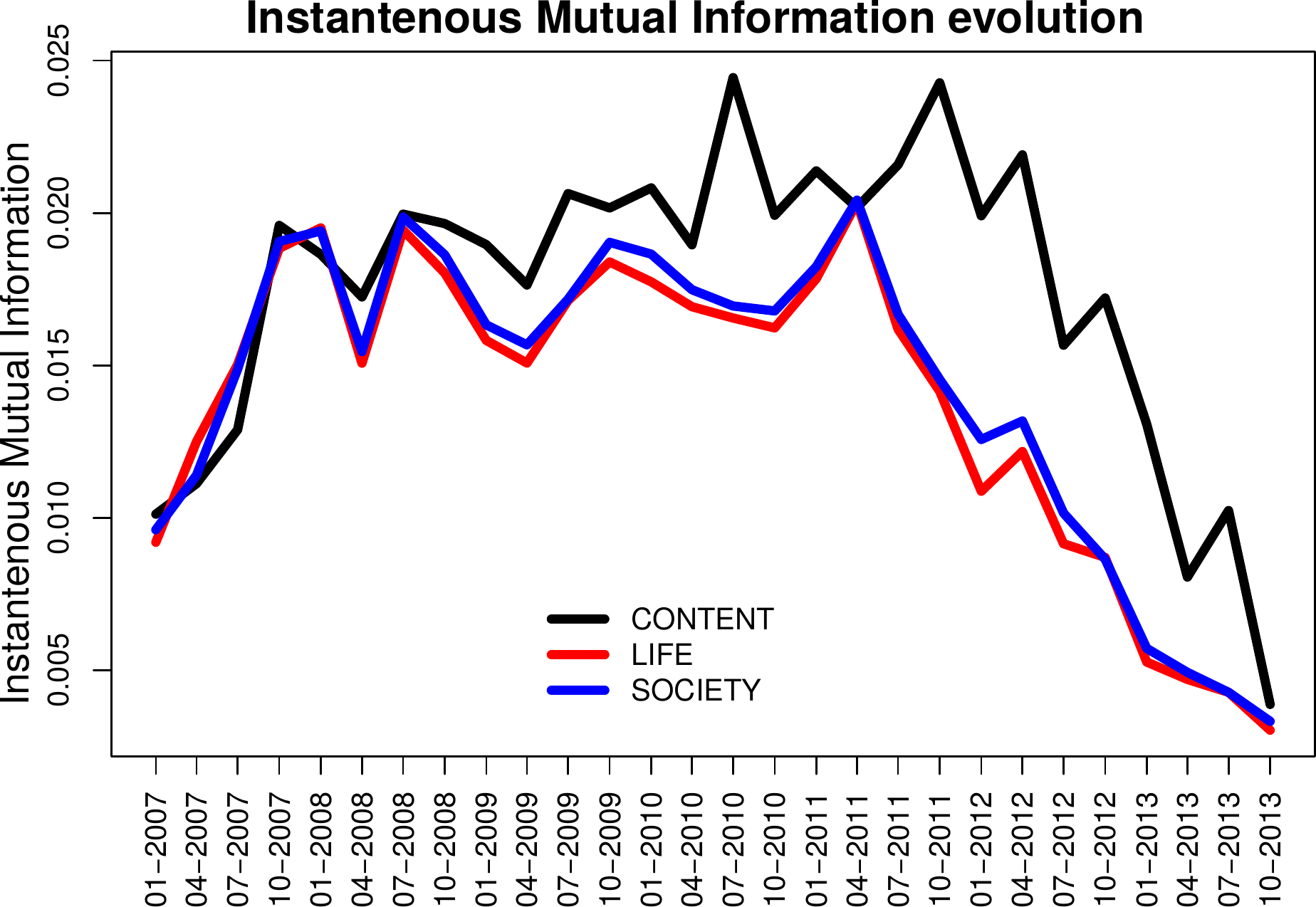}%
		\label{subfig:info-theo-instantaneous-mutual-information-new-entry}
	}
	\hfill
	\subfloat[]{
		\includegraphics[height=0.163\textheight]{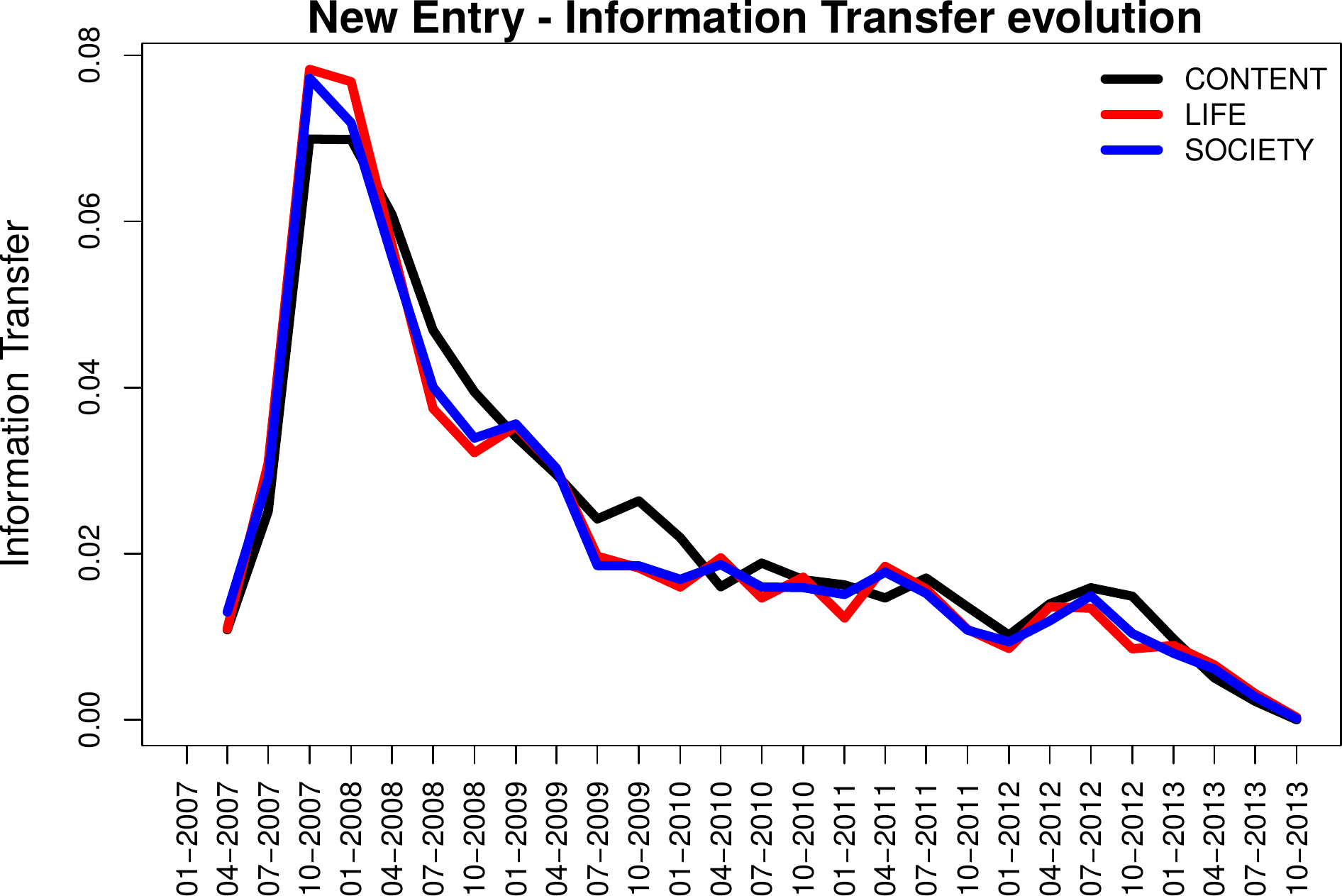}%
		\label{subfig:info-theo-information-transfer-evolution-new-entry}
	}
	\hfill
	\subfloat[]{
		\includegraphics[height=0.163\textheight]{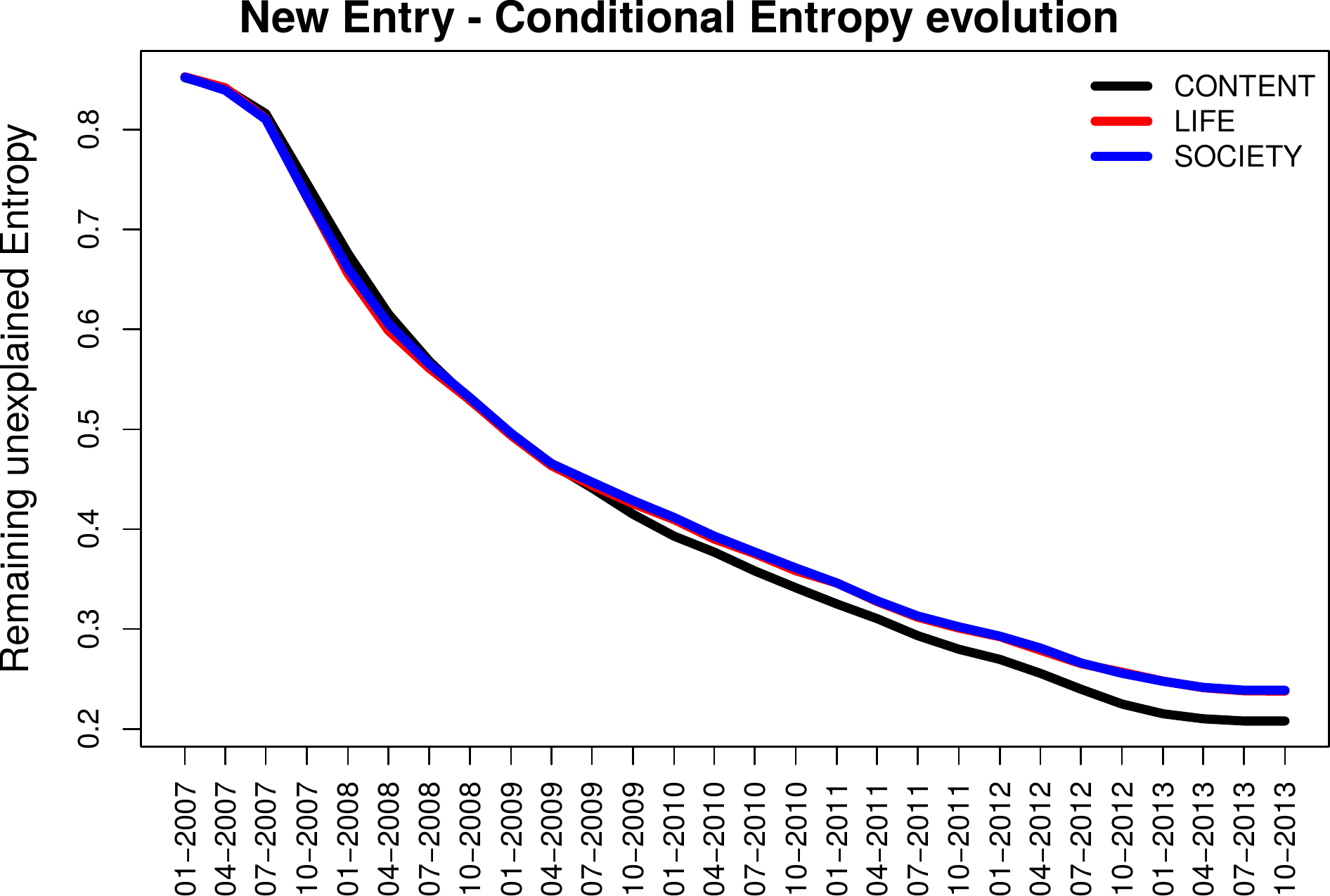}%
		\label{subfig:info-theo-remaining-unexplained-entropy-evolution-new-entry}
	}

	\caption{
	%gender - Analyzing the ``Fixed population'' and ``New entry'' using Temporal Information Theory measures. 
	(a) Conditional entropy after conditioning on each feature (with New Entry \texttt{NE}). 
	Feature are \rva{ordered by the conditional entropy $H(Y|X)$, here $Y = 0/1$ is the gender attribute, $X$ is each feature.
	Lower is better.}
%	The lower the conditional entropy, the more informative the feature is.
	\emph{Basic} features are shown in red, \emph{extended} features in blue. 
	(d) Mutual Information between each feature at time $t$ and gender (on \texttt{NE}). 
	\emph{Information Transfer} on Fixed Population \texttt{FP} (b) and \texttt{NE} (e).
	Conditional entropy evolution on \texttt{FP} (c) and \texttt{NE} (f). We can see that while later edits contains just as much information about a user's privacy as the earlier edits, they \revOLD{contribute less to prediction gain}, since most of the information they bring was already known. %been learned
	}
	\label{fig:info-theory-measures} \captionmoveup
	\vspace{-0.2in}
\end{figure*}

\textbf{Sources of privacy loss: the \emph{online breadcrumbs} and newcomers.}
\rev{We hypothesize that}
%Our hypothesis is that 
the temporal privacy loss\eat{detected earlier} is caused by \rva{a joint effect of} two factors:
i) the information learned from new users who enter the population and
ii) \rev{online breadcrumbs --} 
%the information learned from the user's own behavior.
\rva{a more accurate estimate of how the behavioral features correspond to personal traits.}
To separate these factors, we 
%construct two scenarios of populations: 
\rva{study two scenarios defined by subsets of the editor population:}
the \emph{``New Entry''} (\texttt{NE}) in which new users can enter freely throughout time and \emph{``Fixed population''} (\texttt{FP}), which is limited only to users active in the first timeframe (\textit{i.e.}, first quarter of 2007).
%In order to separate the impact of each factor, we perform a similar experiment on a fixed population of users. 
%In the ``Fixed population'' (\texttt{FP}), we limit the studied population to only the users in a activity in the first timeframe. 
%No new users are allowed to enter the population. 
In Fig.~\ref{fig:new-entry-vs-fixed}, we plot the \rev{AUC over time} when studied on \texttt{FP} and we compare it to \texttt{NE}. 
\rev{The curve corresponding to \texttt{FP} increases over time, though slower in later timeframes.
The difference between the first and last timeframe is statistically very significant -- t test $p < 0.01$.}
%during the first timeframes \rev{(t test $p < 0.01$)} and becomes constant afterwards. 
%We attribute this initial burst of privacy loss to the user's own activity.
\rev{Intuitively, only the online breadcrumbs could cause this privacy decay for \texttt{FP}.}
By comparison, predictions on \texttt{NE} are constantly \rev{more accurate} and continue to \rev{notably} improve \rev{beyond} the initial burst detected for \texttt{FP}.
We attribute this \rev{additional} improvement to information relating to new users entering the system.

%Why just an initial burst? 
% Does it mean that 
\textbf{\revOLD{Are} later edits less harmful to one's privacy?}
Intuitively, the longer a user edits, the more she discloses about herself.
We approach this issue using the Temporal Information Theory measure detailed in Sec.~\ref{subsec:measure-privacy-loss}, on both \texttt{FP} and \texttt{NE}  populations.
%Consider 
\rev{Let $T$ be} the number of
%3-months temporal datasets that we construct.
\rev{timeframes}.
Fig.~\ref{subfig:info-theo-total-remaining-unexplained-entropy-new-entry} shows the conditional entropy of the class variable $Y$ (here gender), after conditioning on all the temporal instantiations of a given feature $X$ (\textit{i.e.} $H(Y | X_{1:T})$), \rev{on the \texttt{NE} dataset.}
%In other words, $H(Y | X_{1:T})$.
We \rva{can see which} features which give out the most information about a user's gender.
\rva{Consistent with} the data profile in Sec.~\ref{sec:data}, \texttt{CONTENT} differentiates the most \textit{males} from \textit{females}.
The thematic features, like \texttt{Life}, \texttt{Society}, \texttt{Nature} or \texttt{Culture} follow, having very similar scores.
%One surprise is 
\rev{Surprisingly}, \texttt{USER} distinguishes gender less than \rva{seen from} the profiling analysis.
%The presented results were obtained on the \texttt{NE} population.
An almost identical ordering of importance of features is obtained on \texttt{FP}.
Fig.~\ref{subfig:info-theo-instantaneous-mutual-information-new-entry} presents the instantaneous mutual information \rev{over time} between the gender variable and the three most important features.
All three present very similar dynamics (both on \texttt{NE} and on \texttt{FP}).
The information overlap between the temporal instantiation of features and the private trait remains almost constant until close to the end of the studied period.
\rev{This answers the questions whether later edits are less harmful, by showing} that later activity \rev{hurts privacy as much as} the initial activity, as it discloses similar quantities of information.
\rev{We further study the amount of \emph{new} information introduced features in later timeframes.}
%\textbf{The marginal utility of newer features decreases over time.}
Fig.~\ref{subfig:info-theo-information-transfer-evolution-fixed-population} and~\ref{subfig:info-theo-information-transfer-evolution-new-entry} show that the \emph{Information Transfer} \rev{over time on respectively \texttt{FP} and \texttt{NE}, for the same three features}.
\rev{All series} present an initial burst, after which \rev{they} drop quickly.
Note that, due to differences in the effectives of the \texttt{FP} and \texttt{NE} populations, the absolute values 
%of measures obtained on each population 
are not directly comparable and only their evolutions are meaningful.
\rva{We can see} \rev{that later features $X_t$ bring few \emph{new} information not already disclosed by the earlier features $X_{t-1}, X_{t-2} \ldots$.}
\rva{The take-home message for this subsection is:}
{\em While later edits contain just as much information about a user's privacy as the earlier edits, they tend to be less harmful since most of the information they bring has already been learned}.

%This behavior, together with the conclusion of the previous paragraph, signals that, while each $X_t$ brings as much information as $X_{t-1}, X_{t-2} \ldots$, it brings few new information not already disclosed by the earlier features.

\textbf{\rev{The continuous impact of newcomers on privacy loss.}}
\rev{Information Theory measures provide means for separately quantifying the privacy loss due ``online breadcrumbs'' and newcomers.}
\rev{For} \texttt{FP} (Fig.~\ref{subfig:info-theo-information-transfer-evolution-fixed-population}), the utility of later edits drops to virtually zero, whereas \rev{for} \texttt{NE} (Fig.~\ref{subfig:info-theo-information-transfer-evolution-new-entry}) they decrease to a non-negligible score.
%This is the impact of 
The information inferred from newcomers \rev{seems to be} moderate, but constant in time.
This information is also responsible for the continuous increase of prediction performance detected in Fig.~\ref{fig:new-entry-vs-fixed} for the \texttt{NE} population.
\rev{Similar} conclusions can be drawn by studying the conditional entropy \rev{over time for the two populations}.
For \texttt{FP} (Fig.~\ref{subfig:info-theo-remaining-unexplained-entropy-evolution-fixed-population}) it decreases rapidly and remains constant afterwards, showing that virtually no new information is learned after the initial burst.
For \texttt{NE} (Fig.~\ref{subfig:info-theo-remaining-unexplained-entropy-evolution-new-entry}), it continues to decay even after the initial burst, though at a slower pace.

\secmoveup
\subsection{\rva{Predicting the attributes of exited users}} 

\begin{figure*}[tbp]
\centering
	\subfloat[] {
		\includegraphics[height=0.154\textheight]{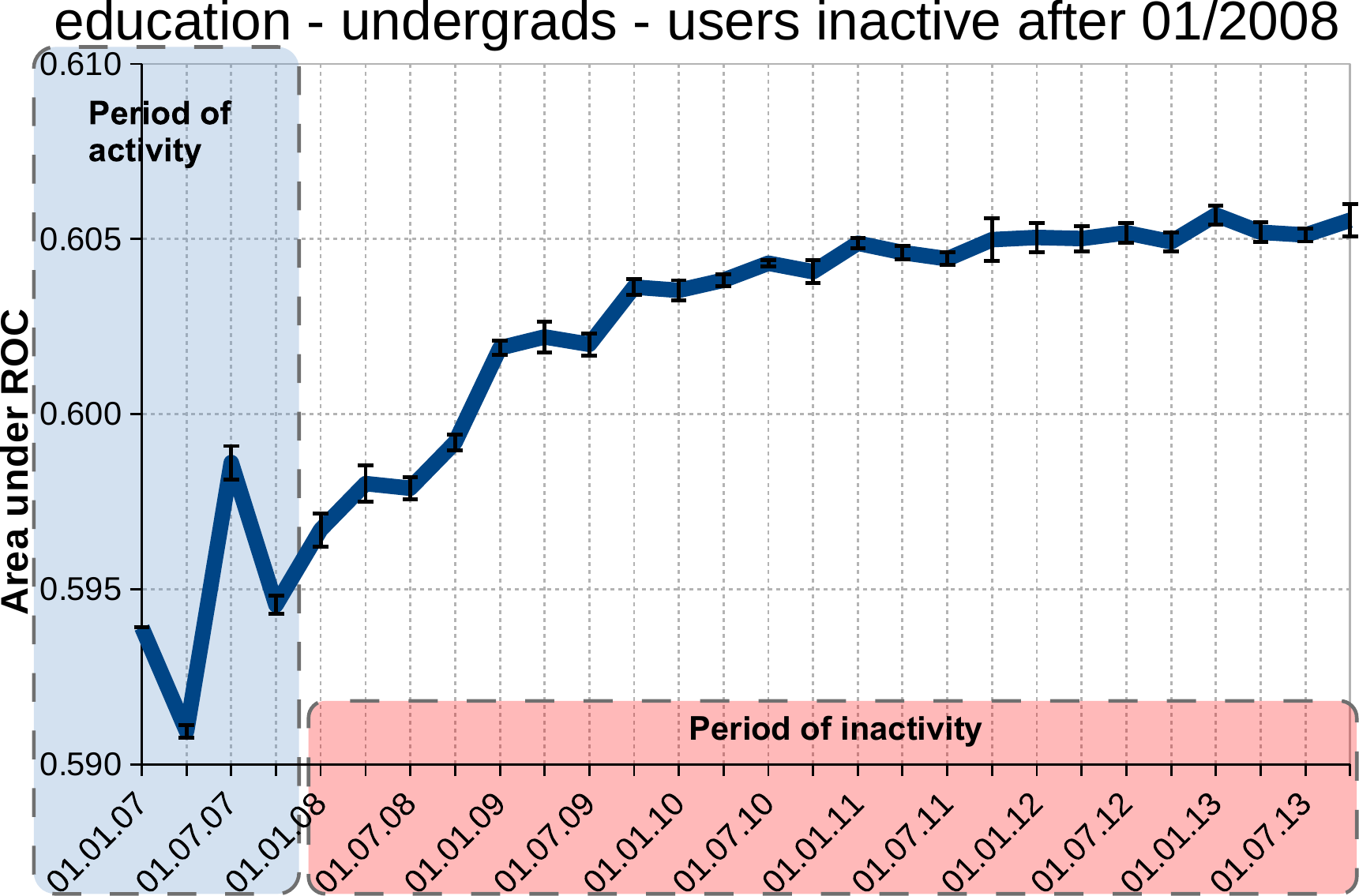}%
		\label{fig:privacy-users-stopped-editing-educations}
	}
	\hfill
	\subfloat[]{
		\includegraphics[height=0.154\textheight]{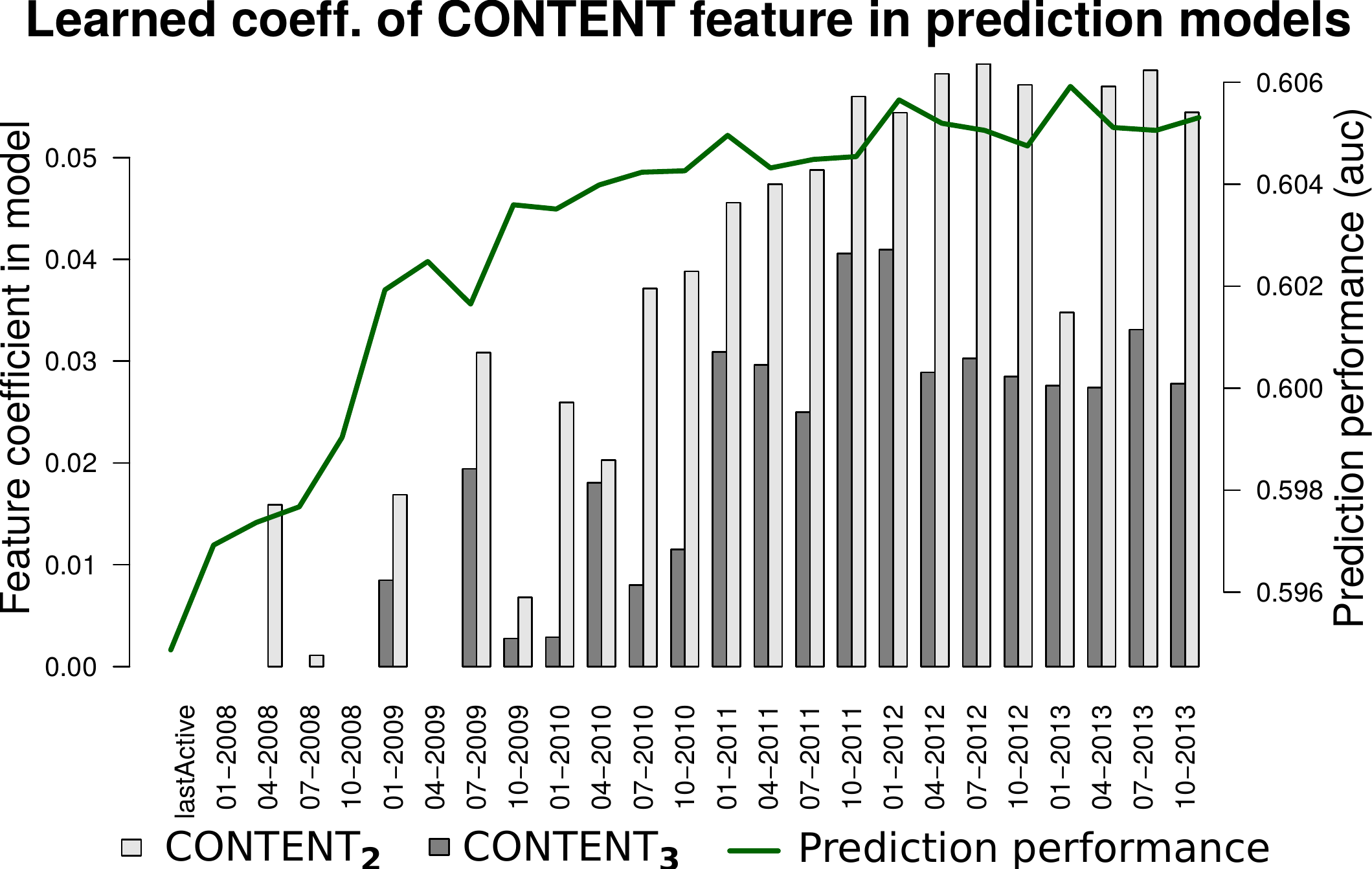}%
		\label{fig:privacy-users-stopped-editing-CONTENT-coeff}
	}
	\hfill
	\subfloat[]{
		\includegraphics[height=0.154\textheight]{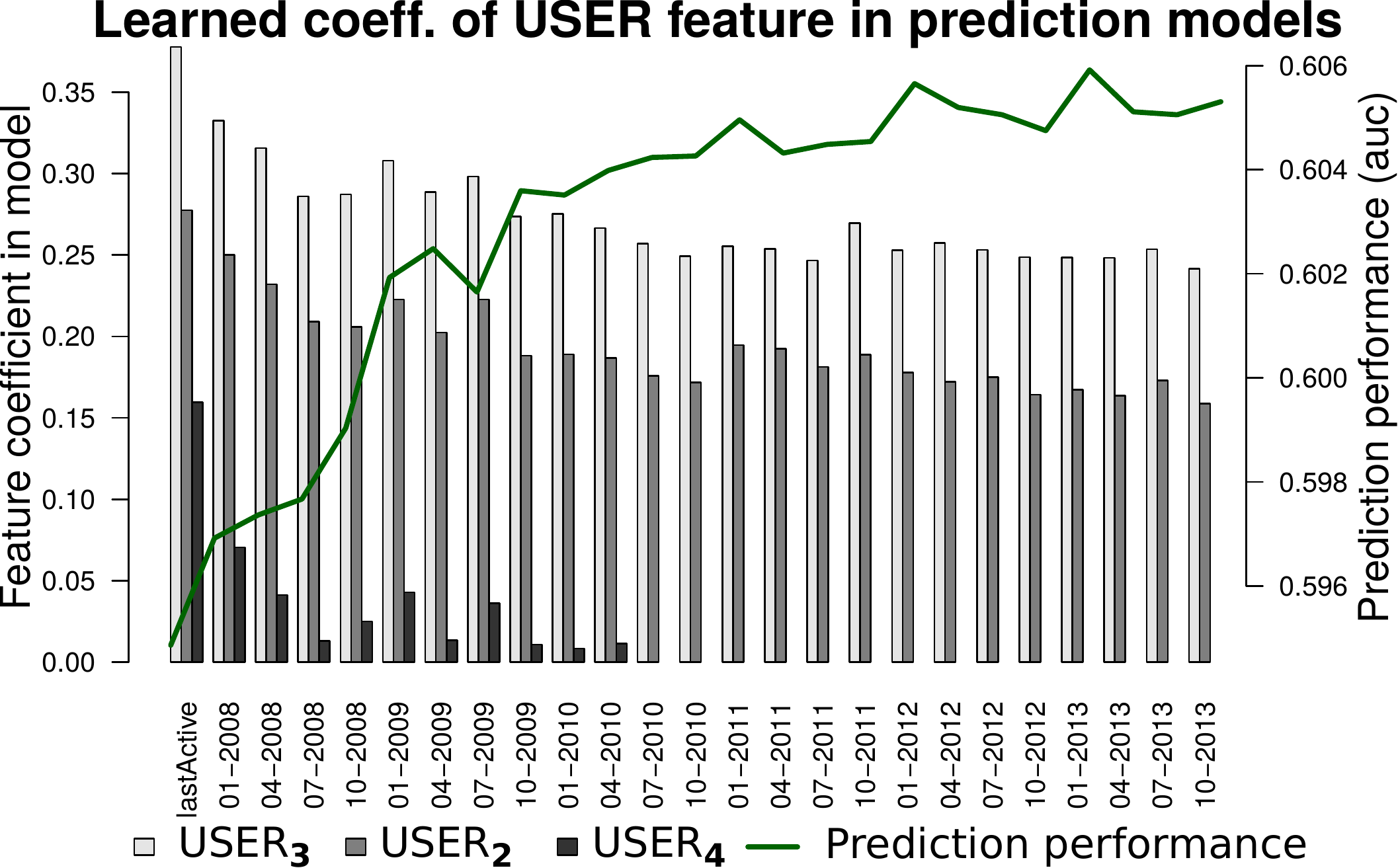}%
		\label{fig:privacy-users-stopped-editing-USER-coeff}
	}
	\caption{
	(a) Increase of prediction performance for for users retired after 01.2008 -- education/\textit{undergrads} \rev{(other in SI~\cite{supplemental})}. 
	Coefficients \rev{one model per timeframe}:
	(b) \texttt{CONTENT} coefficient (absent in the model corresponding to the dataset where users were last active) increase in importance.
	(c) \texttt{USER} relating features decrease in importance.}
	\captionmoveup
\end{figure*}

%\begin{figure}[tbp]
%	\centering
%	\includegraphics[height=0.135\textheight]{exited-users-religion-christians}%
%
%	\caption{The temporal privacy loss (religion/\textit{christian}) for users retired after 01.2008. Retired users have edited prior to 01.2008, but they did no more edits afterwards. The model is trained on all users having declared their religion and tested on the retired users.}
%	\label{fig:privacy-users-stopped-editing-religion}
%%	\vspace{-0.2in}
%\end{figure}

%\textcolor{ForestGreen}{[MC: SHOULD WE COME UP WITH AN ILLUSTRATIVE CARTOON FOR THE MANY WAYS IN WHICH WE EVALUATE PRIVACY-LOSS? I THINK THE READER CAN BE CONFUSED, EVEN IF IT IS WELL EXPLAINED IN THE TEXT]
%}

\textbf{Privacy continues to erode even for retired users.}
%We perform a follow-up study on 
\rev{We further analyze}
what happens to the privacy of users who left the system. 
After their retirement, no more user-originating information \rev{(``online breadcrumbs'')} is available to disclose private traits.
%In these circumstances, the question is 
%\rev{We seek to find} whether new information can be inferred about these users \rev{from newcomers only}.
We quantify the \rev{prediction performance} on a \rev{user population} who have edited prior to 01.01.2008, but stopped after this date.
Therefore, any information introduced in the system by the users themselves is restricted to the timeframes before 2008.
\rev{In Fig.~\ref{fig:privacy-users-stopped-editing-educations}, we plot the AUC over time for the education/\textit{undergrads} binary classifier.}
%We detect an increase of performance when predicting whether users are 
%\textit{christian} (Fig.~\ref{fig:privacy-users-stopped-editing-religion}) and if they are \textit{undergrads}.
We \rev{observe} an \rev{constant} increase of \rev{prediction} performance, \rev{even though it shows a \rva{saturation} in later timeframes}.
%The increase in prediction score seems to be gradual and constant.
%predicting religion seems to have a step around the end of 2010, the increase in predicting education 
\rev{An increase of prediction performance for for retired users is also observable for religion/\textit{christian} (see the SI~\cite{supplemental}), but not for any of the other binary classifiers.}
No longer being in activity, the loss of privacy after 01.2008 is not the result of \rev{the users'} actions.
A Temporal Information Theory analysis performed on this exited population shows \emph{Information Transfer} values of zero after 01.2008 -- \rev{i.e. no information originating with ``online breadcrumbs''}.
\rev{As far as we know, these are the first results to show that certain private traits could be predicted increasingly better even after the users exited the system.}

%\textbf{How the model learns to better predict the undergrads.}
\textbf{Why does privacy degrade for exited users?}
\rev{Intuitively, user originating information is available only until the exit of the users, afterwards} the source of new information \rva{are} \rev{in the actions of other users}.
\rev{Predictions about unseen retired users can be made only using} features relating to their period of activity (\rev{here} 01/2007 - 12/2007).
%Since any meaningful prediction that can be made for retired users can only use the features relating to their period of activity (01/2007 - 12/2007),
%, on some of the descriptive features
%We study in more detail the prediction user being \textit{undergrads} by
\rev{In the logistic regression models learned at each timeframe, the strength of the links between features and the class variable are given by the corresponding coefficients.}
\rev{We study} the coefficients of features \rev{which encode the activity of users prior to their retirement (i.e. $X^u_{1:4}$).}
% (which are the only ones useful for describing the activity of exited users).
\rev{We show the coefficients relating to \texttt{CONTENT} (in Fig.~\ref{fig:privacy-users-stopped-editing-CONTENT-coeff}) and to \texttt{USER} (in Fig.~\ref{fig:privacy-users-stopped-editing-USER-coeff}), in models learned for education/\textit{undergrads} in each timeframe.}
\rev{In later timeframes, \texttt{CONTENT} features observe an} increase in importance, \rev{with} both $\texttt{CONTENT}_2$ (number of \texttt{CONTENT} revisions \rev{in} the $2^{nd}$ quarter of 2007) and $\texttt{CONTENT}_3$ \rev{steadily increasing from being} completely absent in the initial models.
\rev{$\texttt{CONTENT}_4$ remains absent for all timeframes.}
Simultaneously, \texttt{USER} features decrease in importance, with $\texttt{USER}_4$ disappearing completely.
\rev{We hypothesize that} the AUC increase \rev{observed  after 01.2008} originates with the currently active users \emph{whose activity overlapped with the exited users}:
%This can be interpreted as follows: 
\rev{classifier} learns from users active both before and after 01.2008, by modifying the weights of features, including those before 01.2008. 
\rev{In this case,} they learn that \texttt{CONTENT} features should have more importance, while \texttt{USER} features should have less.
%Consequently, the prediction performance is \rev{increased even} for retired users, \rev{who are} only described by features prior to 01.2008.
\rva{In summary, predicting of personal traits increase even for retired users, and the key factor for this improvement is better estimates on a subset of important features such as \texttt{CONTENT}.}

%Any increase in the performance after this date would be the direct result of the information introduced in the system by the other users. 
%Fig.~\ref{fig:privacy-users-stopped-editing} shows the evolution of prediction performance for retired users. 
%Similar to the previous analysis, two scenarios are studied. 
%In the first one, we study a fixed population of users, which includes the 650 retired users. 
%No newcomers are allowed to enter. 
%In the second scenario, new users are allowed to enter. 
%As expected, 
%Starting from their retirement, the classifier cannot learn any new information, since these users do not introduce any new data in the system. 
%However, the \texttt{NE} population exhibits a burst of learning performance after 01.10.2012, whereas the fixed population shows only a flat line. 
%This result suggests that additional information was learned by the classifier from newly entered users. 
%This information was used, in turn, to improve the classification for retired users.

%!TEX root = WSDM_2015_wikipedia.tex

\secmoveup
\section{Discussion}
%% LX: overall: 
%% LX: the 1st and 2nd paragraph in the previous version are repetitive, merge+consolidate
%% LX: the discussions/bold headings needs to be more structured

\eat{ %%LX: remove leading sentence, get to the point directly
\eat{We already know}\rva{It is known} 
%%LX: not just "we" know, but everyone in the community does
that leaving information behind in the online environment erodes privacy, 
\st{but the \eat{speed and} extent at which this happens over long temporal 
scales has not been quantified.}
}
%LX: use a positive statement
\rva{We present a first study to quantify the extent of gradual privacy erosion over six years.
First, we set up a large scale evaluation using Wikipedia editing behavior to predict private traits over time.}
We analyze a 13 year \eat{long }history of Wikipedia \rva{edits made} by more than 117 \rev{thousand} users. 
\rva{Our descriptive analysis showed that as Wikipedia evolves, editors of different personal traits shows distinct patterns in their volume of edits and topical preferences.}
\rva{Second, we provide experimental evidence that time has an adverse effect on privacy. 
We show that prediction performance on private traits, such as gender, education and 
religion, increases with longer Wikipedia editing history.}
\rva{Third, we show that prediction of private traits improves even for users who have stopped 
editing Wikipedia. We further quantify the effect of predictability, and found that the improved performance can be attributed to two factors: new editors of Wikipedia, and better estimate of feature relevance - with the first having a larger effect.}
%%% -- significance statement
\rva{To the best of our knowledge, this is the first study to quantify the change of private traits over time, using Wikipedia, an open online dataset containing behavior breadcrumbs.} 
%The conclusions of our study quantify this privacy loss and show that it can increase over time, 
%even if a user's activity has ceased. 
\rva{This work shall raise awareness in the public
on the privacy implications of online activity over long periods of time.} The fact that the information of newcomers can be used to learn more about existing members has profound implications: users do not have complete control over the consequences of the information they release. 
\rva{Reflecting on this work, we would like to discuss a few of its limitations, practical implications and connections to other areas of research.}

\textbf{What does it really mean for Wikipedia users, should they be worried?} 
To the best of our knowledge, no studies have shown that the real identity of Wikipedia users can be revealed based on their editing activity. 
However, this may not the case for other social networks. % platforms. 

\rev{
\textbf{User disclosure bias.}
% One of the assumptions of the current work is that building models over the 
% revealed information of some Wikipedia editors is generalizable to other editors.
% This makes our framework sensitive to ground truth bias, like most classification-based methods.
\rva{This work uses self-disclosed personal traits on users' public profile. 
It is well-known that such a data source is prone to users' disclosure biase.} The set of users who voluntarily disclose private information might be biased towards users less concerned with their privacy, who in turn has a distinct behavioral pattern. \rva{Validating the effect of such a bias would require an alternative source of groundtruth and maybe even behavioral data, and is beyond the scope of this study.}
\eat{Using publicly revealed user information, we run the risk of \emph{user disclosure bias}.
 vocal minority groups etc.
These biases may affect the ability to generalize to the entire population the conclusions inferred on the ground truth population.
}}

\eat{ %%LX: this was said above, remove.
\textbf{The way ahead.}
While posing the framework for an empirical case study, this paper does not completely solve the problem of the evolution of privacy loss. 
A number of questions remain to be answered. 
}
\textbf{What about other online platforms?}
\rev{Although \eat{our findings have been reported for the specific case of }\rva{we predict predict private traits from }Wikipedia, \rva{similar prediction results (on a static snapshop) was reported for other platforms such as Facebook~\cite{KOS13}}. 
%we have no reason to believe why they couldn't potentially extend to other manifestations of online behavior.
}
Being a collaborative encyclopedia, Wikipedia \rev{records} \rva{ relatively small amount of} information about its users. More detailed longitudinal analyses could be performed on platforms such as Facebook, and it is likely to also see an increasing trend for predicting personal traits. % out to correlate more with user information.

%%LX: I'm not sure that this question is well-posed. consider remove? 
\textbf{A natural law of evolution of privacy loss.}
This study provides empirical analysis about \eat{the }privacy loss. 
%Starting from the data, we detected a privacy decay tendency, exemplified on features like gender, religious view and educational status.
%Our main target is to propose a physical model of the privacy loss, which could 
\rva{A longer-term open challenge is a physical model for privacy loss}, i.e., predict the de-anonymization rate of a given anonymized dataset.  

\textbf{Effective conditions for preserving privacy?} 
\rev{There is a growing literature on \eat{the problem of }characterizing which privacy guarantees can be obtained under a given information release protocol~\cite{WAS10}. 
We hope our findings invite new empirical and theoretical investigation into \eat{this problem in }the case in which data release is spatio-temporal and heterogeneous across different entities.}
In light of \eat{the conclusions of }this study, we advocate that new means should be found to tackle the issue of \rev{online privacy}.
We argue \eat{that another means of} one feasible means to preserving privacy is to \eat{push law-makers to} construct laws which would enable erasing the recorded activity in the online environment.

%\textcolor{ForestGreen}{[MC: GREAT END! AS IT CONNECTS VERY WELL WITH GOOGLE'S RIGHT TO BE FORGOTTEN]}

%ACKNOWLEDGMENTS are optional
\vspace{0.2cm}
{ \small
\textbf{Acknowledgments}
NICTA is funded by the Australian Government through the Department of Communications and the Australian Research Council through the ICT Centre of Excellence Program. 
This material is based on research sponsored by the Air Force Research Laboratory, under agreement number FA2386-15-1-4018. 
%The U.S. Government is authorized to reproduce and distribute reprints for Governmental purposes notwithstanding any copyright notation thereon.
%The views and conclusions contained herein are those of the authors and should not be interpreted as necessarily representing the official policies or endorsements, either expressed or implied, of the Air Force Research Laboratory or the U.S. Government.
}

{ \small
%% ANDREI NOTE: when updating ref list, you need to use bibtex.
%% otherwise, just use the compiled new-refs.bbl for column balancing.
%\bibliographystyle{abbrv}
%\bibliography{new-refs}  % sigproc.bib is the name of the Bibliography in this case

}
% ACM needs 'a single self-contained file'!
%

%\includepdf[pages=-]{../supplementary-materials/WSDM_2015_wikipedia-supplementary.pdf}


\section*{Supplementary material}

# *Supplementary Material*:
# Evolution of privacy loss in Wikipedia

**Marian-Andrei Rizoiu** [1,*], **Lexing Xie** [2], **Tiberio Caetano** [3] **and Manuel Cebrian** [4]

[1] *NICTA & Australian National University, Canberra, Australia*
[2] *Australian National University & NICTA, Canberra, Australia*
[3] *Ambiata, ANU and UNSW, Sydney, Australia*
[4] *NICTA & University of Melbourne, Melbourne, Australia*

Correspondence*:
Marian-Andrei Rizoiu
NICTA Research Lab, 7 London Circuit, Canberra, Australia,
Marian-Andrei.Rizoiu@nicta.com.au

We provide in this document detailed information about:

- the construction of the Wikipedia dataset used in the Main Text. Additionally, we provide links for downloading the user dataset (useful for reproducing the prediction results) as well as the complete edit dataset;
- additional data profiling figures, completing the narrative in the Main Text and provided for completeness;
- additional prediction performance measuring.

## CONTENTS

## 1  THE CONSTRUCTION OF THE WIKIPEDIA DATASET

Wikipedia uses internally a revision system, which records every edit or modification made to a page by a user. Such atomic operations are called "revisions" in Wikipedia's vocabulary. The version of a page at any given moment in time can be obtained by over-imposing all the revisions made to the page from the beginning of time until the given moment. The entire history of revisions is publicly available for download [1]. The user descriptive features introduced in the main article were constructed starting from the July 2013 English Wikipedia stub dump. This dump contains all the history of Wikipedia,

---

[1] Download Wikipedia dumps: `https://dumps.wikimedia.org/enwiki/latest/`

starting from its creation in January 2001 until July 2013. We record for each individual revision: its editor, target page and timestamp. For purposes of internal organization, Wikipedia page are assigned into namespaces, which are categories based the intended purposes of pages: main articles, talks around article, user pages, user talks, community pages, project pages *etc*. The basic descriptive features for users are constructed based on the Wikipedia namespaces of edited pages, as shown in the main article.

## 1.1 THE WIKIPEDIA CATEGORIZATION SYSTEM AND THE EXTENDED FEATURES.

We construct the extended descriptive features, based on the Wikipedia categories[2], which are a categorization system, based on the theme of the articles. This categorization system defines main categories, such as Geography, History, Arts, which can be further divided into finer subcategories. This forms a shallow hierarchy, *i.e.*, a hierarchy with a low number of levels. Loops are allowed inside the hierarchy – a given category can be a subcategory of multiple parent categories –, though discouraged, and the resulting category graph does not posses a strict tree structure. We select the 23 main categories of Wikipedia to serve as thematic features in a user's editing description. Each Wikipedia article can be placed under one or multiple (sub-)categories. Every time a user edits a page, we propagate the resulted revision through the category graph by following the category parent relation. The user's revision counts for all reachable main topics are incremented. We avoid the infinite propagation through the loops using a propagation threshold, equal to the average "height" of the hierarchy. While not a proper hierarchy, we define the "up" propagation direction as towards the main categories.

## 1.2 CONSTRUCTING THE EDITOR'S PRIVATE DATA: THE PRIVATE TRAITS.

User pages are similar to regular Wikipedia pages, with the difference that they are dedicated to users. Each user has, by default, a user page in Wikipedia, which serves the role of a "social profile" as in an online social network. The associated talk pages are employed for private discussions. The user pages and the user talk pages form Wikipedia's social aspect, which has been shown to be a prolific environment for activities such as campaign for internal elections (**Danescu-Niculescu-Mizil et al.**, 2012). All information that we use as private information is provided by some of the editors themselves, on their personal user pages. By adding labels to their user pages, editors give information about their geographic location, nationality, religion, profession, education level, philosophy or even sexual preferences. Similar to the aforementioned page categories, user categories are re-constructed bottom-up, based on the label information scrapped from the user pages using the Wikipedia API. As an example about the type of information that we can obtain about editors, we show the visualization of the first level child nodes of the categories providing information about the editor's location (Figure 2a), profession (Figure 2b) and religion (Figure 3). Some of the categories are further divided into subcategories, which provide increasingly fine-grained information. The three categories selected as private traits in the main article (*i.e.*, gender, religion and education) were chosen for being closer to what humans perceive as private information, as opposed to spoken languages or geographic location. We selected only a subset of subcategories out of all available subcategories (*e.g.*, *christian*, *muslim*, *atheist* and *jewish* for religion) in order to have a reasonable number of editors to perform the analysis on.

We restrict our dataset to only registered users for which at least one private information is retrieved from their user page. Summarizing, the dataset includes for each user:

- a description of the user's editing activity, recorded down to revisions on individual pages, together with their timestamp;

---

[2] Wikipedia categories descriptions: `https://en.wikipedia.org/wiki/Portal:Contents/Categories`

- some private information, retrieved from their own user pages, under the form of user categories.

The dataset contains:

- 188,805,088 revisions
- 117,523 users
- 8,679 user categories (and their hierarchical relations)
- 22,172,813 edited pages
- 430,410 page categories (and their hierarchical relations)
- extent of time: beginning of Wikipedia (January 2001) until July 2013.

## 1.3 DOWNLOADING THE WIKIPEDIA DATASET

We make available for the public to download the Wikipedia dataset used in this work. We provide two versions of the dataset: the *temporal user editing behavior dataset* and *complete editing dataset*.

**Temporal user editing behavior dataset**
**Download (∼82 MB):** `http://goo.gl/Tx5SoI`
**Description:** Dataset containing user activity over time and the three private traits used in this work: gender, education and religion.
**Usage:** This dataset is provided so that any third party can reproduce the experiments described in this work.
**Technical description:** This dataset captures the editing activity of Wikipedia users from 01.2002 until 07.2013. It is comprised of 47 CSV ("Comma Separated Values" format) files, one for each quarterly timeframe. Each file contains 117,523 lines (plus a header line), describing the editing activity of each of the users in our dataset, from the beginning of Wikipedia until the beginning of the timeframe corresponding to the given file. The columns in the CSV file (starting from the $6^{th}$ column onwards) correspond to the *basic* and *extended* features, described in the Main Text, Sec. 4.1. For example, column `AGRICULTURE` in file "2003_07_dataset.csv" gives, for each editor, the total number of revisions relating to `AGRICULTURE`, performed from the beginning of Wikipedia until June $31^{st}$ 2003. Columns `ALL` gives the total number of revisions performed by an editor. Columns `gender`, `education` and `religion` correspond to the private traits of the editors. The content of these columns does not vary between the different temporal datasets. Fields in these columns take values when information is known about the users, otherwise they take the value `NA` (not available).

**Complete editing dataset**
**Download:**
   **1% Sample (∼495 MB):** `http://goo.gl/T47UVj`
   **Complete (∼3.6 GB):** `http://goo.gl/2iLH7A`
**Description:** Dataset containing information about Wikipedia pages, categories, user, user categories and revisions, under a relational database format.
**Usage:** This dataset is provided to allow extensions to current work and/or to learn new insights for this data.
**Technical description:** The *complete editing dataset* is provided as a SQL relational database, with the table schema presented in Fig. 1. It contains three main tables:

- `user` – gives the names and total number of edits for each used in the dataset;
- `page` – provides information about the Wikipedia pages in the dataset (title, namespace, creation date *etc.*);

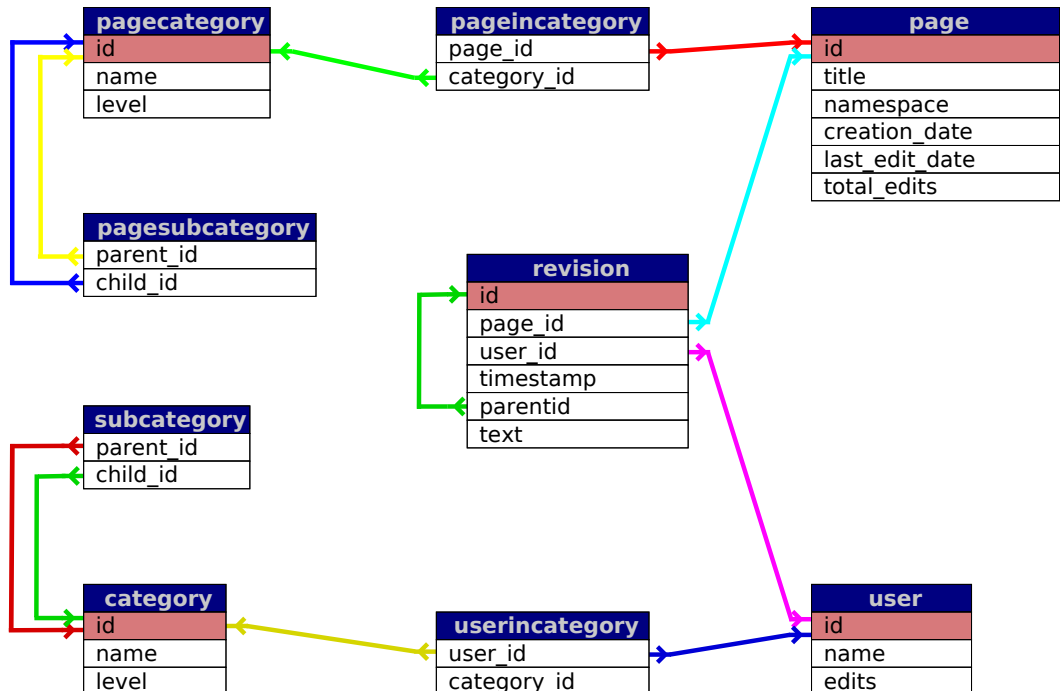

**Figure 1.** SQL table schema, showing the relations between the tables in the "complete editing dataset". Table names are shown on blue background, primary keys are shown on red background. Arrows indicate foreign key relations, with the direction from a given field to its corresponding foreign key.

- `revision` – records atomic edits (revisions) made on page *page_id* by user *user_id* at time *timestamp*. The field *text* is always empty since we work with the structure of Wikipedia only and not with the text. This field exists here to allow future enhancements. With more than 188 million revisions, this table is the bulk of the dataset. For easier handling and speed of prototyping, we provide the "1% Sample" dataset, which contains only 1% of the revisions in the "Complete" dataset (the content of all the other tables is identical).

The remaining 6 tables describe the page/user categories and their relations. Table `pagecategory` gives information about page categories, which are constructed based on the category system described in Sec. 1.1. Similarly, table `category` provides information about user categories (described in Sec. 1.2), which serve as private traits. Table `pageincategory` (`userincategory`) gives the n-to-m relations between pages (users) and their respective categories. Table `pagesubcategory` (`subcategory`) encodes the hierarchic relation between categories.

## 2  ADDITIONAL DATA PROFILING GRAPHS

As indicated in the main article, we present hereafter additional graphs showing:

- Fig. 4 and 5: The decline of editorship and the rise of maintenance, number of total revisions, new and Thematic features all present very similar evolutions to those presented here and in the main article.
- Evolution of editing patterns for gender (Fig. 6 and 7), education (Fig. 8 and 9) and religion (Fig. 10 and 11). Pairs (feature, private trait) for all the *basic* and *extended* features. The number of revisions performed by editors during each each quarterly timeframe on each feature are computed as percentages of the total number of revisions (`ALL`), and the mean value over all users is presented.

## 3  STATISTICAL TESTING OF INCREASING PRIVACY LOSS

In the Main Text Sec. 6.1, we show that Privacy Loss follows an increasing trend. The prediction performances, measured using the AUC ROC indicator **James et al.** (2013), obtained for later timeframes are higher compared to those of earlier timeframes. This evolution is shown in Main Text Fig. 6 and completed in SI Fig. 13. In order to rigorously test this increasing trend, we perform statistical testing analysis. For each timeframe, we train the Logistic Regression 20 times (cf. testing protocol described in Main Text Sec 4.2). We perform a *2-sample two-sided t test* between the AUC ROC results obtained for the first and the last timeframe, for each binary predictor. The null hypothesis states that the measure for both timeframes has the same mean value (i.e. no Privacy Loss), whereas the alternative hypothesis states that the mean values for the two timeframes are different. Table 1 shows the p-values of the performed t tests, as well as the significance level: * significant ($p < 0.05$), ** very significant ($p < 0.01$) and *** highly significant ($p < 0.001$). With the exception of religion/*jewish*, the increase of prediction performance for all binary predictors is statistically significant. Furthermore, we study how significant is the increase of prediction performance of the Fixed Population (`FP`) described in Main Text Sec. 6.1. The evolution of the AUC ROC measure, depicted in Main Text Fig. 8, is increasing in early timeframes and plateaus afterwards. By performing statistic testing, we show in Table 1 that even the loss of privacy due to the users' own behavior is statistically highly significant — line "gender `FP`".

**Table 1.** Hypothesis testing of the statistical significance of the increase of learning performance. For each trained predictor, we show the p-value and the significance level.

|          | class         | p-value                  | Significance level |
|----------|---------------|--------------------------|--------------------|
|          | gender        | $3.219 \times 10^{-14}$  | ***                |
|          | gender `FP`   | $1.262 \times 10^{-4}$   | ***                |
| educat.  | *undergrads*  | $5.882 \times 10^{-13}$  | ***                |
|          | *grads*       | $2.016 \times 10^{-6}$   | ***                |
|          | *PhD*         | $1.496 \times 10^{-2}$   | *                  |
| religion | *atheist*     | $1.018 \times 10^{-3}$   | **                 |
|          | *christian*   | $6.916 \times 10^{-12}$  | ***                |
|          | *jewish*      | $1.064 \times 10^{-1}$   |                    |
|          | *muslim*      | $4.055 \times 10^{-12}$  | ***                |

# 4 COMPUTING OTHER PERFORMANCE MEASURES: FSCORE AND "EQUAL PRECISION AND RECALL"

In statistics, the receiver operating characteristic (ROC curve) is a graphical plot that illustrates the performance of a binary classifier system, while varying its discrimination threshold. When using normalized units, the area under the curve (the AUC measure) is equal to the probability that a classifier will rank a randomly chosen positive instance higher than a randomly chosen negative one (assuming 'positive' ranks higher than 'negative') (**Fawcett**, 2006). One of the advantages of the AUC measure, used in the Main Text to evaluate learning performances over each private trait, is that it does not require setting a decision threshold. Other measures, such as Accuracy, Precision, Recall or F-Measure, require establishing a decision boundary in order to decide if a given example belongs (or not) to a given class.

We devise the following protocol: for the predictions of a given model, we perform a line search for the decision threshold. For F-Measure, we search for the point which maximizes the measure. For the *point of equal Precision and Recall* (ePR), we choose the boundary for which Precision equals Recall. This measure is also known as the Precision-recall breakeven point. Once a boundary has been chosen, the given prediction is scored using the value of the measure at that point (i.e. the F-Measure or Precision/Recall). According to the testing protocol described in the Main Text Sec 4.2, each classifier is trained 20 times. We calculate and report the mean value and the standard deviation.

Table 2 presents these statistics, computed for the latest available snapshot (July 2013), alongside with each class prior (for gender, we present the prior of *male*). Three traits seem to be inferred particularly accurate: gender, religion/*jewish* and religion/*muslim*. The evolution over time of the equal Precision/Recall of each attribute is presented in Fig. 12. Two attributes present a noteworthy evolution (clearly outside the standard deviation denoted by error bars): gender can be predicted increasingly accurate overtime, while religion/*muslim* decreases slightly, probably due to the increase of active self-declared muslim editors.

**Table 2.** Performance of prediction of each private traits in the last timeframe (10/2013). Mean and standard deviation (in *italic fontface*) of F-Measure and "point of equal Precision and Recall" are calculated over 20 random stratified splits (each class prior is given in the "Class prior" column).

|  | class | Class prior | F-Measure | | ePR | |
|---|---|---|---|---|---|---|
|  | gender (*m*) | 0.719 | 0.840 | *±0.001* | 0.791 | *±0.006* |
| educat. | *undergrads* | 0.420 | 0.605 | *±0.007* | 0.535 | *±0.012* |
|  | *grads* | 0.493 | 0.701 | *±0.001* | 0.607 | *±0.008* |
|  | *PhD* | 0.086 | 0.239 | *±0.018* | 0.175 | *±0.030* |
| religion | *atheist* | 0.324 | 0.796 | *±0.001* | 0.693 | *±0.008* |
|  | *christian* | 0.451 | 0.704 | *±0.000* | 0.555 | *±0.008* |
|  | *jewish* | 0.107 | 0.952 | *±0.000* | 0.913 | *±0.002* |
|  | *muslim* | 0.116 | 0.937 | *±0.000* | 0.900 | *±0.002* |

## 5 ADDITIONAL PREDICTION PERFORMANCE GRAPHS

- Figure 13: Editing Wikipedia discloses private information. AUC evolution on the remaining classes of private traits, apart from those presented in the Main Text.
- Figure 14: In addition to education/*undergrads* information revealed about editors retired after 01.2008 (shown in Main Text Fig. 10a), we present also the increase of prediction performance for religion/*christian*. No other binary classifier presents an increase of AUC score over time.

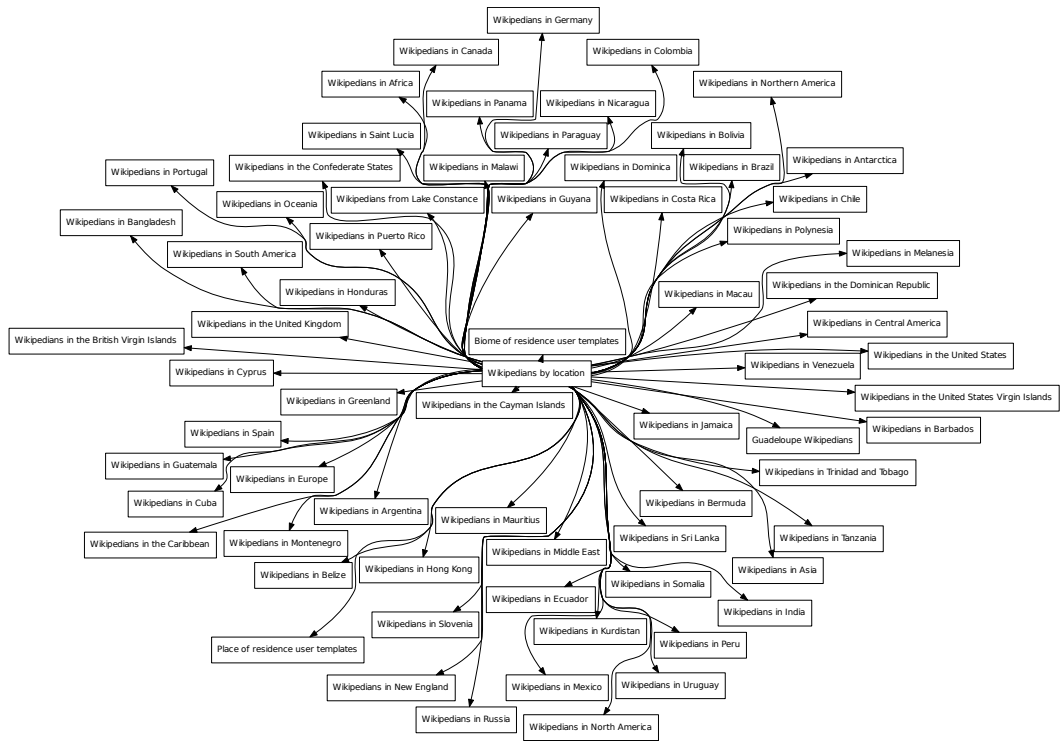

(a)

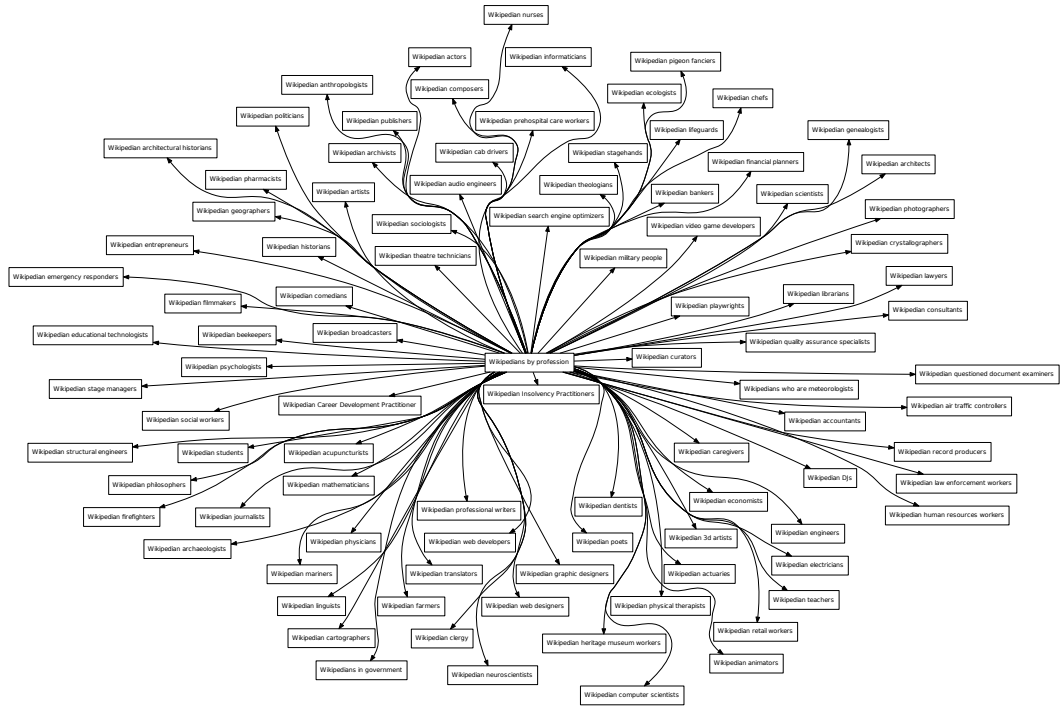

(b)

**Figure 2.** Visual representation of first level user categories that can be extracted from the Wikipedia user pages, relating to geographic location (a) and profession (b). Some categories are further subdivided into subcategories (not shown in this figure)

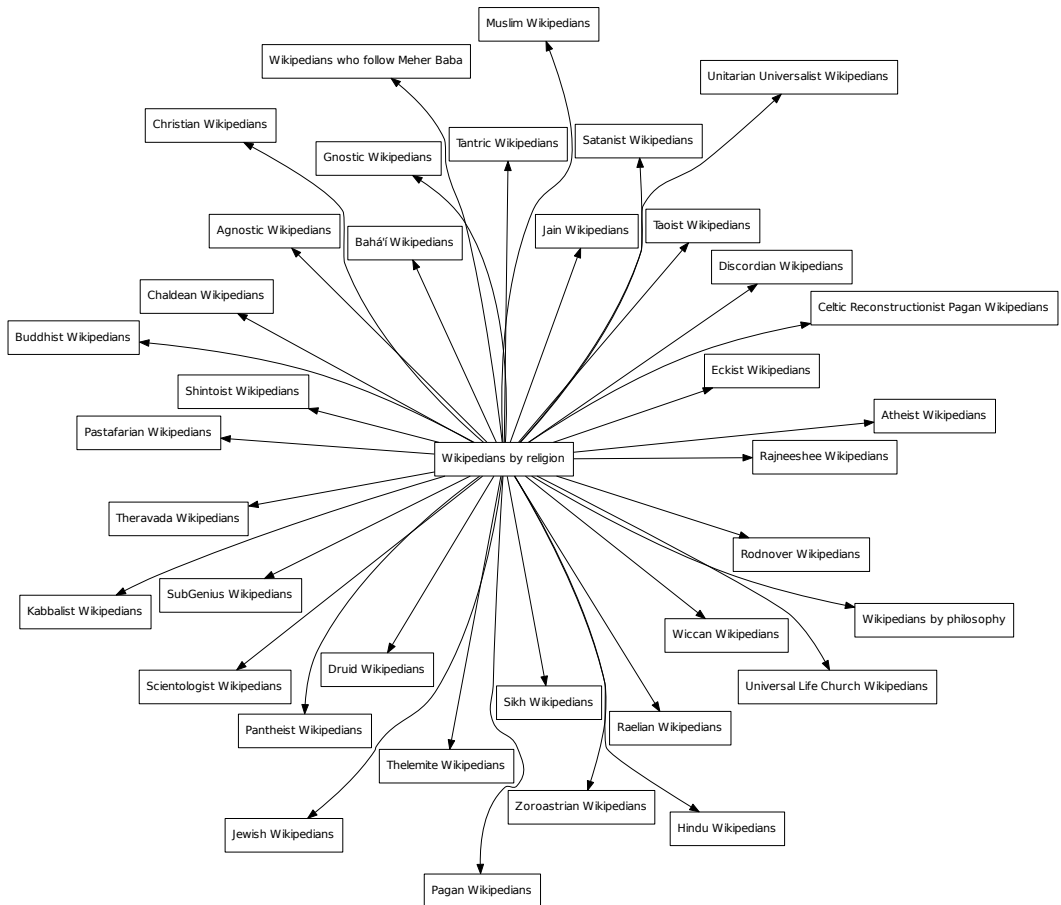

**Figure 3.** (cont. of Figure 2) Visual representation of user categories that can be extracted from the Wikipedia user pages, relating to editors' religion.

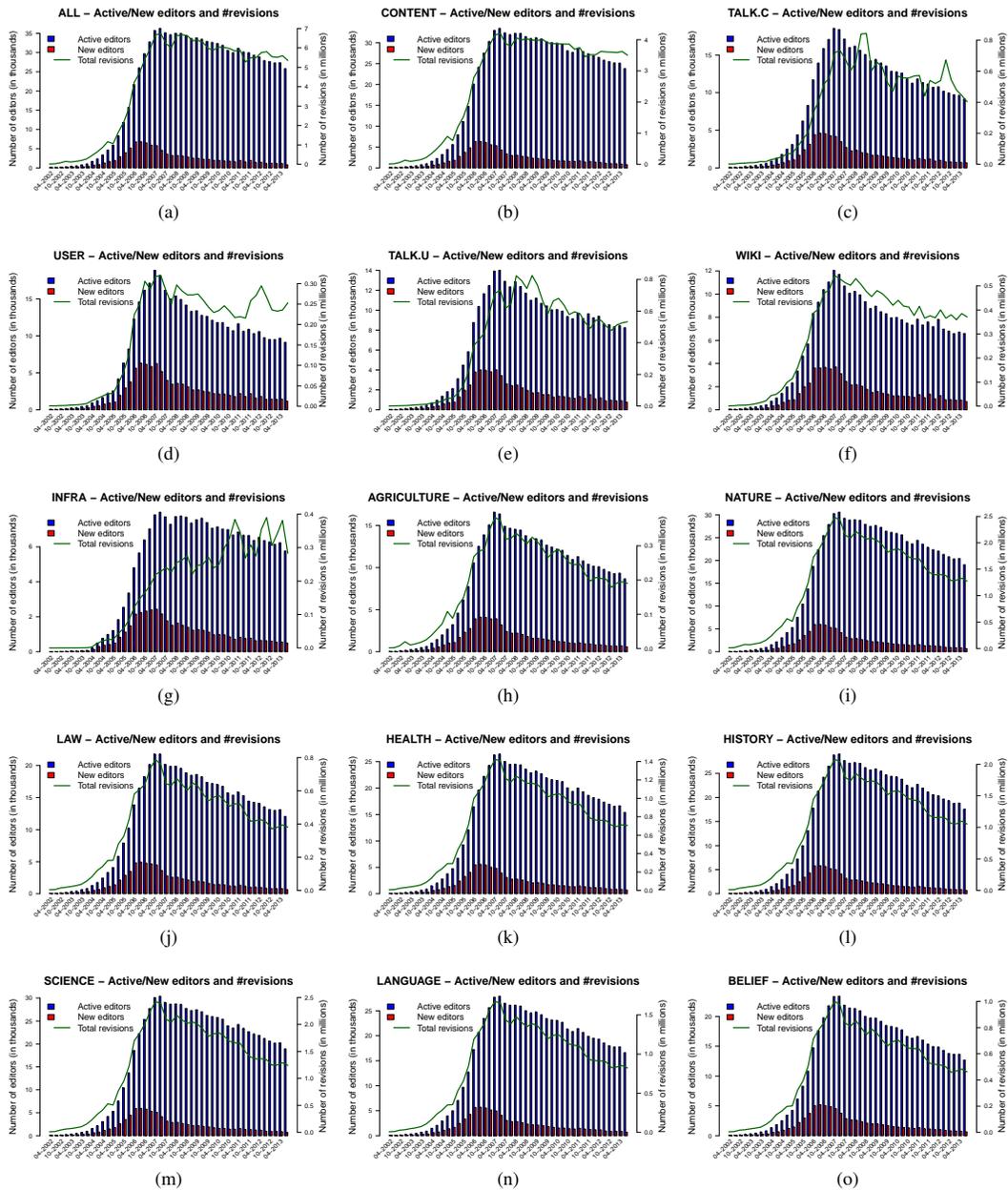

**Figure 4.** The number of revisions added to Wikipedia is constantly descending, as well as the number of active editors and new editors. Showing the evolution, broken down by feature.

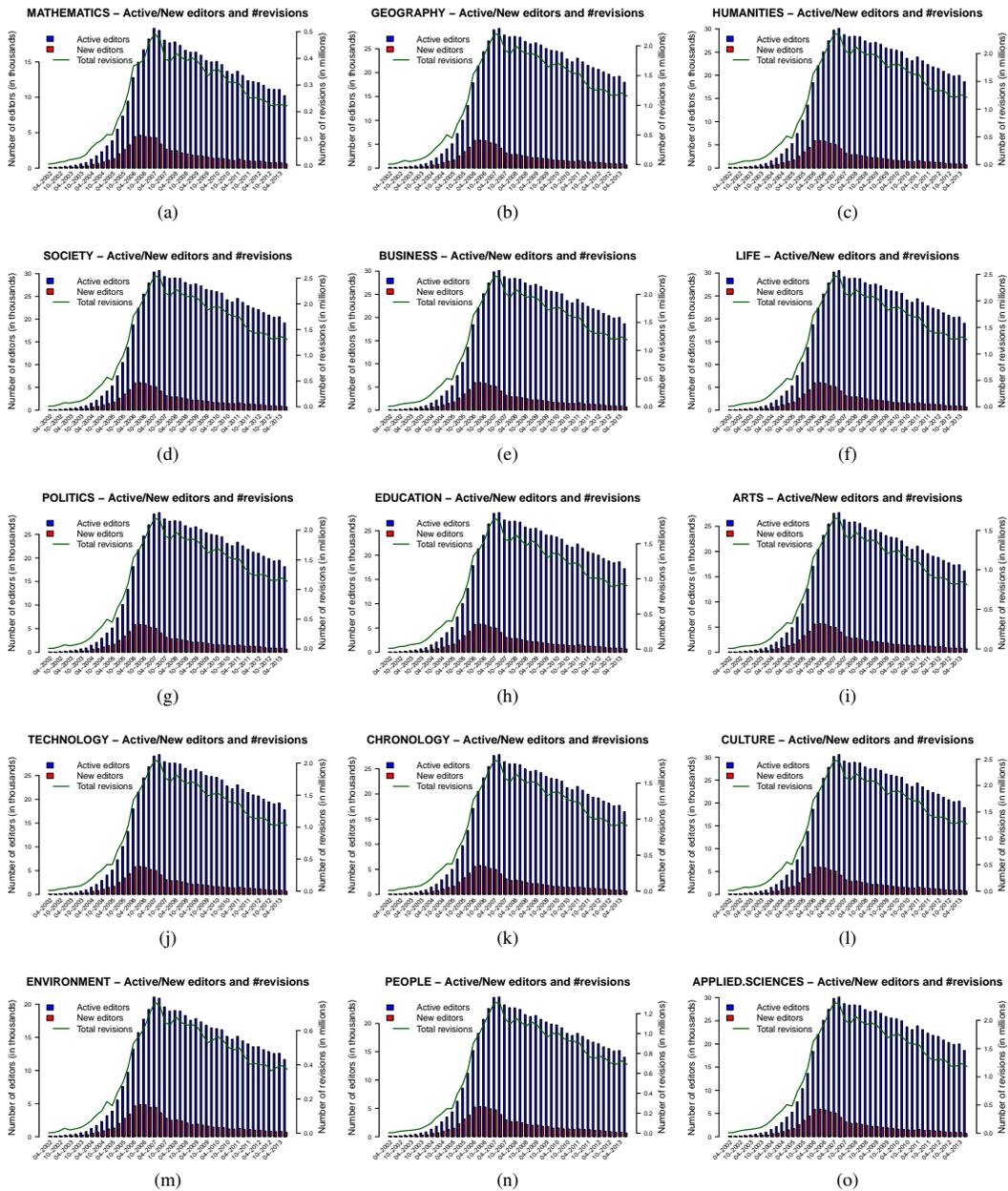

**Figure 5.** (cont. of Fig. 4) The number of revisions added to Wikipedia is constantly descending, as well as the number of active editors and new editors. Showing the evolution, broken down by feature.

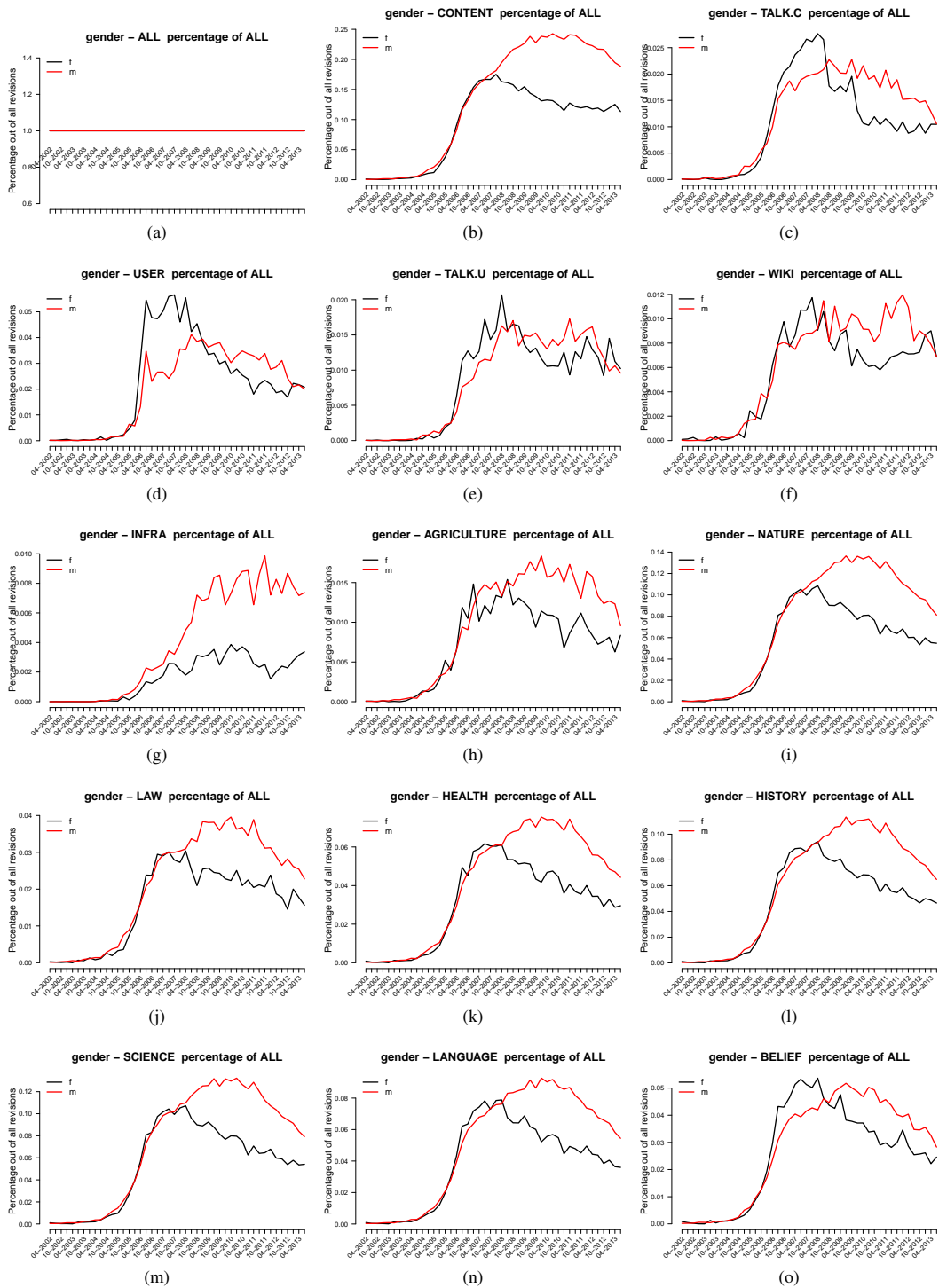

**Figure 6.** Temporal evolution of mean *basic* and *extended* values of features, broken down by gender. All features are expressed as percentages of `ALL`.

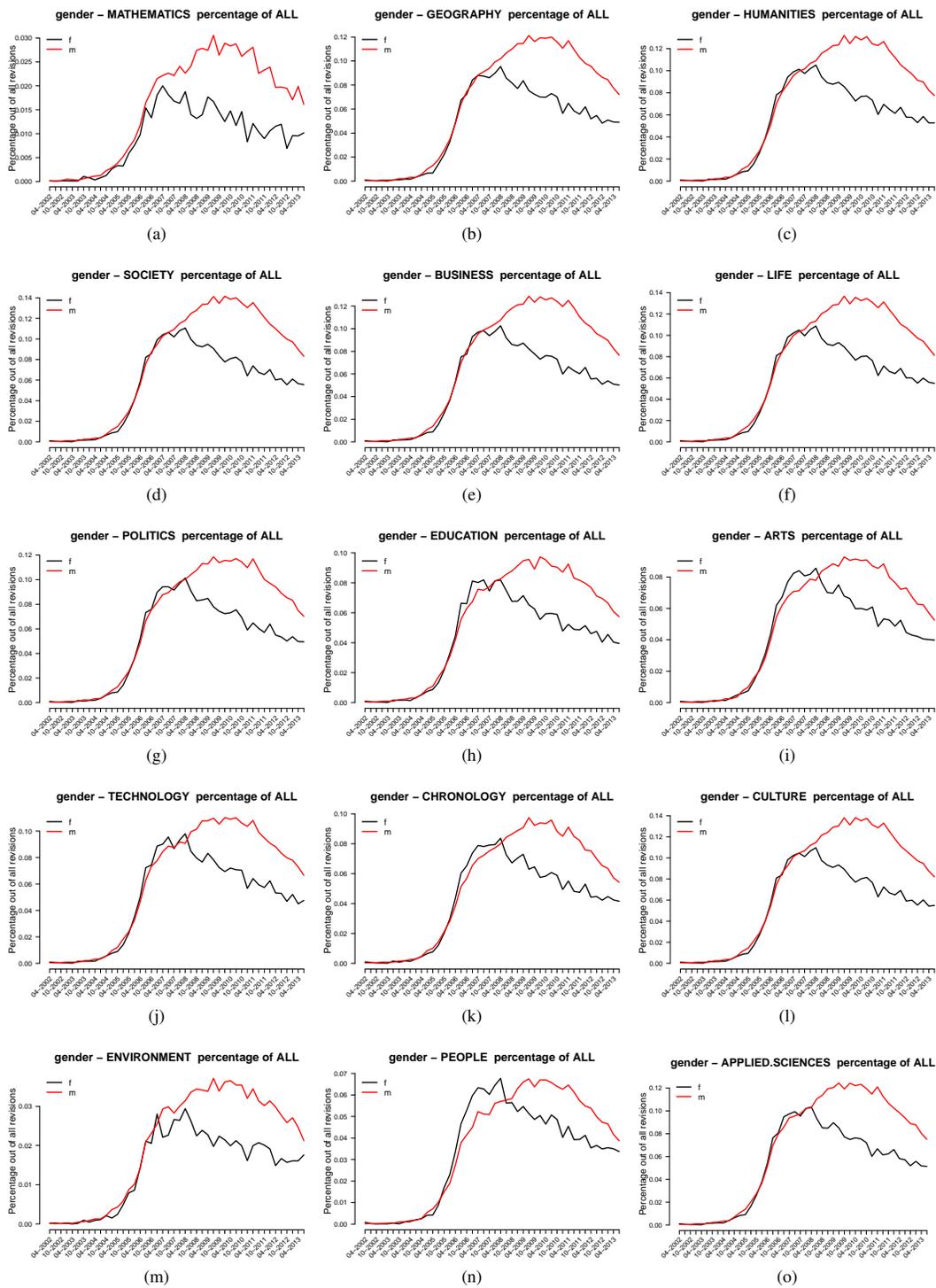

**Figure 7.** (cont. Fig 6) Temporal evolution of mean *basic* and *extended* values of features, broken down by gender. All features are expressed as percentages of `ALL`.

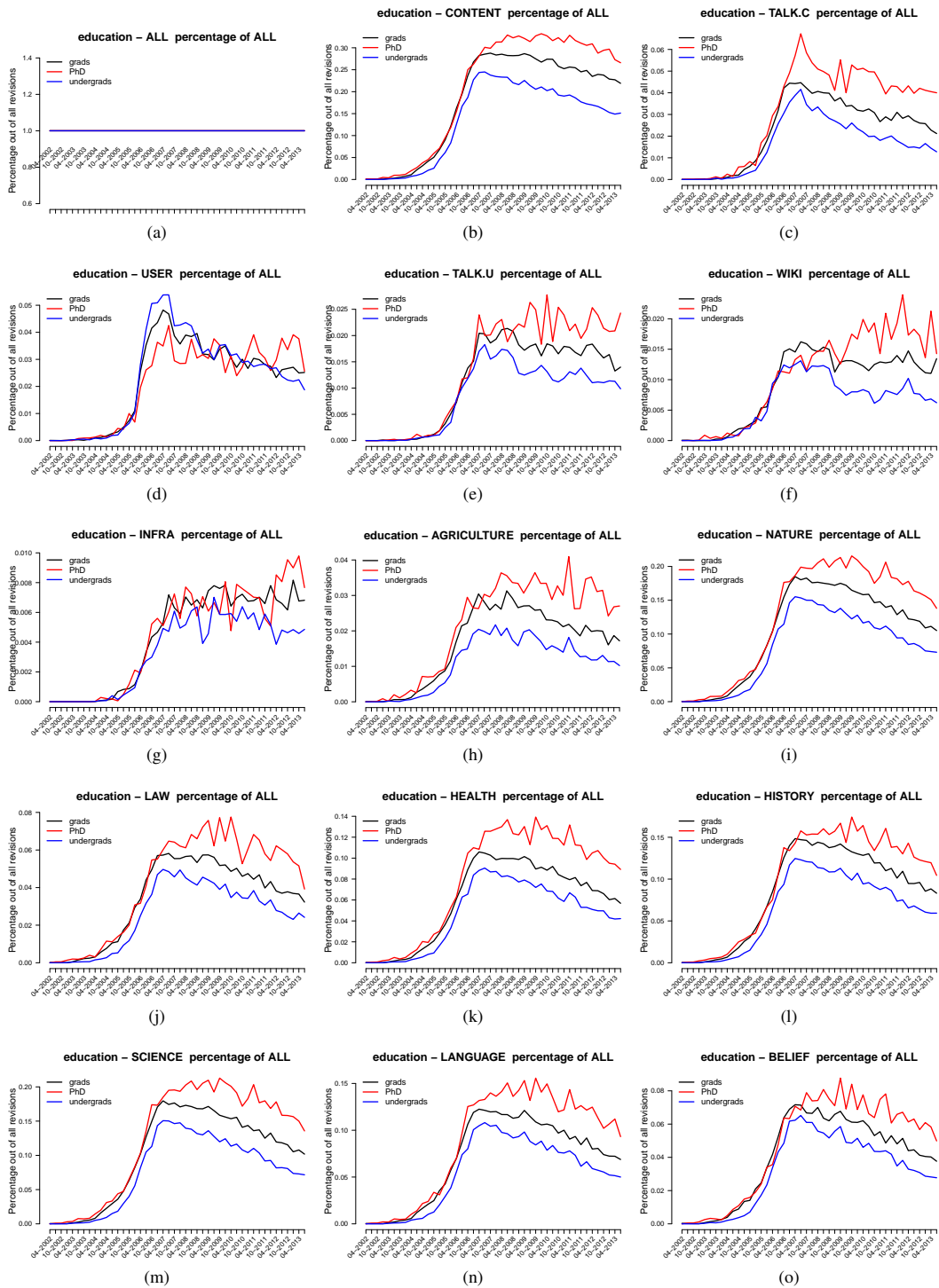

**Figure 8.** Temporal evolution of mean *basic* and *extended* values of features, broken down by **education**. All features are expressed as percentages of `ALL`.

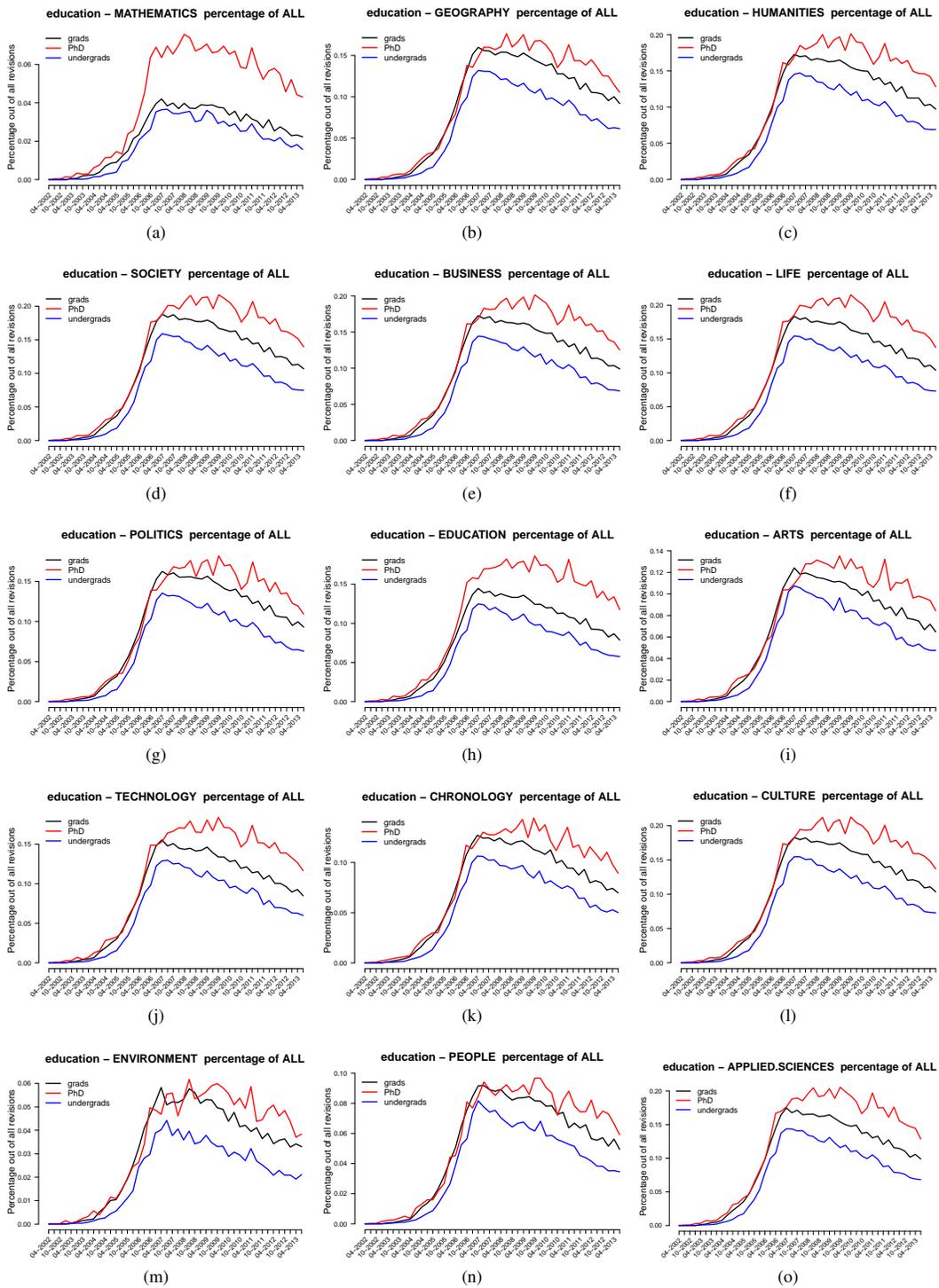

**Figure 9.** (cont. Fig 8) Temporal evolution of mean *basic* and *extended* values of features, broken down by **education**. All features are expressed as percentages of `ALL`.

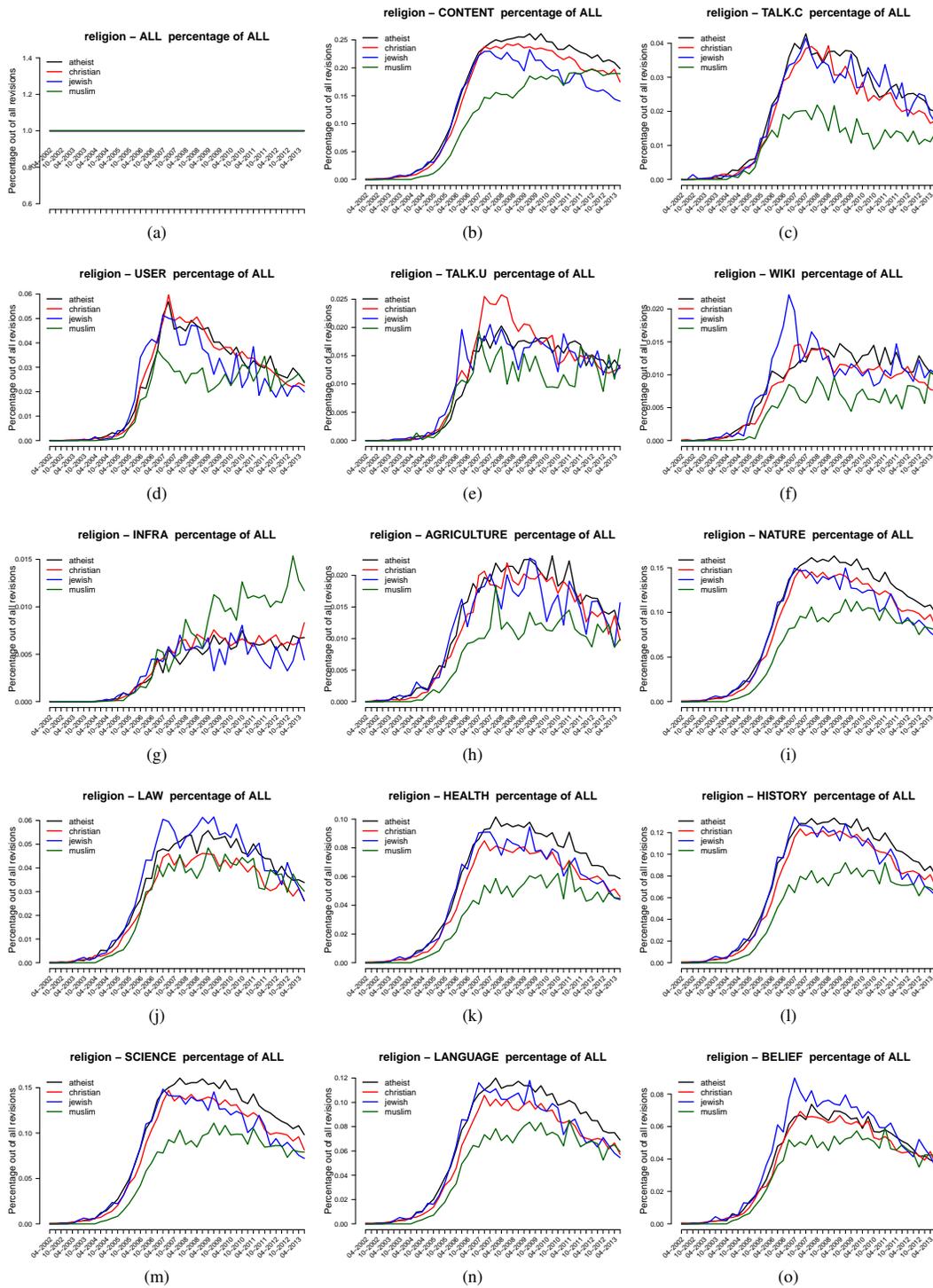

**Figure 10.** Temporal evolution of mean *basic* and *extended* values of features, broken down by **religion**. All features are expressed as percentages of `ALL`.

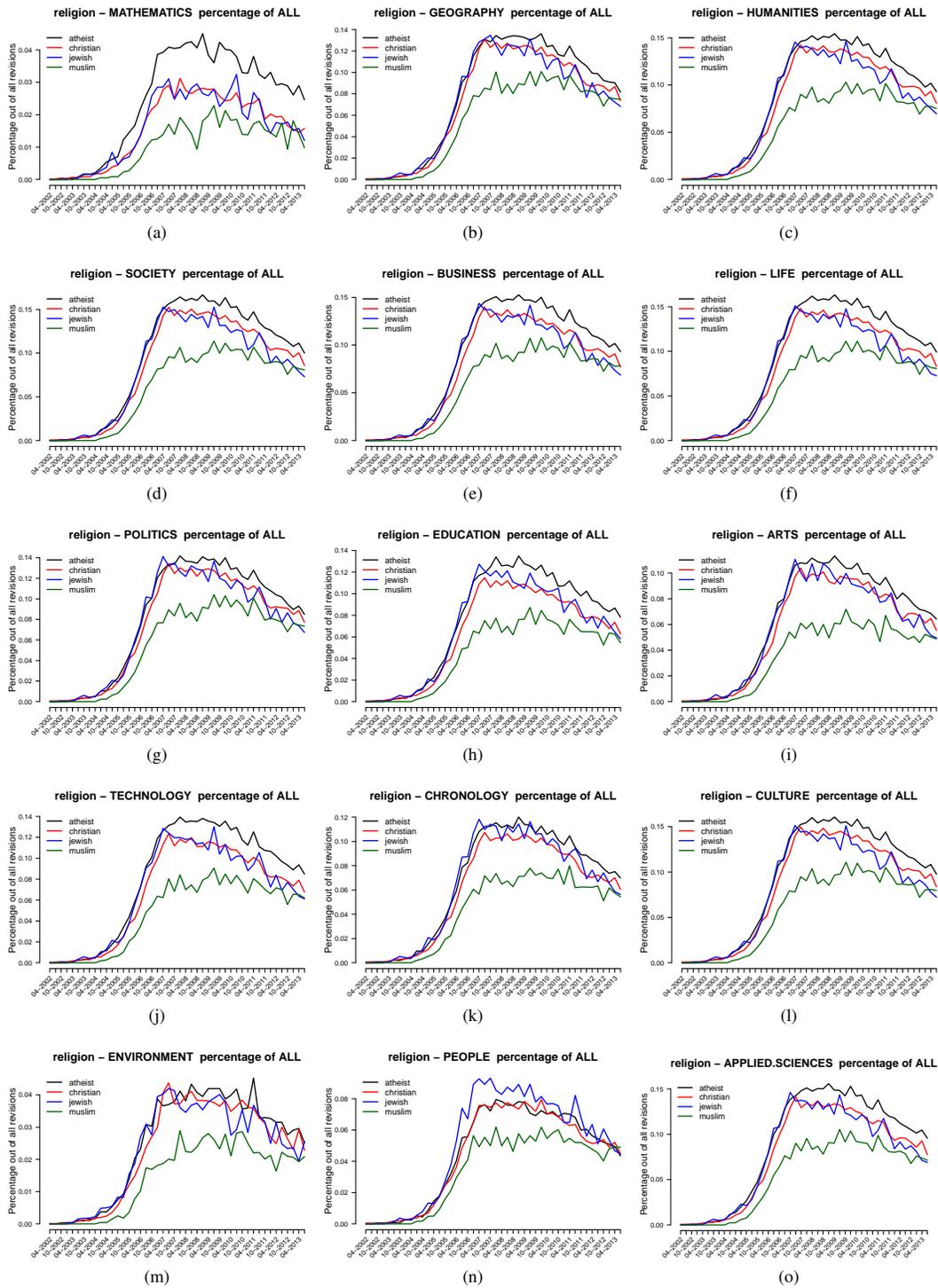

**Figure 11.** (cont. Fig 10) Temporal evolution of mean *basic* and *extended* values of features, broken down by **religion**. All features are expressed as percentages of `ALL`.

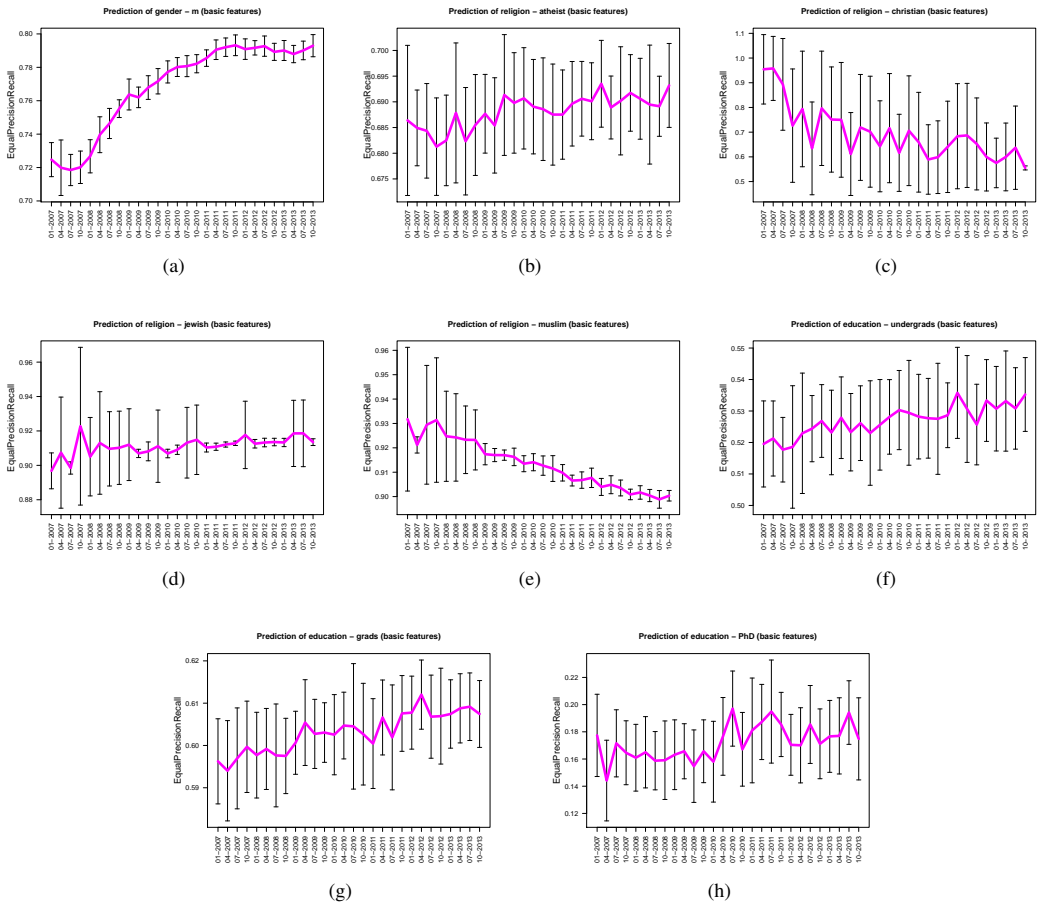

**Figure 12.** Evolution of prediction over time, measure using the *Equal Precision Recall point*. The lines represents the mean value and the error bars represent the standard deviation over 20 random split. In order from (a) to (h), the graphics present respectively gender, religion/*atheist*, religion/*christian*, religion/*jewish*, religion/*muslim*, education/*undergrads*, education/*grads* and education/*PhD*. The values for the rightmost timeframe are given in Table 2.

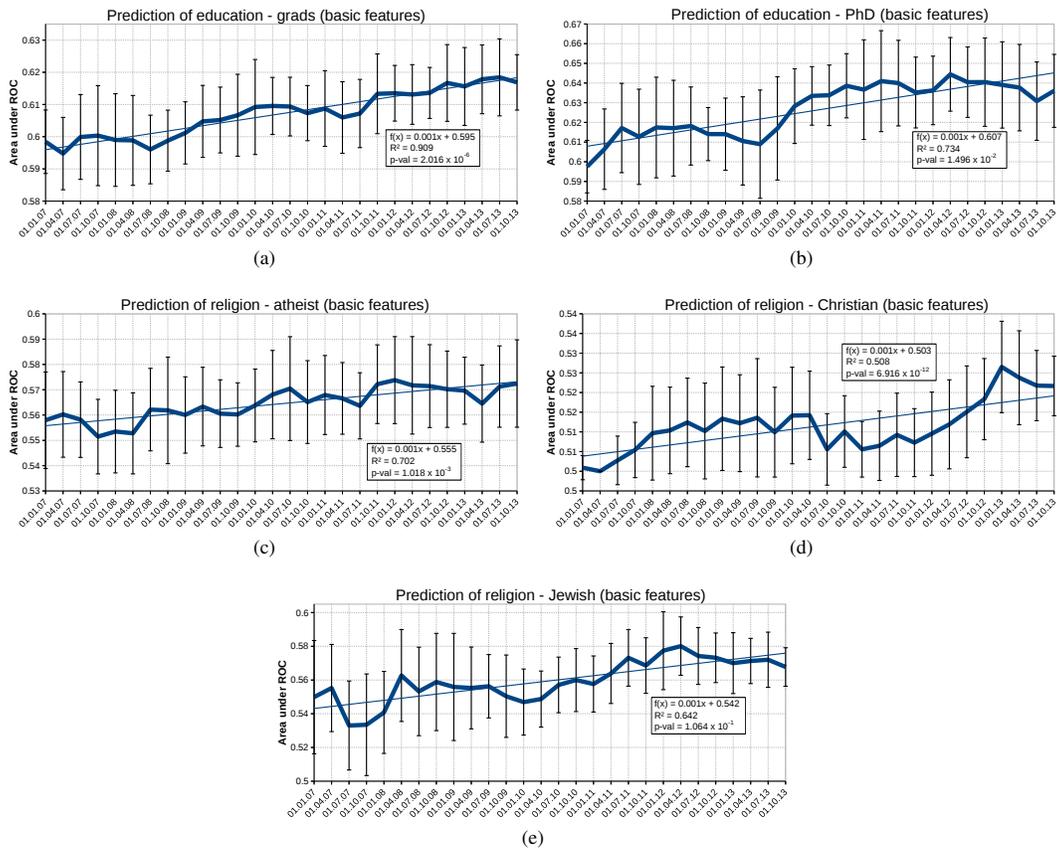

**Figure 13.** Results on the remaining binary predictors: Temporal evolution of the AUC, measuring privacy loss. Result of inferring, using binary predictors on the basic feature set, of education (*graduates* and *PhD*) and religion (*atheist*, *christian* and *jewish*).

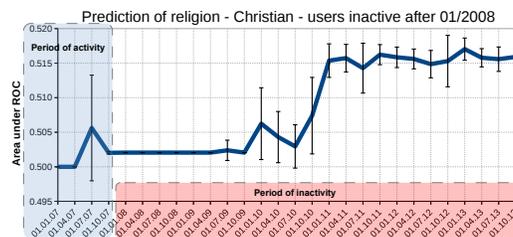

**Figure 14.** Increase of prediction performance for for users retired after 01.2008 – religion/*christian*.



\end{document}